\documentclass[a4paper]{article}
\usepackage[a4paper,top=2.5cm, bottom=2.5cm, left=2.5cm, right=2.5cm]{geometry}

\usepackage[T1]{fontenc}
\usepackage[utf8]{inputenc}
\usepackage{array}
\usepackage{multirow}
\usepackage{float}
\usepackage{url}
\usepackage{amsthm}
\usepackage{amsmath}
\usepackage{amssymb}
\usepackage{graphicx}
\usepackage{color}
\usepackage[english]{babel}
\usepackage{dsfont} 
\usepackage{algpseudocode}
\usepackage{algorithm}

\usepackage{chngcntr}
\counterwithin{paragraph}{section}
\counterwithin{paragraph}{subsection}

\newcommand{\G}{\mathcal{G}} 
\newcommand{\E}{\mathcal{E}} 

\newcommand{\Larg}{L} 
\newcommand{\V}{\mathcal{V}} 
\newcommand{\W}{W} 
\newcommand{\D}{D} 
\newcommand{\x}{u} 
\renewcommand{\b}[1]{{\bf #1}} 
\newcommand{\T}{\mathcal{T}}

\renewcommand{\O}{\mathcal{O}}

\newcommand{\lmax}{\lambda_{\rm max}}

\let\oldll\ll
\renewcommand{\ll}{\lambda_\ell}

\newcommand{\Rbb}{\mathbb{R}} 

\newcommand{\Esp}{\mathbb{E}} 

\newcommand{\scp}[2]{\langle #1, #2 \rangle}

\DeclareMathOperator*{\argmin}{arg\,min}
\DeclareMathOperator*{\argmax}{arg\,max}

\newtheorem{theorem}{Theorem} \newtheorem{definition}{Definition}

\newtheorem{example}{Example}

\providecommand{\keywords}[1]{\textbf{\textit{Index terms---}} #1}
\begin{document}

\title{Stationary signal processing on graphs}

\author{Nathana\"el Perraudin and Pierre Vandergheynst 
\thanks{
EPFL, Ecole Polytechnique Fédérale de Lausanne,
LTS2 Laboratoire de traitement du signal, CH-1015 Lausanne, Switzerland}
}



\maketitle

\begin{abstract} 
Graphs are a central tool in machine learning and information processing as they allow to conveniently capture the structure of complex datasets. In this context, it is of high importance to develop flexible models of signals defined over graphs or networks. 
In this paper, we generalize the traditional concept of wide sense stationarity to signals defined over the vertices of arbitrary weighted undirected graphs. We show that stationarity is expressed through the graph localization operator reminiscent of translation. We prove that stationary graph signals are characterized by a well-defined Power Spectral Density that can be efficiently estimated even for large graphs. We leverage this new concept to derive Wiener-type estimation procedures of noisy and partially observed signals and illustrate the performance of this new model for denoising and regression. 
\end{abstract}


\section{Introduction}

\begin{figure}[htb!]
\begin{center}
\includegraphics[width=0.7\linewidth]{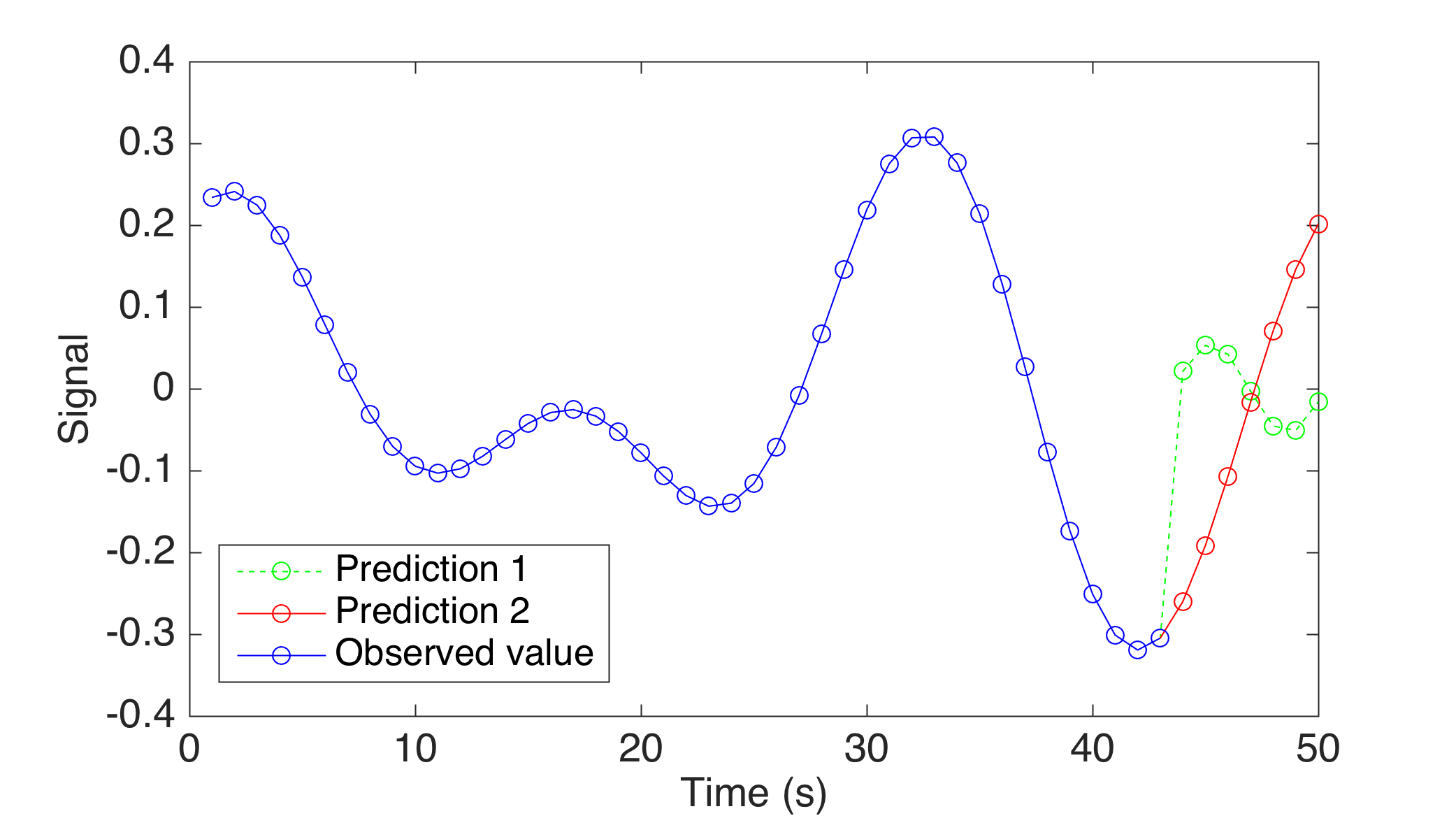} 
\end{center}
\caption{Signal prediction. The red curve is more likely to occur than the green curve because it respects the frequency statistics of the blue curve.}
\label{fig:intro_stationarity_time}
\end{figure}

Stationarity
 is a traditional hypothesis in signal processing used to represent a special type of statistical relationship between samples of a temporal signal. The most commonly used is wide-sense stationarity, which assumes that the first two statistical moments are invariant under translation, or equivalently that the correlation between two samples depends only on their time difference. 
Stationarity is a corner stone of many signal analysis methods. The expected frequency content of stationary signals, called Power Spectral Density (PSD), provides an essential source of information used to build signal models, generate realistic surrogate data or perform predictions. 
In Figure~\ref{fig:intro_stationarity_time}, we present an example of a stationary process (blue curve) and two predictions (red and green curves).
As the blue signal is a realization of a stationary process, the red curve is more probable than the green one because it respects the frequency content of the observed signal.  

\begin{figure}[htb!]
\begin{center}
\includegraphics[width=0.45\linewidth]{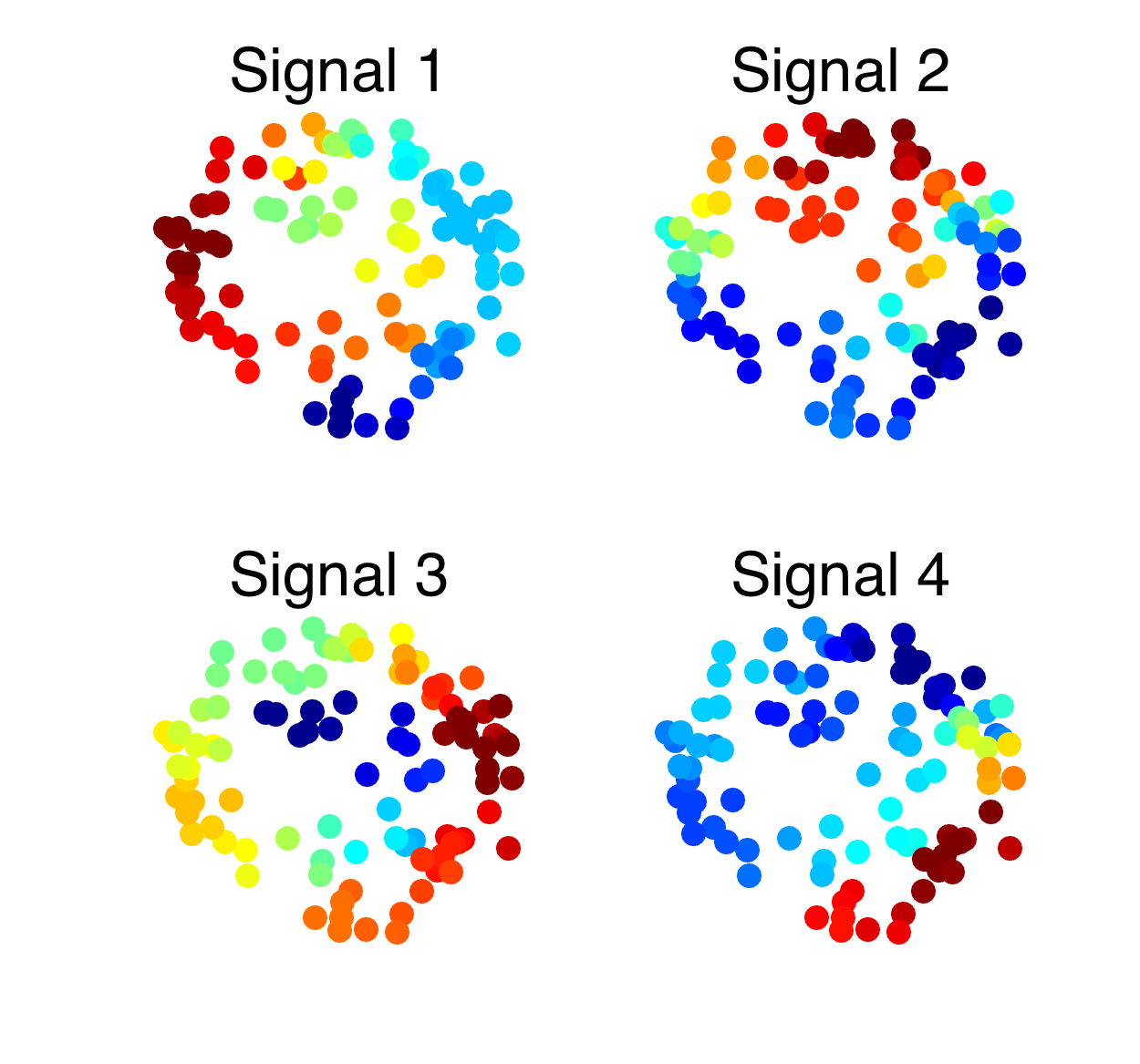} 
\includegraphics[width=0.45\linewidth]{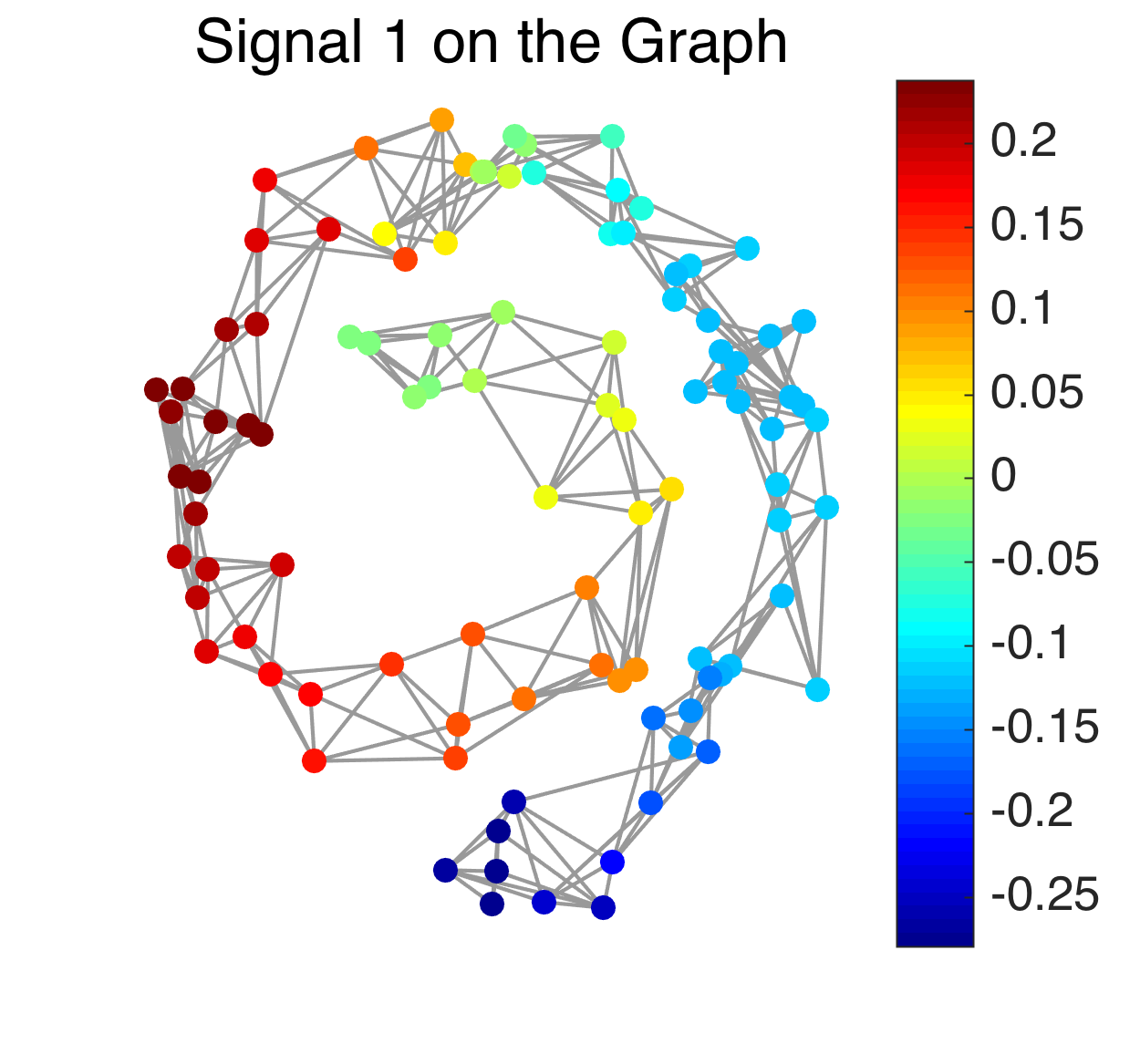} 
\end{center}
\caption{Example of stationary graph signals. The graph connections express relationships between the different elements of one signal. In this case, the signal varies smoothly along the snail shape of the graph.}
\label{fig:intro_stationarity_graph}
\vspace{-0.5cm}
\end{figure}

Classical stationarity is a statement of statistical regularity under arbitrary translations and thus requires a regular structure (often "time"). However many signals do not live on such a regular structure. For instance, imagine that instead of having one sensor returning a temporal signal, we have multiple sensors living in a two-dimensional space, each of which delivers only one value. 
In this case (see Figure~\ref{fig:intro_stationarity_graph} left), the signal support is no longer regular. Since there exists an underlying continuum in this example (2D space), one could assume the existence of a 2D stationary field and use Kriging \cite{williams1998prediction} to interpolate observations to arbitrary locations, thus generalizing stationarity for a regular domain but irregularly spaced samples. 

On the contrary, the goal of this contribution is to generalize stationarity for an irregular domain that is represented by a graph, \emph{without resorting to any underlying regular continuum}. 
Graphs are convenient for this task as they are able to capture complicated relations between variables.
In this work, a graph is composed of vertices connected by weighted undirected edges and signals are now scalar values observed at the vertices of the graph.
Our approach is to use a weak notion of translation invariance, define on a graph, that captures the structure (if any) of the data. Whereas classical stationarity means correlations are computed by translating the auto-correlation function, here correlations are given by localizing a common graph kernel, which is a generalized notion of translation as detailed in Section~\ref{sec:localization_operator}.


Figure~\ref{fig:intro_stationarity_graph} (left) presents an example of  random multivariate variable living in a 2-dimensional space. Seen as scattered samples of an underlying 2D stochastic function, one would (rightly) conclude it is not stationary. However, under closer inspection, the observed values look stationary \emph{within} the spiral-like structure depicted by the graph in Figure~\ref{fig:intro_stationarity_graph} (right). 
The traditional Kriging interpolation technique would ignore this underlying structure and conclude that there are always rapid two dimensional variations in the underlying continuum space. This problem does not occur in the graph case, where the statistical relationships inside the data follow the graph edges resulting in this example in signals oscillating smoothly over the graph. 
A typical example of a stationary signal on a graph would be the result of a survey performed by the users of a social network. If there is a relationship between a user’s answers and those of his neighbors, this relationship is expected to be constant among all users. Using stationarity on the graph, we could predict the most probable answer for users that never took the survey.

\subsection{Contributions}
We use spectral graph theory to extend the notion of stationarity to a broader class of signals. Leveraging the graph localization operator, we establish the theoretical basis of this extension in Section~\ref{sec:theory}. We show that the resulting notion of stationarity is equivalent to the proposition of Girault~\cite[Definition 16]{girault2015signal}, although the latter is not defined in terms of a localization operator. Localization is a very desirable feature, since it naturally expresses the scale at which samples are strongly correlated.

Since our framework depends on the power spectral density (PSD), we generalize the Welch method~\cite{welch1967use,bartlett1950periodogram} in Section~\ref{sec:psd_estimation} and obtain a scalable and robust way to estimate the PSD. It improves largely the covariance estimation when the number of signals is limited.

Based on the generalization of Wiener filters, we propose a new regularization term for graph signal optimization instead of the traditional Dirichlet prior, that depends on the noise level and on the PSD of the signal. The new optimization scheme presented in Section~\ref{sec:wiener} has three main advantages: 1) it allows to deal with an arbitrary regularization parameter, 2) it adapts to the data optimally as we prove that the optimization model is a Maximum A Posteriori (MAP) estimator, and 3) it is more scalable and robust than a traditional Gaussian estimator.

Finally, in Section~\ref{sec:USPS}, we show experimentally that common datasets such as USPS follow our stationarity assumption. In section~\ref{sec:experiments}, we exploit this fact to perform missing data imputation and we show how stationarity improves over classical graph models and Gaussian MAP estimator.


\subsection{Related work} 
Graphs have been used for regularization in data applications for more than a decade~\cite{shuman2013emerging,smola2003kernels,zhou2004regularization,peyre2008non} and two of the most used models will be presented in Section~\ref{sec:convex_model}. 
The idea of graph filtering was hinted at by the machine learning community~\cite{smola2003kernels} but developed for the spectral graph wavelets proposed by Hammond et al.~\cite{hammond2011wavelets} and extended by Shuman et al. in~\cite{shuman2016vertex}. While in most cases, graph filtering is based on the graph Laplacian, Moura et al.~\cite{sandryhaila2013discrete} have suggested to use the adjacency matrix instead. 

We note that a probabilistic model using Gaussian random fields has been proposed in~\cite{gadde2015probabilistic,zhang2015graph}. 
In this model, signals are automatically graph stationary with an imposed covariance matrix. Our model differentiates itself from these contributions because it is based on a much less restrictive hypothesis and uses the point of view of stationarity. A detailed explanation is given at the end of Section \ref{sec:theory}.

Finally, stationarity on graphs has been recently proposed in~\cite{girault2015stationary,girault2015signal} by Girault et al. These contributions use a different translation operator, promoting energy preservation over localization. While seemingly different, we show that our approach and Girault's result in the same graph spectral characterization of stationary signals. Girault et al \cite{girault2014semi} have also shown that using the Laplacian as a regularizer in a de-noising problem (Tikhonov) is equivalent to applying a Wiener filter adapted to a precise class of graph signals. In \cite[pp 100]{girault2015signal}, an expression of graph Wiener filter can be found.

After the publication of the first version of this contribution, additional work was done on the topic. First some PSD estimation methods were proposed in~\cite{marques2016stationary,chepuri2016subsampling}. Then stationarity has been extended to time evolving signals on graphs in~\cite{loukas2016stationary,perraudin2016towards}.

\section{Background theory}

\subsection{Graph signal processing}

\paragraph{Graph nomenclature}
A graph $\G=\{ \V,\E,\mathcal{W}\}$ is defined by two sets: $\V,\E$ and a weight function $\mathcal{W}$.
 $\V$ is the set of vertices representing the nodes of the graph and
 $\E$ is the set of edges that connect two nodes if there is a particular relation between them. In this work all graphs are undirected. To obtain a finer  structure, this relation can be quantified by a weight function $\mathcal{W} : \V \times \V \rightarrow \Rbb$ that reflects to what extent two nodes are related to each other. Let us index the nodes from $1,\dots, N=|\V|$ and  construct the weight matrix $\W \in \Rbb^{N \times N}$ by setting $W[i,n] = \mathcal{W}(v_i,v_n)$ as the weight associated to the edge connecting the node $i$ and the node $n$. When no edge exists between $i$ and $n$, the weight is set to $0$. For a node $v_i\in \V$, the degree $d[i]$ is defined as $d[i]=\sum_{n=1}^N \W [i,n]$. 
In this framework, a signal is defined as a function $f: \V \rightarrow  \mathbb{R}$ (this framework can be extended to $\mathbb{C}$) assigning a scalar value to each vertex. It is convenient to consider a signal $f$ as a vector of size $N$ with the $n^{th}$ component representing the signal value at the $n^{th}$ vertex. 

The most fundamental operator in graph signal-processing is the (combinatorial) graph Laplacian, defined as:
$\Larg = \D-\W,$
where $\D$ is the diagonal degree matrix ($\D[i,i]=d[i]$). 


\paragraph{Spectral theory}
Since the Laplacian $\Larg$ is always a symmetric positive semi-definite matrix, we know from the spectral theorem that it possesses a complete set of orthonormal eigenvectors. We denote them by $\{ \x_\ell \}_{\ell=0,1,..., N-1}$. For convenience, we order the set of real, non-negative eigenvalues as follows: $0=\lambda_0 < \lambda_1 \leq \dots \leq \lambda_{N-1} = \lambda_{\rm max}$. When the graph is connected\footnote{a path connects each pair of nodes in the graph},
there is only one zero eigenvalue. In fact, the multiplicity of the zero eigenvalue(s) is equal to the number of connected components. For more details on spectral graph theory, we refer the reader to~\cite{chung1997spectral,chung2005laplacians}. 
The eigenvectors of the Laplacian are used to define a graph Fourier basis~\cite{shuman2013emerging,shuman2016vertex} which we denote as $U$. The eigenvalues are considered as a generalization of squared frequencies. The Laplacian matrix can thus be decomposed as 
$$
\Larg = U\Lambda U^*,
$$
where $U^*$ denotes the transposed conjugate of $U$.
The graph Fourier transform is written $\hat{f} = U^*f$ and its inverse $f = U\hat{f}$. This Graph Fourier Transform possesses interesting properties further studied in~\cite{shuman2016vertex}. 
Note that the graph Fourier transform is equivalent to the Discrete Fourier transform for cyclic graphs. The detailed computation for the "ring" can be found in \cite[pp 136-137]{strang1999discrete}.

\paragraph{Graph convolutive filters}
The graph Fourier transform plays a central role in graph signal processing since it allows a natural extension of filtering operations. In the classical setting, applying a filter to a signal is equivalent to a convolution, which is simply a point-wise multiplication in the spectral domain. 
For a graph signal, where the domain is not regular, filtering is still well defined, as a point-wise multiplication in the spectral domain~\cite[Equation 17]{shuman2016vertex}.
A graph convolutive filter $g(\Larg)$ is defined from a continuous kernel $g:\Rbb_+ \rightarrow \Rbb$. In the spectral domain, filtering a signal $s$ with a convolutive filter $g(\Larg)$ is, as the classical case, a point-wise multiplication written as $\hat{s'}[\ell] =  g(\lambda_\ell) \cdot \hat{s}[\ell]$, where $\hat{s'}, \hat{s}$ are the Fourier transform of the signals $s',s$. In the vertex domain, we have
\begin{equation} \label{eq:graph_filter}
s' := g(\Larg) s =  U g(\Lambda) U^* s ,
\end{equation}
where $g(\Lambda)$ is a diagonal matrix with entries $g(\lambda_\ell)$. For convenience, we abusively call 'filter' the generative kernel $g$. We also drop the term convolutive as we are only going to use this type of filters. 
A comprehensive definition and study of these operations can be found in~\cite{shuman2016vertex}.
It is worth noting that these formulas make explicit use of the Laplacian eigenvectors and thus its diagonalization. The complexity of this operation is in general $\O(N^3)$. In order to avoid this cost, there exist fast filtering algorithms based on Chebyshev polynomials or the Lanczos method~\cite{hammond2011wavelets,susnjara2015accelerated}. These methods scale with the number of edges $|E|$ and reduce the complexity to $\O(|E|)$, which is advantageous in the case of sparse graphs.

\subsection{Localization operator} \label{sec:localization_operator}
As most graphs do not possess a regular structure, it is not possible to translate a signal around the vertex set with an intuitive shift. As stationarity is an invariance with respect to translation, we need to address this issue first.  A solution is present in \cite[Equation 26]{shuman2016vertex}, where Shuman et. al. define the generalized translation for graphs as a convolution with a Kroneker delta. The convolution $\ast$ is defined as the element-wise multiplication in the spectral domain leading to the following generalized translation definition:
$$
T_is [n]  := (s \ast \delta_i)[n] = \sum_{\ell=0}^{N-1} \hat{s}[\ell] \x_\ell^*[i] \x_\ell[n].
$$

Naturally, the generalized translation operator does not perform what we would intuitively expect from it, i.e it does not shift a signal $s$ from node $n$ to node $i$ as this graph may not be shift-invariant. Instead when $\hat{s}$ changes smoothly across the frequencies (more details later on), then $T_is$ is localized around node $i$, while $s$ is in general not localized at a particular node or set of nodes.

In order to avoid this issue, we define the localization operator as follows
\begin{definition} \label{def:localization_operator}
Let $\mathcal{C}$ be the set of functions $ \Rbb^+ \rightarrow \Rbb$. For a graph kernel $g \in \mathcal{C}$ (defined in the spectral domain) and a node $i$, the localization operator  $\T_i: \mathcal{C}\rightarrow \Rbb^N$ reads:
\begin{equation}
\label{eq:localization_operator}
\T_i g[n]  := \sum_{\ell=0}^{N-1} g(\lambda_\ell) \x_\ell^*[i] \x_\ell[n] = (g(\Larg)\delta_i)[n] =g(\Larg)[i,n].
\end{equation}
\end{definition}
Here we use the calligraphic notation $\T_i$ to differentiate with the generalized translation operator $T_i$. We first observe from \eqref{eq:localization_operator} that the $i^{\text{th}}$ line of graph filter matrix $g(\Larg)$ is the kernel $g$ localized at node $i$. Intuitively, it signifies 
$
[g(\Larg)s](i) = \scp{s}{\T_ig}.
$
We could replace $g(\lambda_\ell)$ by $ \hat{s}[\ell]$ in Definition \ref{def:localization_operator} and localize the discrete vector $\hat{s}$ instead. We prefer to work with a kernel for two reasons. 1) In practice when the graph is large, the Fourier basis cannot be computed making it impossible to localize a vector. On the other side, for a kernel $g$, there are techniques to approximate $\T_i g$. 2) The localization properties are theoretically easier to interpret when $g$ is a filter. Let us suppose that $g$ is a $K$ order polynomial, then the support of $\T_ig$ is exactly contained in a ball of radius $K$ centered at node $i$. Building on this idea, for a sufficiently regular function $g$, it has been proved in~\cite[Theorem 1 and Corollary 2]{shuman2016vertex} that the localization operator concentrates the kernel $g$ around the vertex $i$. 

Let us now clarify how generalized translation and localization are linked. The main difference between these two operators is the domain on which they are applied. Whereas, the translation operator acts on a discrete signal defined in the time or the vertex domain, the localization operator requires a continuous kernel or alternatively a discrete signal in the spectral domain. Both return a signal in the time or the vertex domain. To summarize, the localization operator can be seen as computing the inverse Fourier transform first and then translating the signal. It is an operator that takes a filter from the spectral domain and localizes it at a given node $i$ while adapting it to the graph structure. 

\begin{figure}[htb!]
\begin{center}
\includegraphics[width=0.95\linewidth]{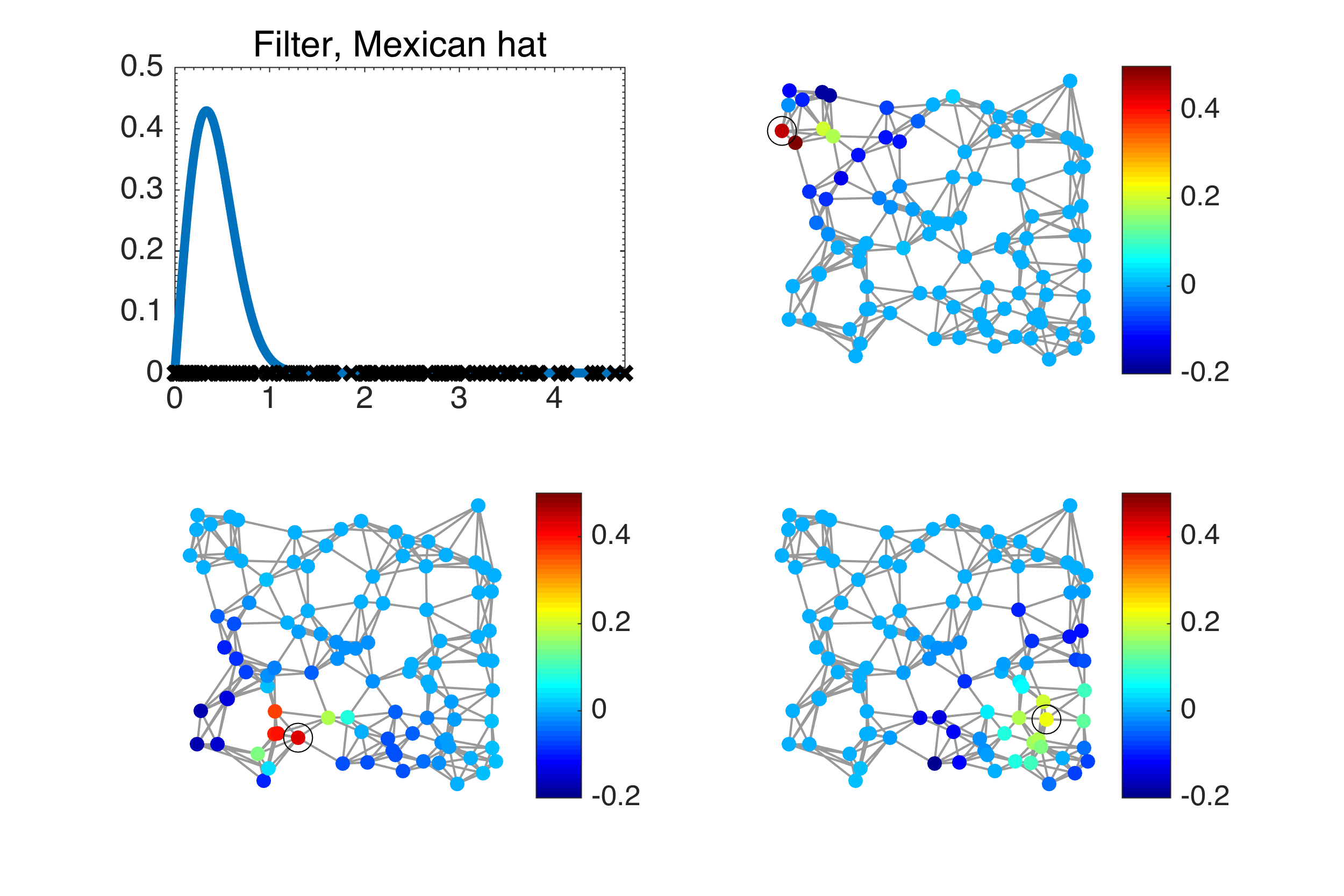} 
\end{center}
\caption{\label{fig:demo_localization} Top left: Mexican hat filter in the spectral domain $g(x) = \frac{5 x}{\lmax}\exp\left(-\frac{25 x^2}{\lmax^2}\right).$ The filter is localized around three different vertices (highlighted by a black circle).}
\vspace{-0.5cm}
\end{figure}

In the classical periodic case ("ring" graph), localization is strongly connected to translation as the localized kernels are translated versions of each other: 
\begin{eqnarray} \label{eq:Tig_time}
\T_ig[n] 
& = & \frac{1}{N}\sum_{\ell=1}^{N} g(\lambda_\ell) e^{-j2\pi\frac{\ell i}{N}} e^{j2\pi\frac{\ell n}{N}} \nonumber\\
& = & \frac{1}{N} \sum_{\ell=1}^{N} g(\lambda_\ell) e^{j2\pi\frac{\ell (n-i)}{N}}
= \T_0g[n-i].
\end{eqnarray}
In this case, localizing a kernel $g$ can be done by computing the inverse discrete Fourier transform of the vector $\hat{s}(\ell) = g(\lambda_\ell)$ and then translating at node $i$.
However, for irregular graphs, localization differs from translation because the shape of the localized filter adapts to the graph and varies as a function of its topology. 
Figure \ref{fig:demo_localization} shows an example of localization using the Mexican hat wavelet filter. The shape of the localized filter depends highly on the graph topology. However, we observe that the general shape of the wavelet is preserved. It has large positive values around the node where it is localized. It then goes negative a few nodes further away and stabilizes at zero for nodes far away.
To summarize, the localization operator preserves the global behavior of the filter while adapting to the graph topology. 
Additional insights about the localization operator can be found in \cite{shuman2016vertex,hammond2011wavelets,perraudin2016global,shuman2015spectrum}.

\subsection{Stationarity for temporal signals}
Let ${\bf x }[t]$ be a time indexed stochastic process. Throughout this paper, all random variables are written in bold fonts. We use $m_{\b{x}} = \Esp \big\{ \b{x} \big\}$ to denote the expected value of $\b{x}$. In this section, we work with the \textbf{periodic discrete} case.
\begin{definition}[Time Wide-Sense Stationarity] \label{def:time-stationarity}
A signal is Time Wide-Sense Stationary  (WSS) if its first two statistical moments are invariant under translation, i.e:
\begin{enumerate}
	\item $ m_{\b{x}}[t] =  \Esp \big\{ \b{x}[t] \big\}  = c \in \Rbb $,
	\item $\Esp \big\{ (\b{x}[t]-m_{\b{x}})(\b{x}[s] - m_{\b{x}})^*\big\} = \eta_{{\bf x }}[t-s]$,
\end{enumerate}
where $\eta_{\b{x}}$ is called the autocorrelation function of $\b{x}$.
\end{definition}
Note that using~\eqref{eq:Tig_time}, the autocorrelation can be written in terms of the localization operator:
\begin{equation} \label{eq:link_auto_loc}
\eta_{\b{x}}[t-s] = \T_s\gamma_{{\bf x }}[t].
\end{equation}
For a WSS signal, the autocorrelation function depends only on one parameter, $t-s$, and is linked to the Power Spectral Density (PSD) through the Wiener-Khintchine Theorem~\cite{Wiener1930generalized}. The latter states that the PSD of the stochastic process $\b{x}$ denoted $\gamma_{\b{x}}[\ell]$ is the Fourier transform of its auto-correlation~:
\begin{equation} \label{theo:Wiener_khintchine}
\gamma_{\b{x}}[\ell] = \frac{1}{\sqrt{N}}\sum_{i=1}^N \eta_{\b{x}}[t]e^{-j 2\pi\frac{\ell t}{N}},
\end{equation}
where $j=\sqrt{-1}$.
As a consequence, when a signal is convolved with a filter $\check{h}$, its PSD is multiplied by the energy of the convolution kernel: for $\b{y} = \check{h} \ast \b{x}$, we have
$$
\gamma_{\b{y}}[\ell] = \left|h[\ell] \right|^2 \gamma_{\b{x}}[\ell],
$$
where $h[\ell] $ is the Fourier transform of $\check{h}$. For more information about stationarity, we refer the reader to~\cite{papoulis2002probability}.

When generalizing these concepts to graphs, the underlying structure for stationarity will no longer be time, but graph vertices. 



\section{Stationarity of graph signals} \label{sec:theory}
We now generalize stationarity to graph signals. While we define stationarity through the localization operator, Girault~\cite{girault2015stationary} uses an isometric translation operator instead. That proposition is briefly described in Section \ref{subsec:Girault}, where we also show the equivalence of both definitions.

\subsection{Stationarity under the localization operator}

Let $\b{x} \in \Rbb^N$ be a stochastic graph signal with a finite number of variables indexed by the vertices of a weighted undirected graph. The expected value of each variable is written $m_{\b{x}}[i] = \Esp \big\{\b{x}[i]\big\}$ and the covariance matrix of the stochastic signal is $\Sigma_{\b{x}} = \Esp \big\{ (\b{x}-m_{\b{x}}) (\b{x}-m_{\b{x}})^* \big\})$. We additionally define $\tilde{\b{x}}=\b{x}-m_{\b{x}}$. 
For discrete time WSS processes, the covariance matrix $\Sigma_{\b{x}}$ is Toeplitz, or circulant for periodic boundary conditions, reflecting translation invariance. In that case, the covariance is diagonalized by the Fourier transform. We now generalize this property to take into account the intricate graph structure. 

As explained in Section~\ref{sec:localization_operator}, the localization operator adapts a kernel to the graph structure. As a result, our idea is to use the localization operator to adapt the correlation between the samples to the graph structure. This results in a localized version of the correlation function, whose properties can then be studied via the associated kernel.
\begin{definition} \label{def:GWSS}
A stochastic graph signal $\b{x}$ defined on the vertices of a graph $\G$ is called Graph Wide-Sense (or second order) Stationary (GWSS), if and only if it satisfies the following properties:
\begin{enumerate}
\item its first moment is constant over the vertex set: $m_{\b{x}}[i] = \Esp\big\{\b{x}[i]\big\} = c \in \Rbb$ and
\item its covariance is the result of localizing a graph kernel: $$\Sigma_{\b{x}}[i,n] = \Esp\big\{ (\b{x}[i] -m_{\b{x}})(\b{x}[n]-m_{\b{x}})\big\} = \T_i\gamma_{\b{x}}[n].$$
\end{enumerate}
\end{definition}

The first part of the above definition is equivalent to the first property of time WSS signals. The requirement for the second moment is a natural generalization where we are imposing an invariance with respect to the localization operator instead of the translation. It is a generalization of Definiton~\ref{def:time-stationarity} using~\eqref{eq:link_auto_loc}. In simple words, the covariance is assumed to be driven by a global kernel (filter) $\gamma_{\b{x}}$. The localization operator adapts this kernel to the local structure of the graph and provides the correlation between the vertices. Additionally, Definition~\ref{def:GWSS} implies that the spectral components of $\b{x}$ are uncorrelated.
\begin{theorem} \label{theo:diag_L}
A signal is GWSS if and only if its covariance matrix  $\Sigma_{\b{x}}$ is jointly diagonalizable with the Laplacian of $\G$ with\footnote{If the graph Laplacian has an eigenspace of multiplicity greater than one, this condition implies that all eigenvalues of the covariance matrix associated with this eigenspace are equal, i.e., if $ \lambda_{\ell_1}=\lambda_{\ell_2}$, then  $u_{\ell_1}^*\Sigma_{\b{x}}{u_{\ell_1}}= u_{\ell_2}^*\Sigma_{\b{x}}{u_{\ell_2}}$. On a ring graph, it ensures 1) that the Fourier transform of the PSD to be symmetric with respect to the $0$ frequency, and 2) that the autocorrelation $\eta_{\b{x}}$ is real and symmetric.}
$\gamma_{\b{x}}(\lambda_\ell)=u_\ell^*\Sigma_{\b{x}}u_\ell$, i.e $\Sigma_{\b{x}} = U \Gamma_{\b{x}} U^*$, where $\Gamma_{\b{x}}$ is a diagonal matrix. 
\begin{proof}
By Definition~\ref{def:localization_operator}, the covariance localization operator can be written as:
\begin{equation}
\T_i \gamma_{\b{x}} [n]= \gamma_{\b{x}}(\Larg)[i,n] = U \gamma_{\b{x}}(\Lambda) U^*[i,n]
\end{equation}
where $\gamma_{\b{x}}(\Lambda)$ is a diagonal matrix satisfying $\gamma_{\b{x}}(\Lambda)[\ell,\ell] = \gamma_{\b{x}}(\ll) $. To complete the proof set $\Gamma_{\b{x}}=\gamma_{\b{x}}(\Lambda)$.
\end{proof}
\end{theorem}

The choice of the filter $\gamma_{\b{x}}$ in this result is somewhat arbitrary, but we shall soon see that we are interested in localized kernels. In that case, $\gamma_{\b{x}}$ will be typically be the lowest degree polynomial satisfying the constraints and can be constructed using Lagrange interpolation for instance. 

Definition~\ref{def:GWSS} provides a fundamental property of the covariance. The size of the correlation (distance over the graph) depends on the support of localized the kernel $\T_i \gamma_{\b{x}}$. In~\cite[Theorem 1 and Corollary 2]{shuman2016vertex}, it has been proved that the concentration of $\T_i \gamma_{\b{x}}$ around $i$ depends on the regularity\footnote{A regular kernel can be well approximated by a smooth function, for instance a low order polynomial, over the spectrum of the laplacian.} of $\gamma_{\b{x}}$. For example, if $\gamma_{\b{x}}$ is a polynomial of degree $K$, it is exactly localized in a ball of radius $K$. Hence we will be mostly interested in such low degree polynomial kernels. 

The graph spectral covariance matrix of a stochastic graph signal is given by $\Gamma_{\b{x}} = U^*\Sigma_{\b{x}} U$. For a GWSS signal this matrix is diagonal and the graph power spectral density (PSD) of $\b{x}$ becomes:
\begin{equation} \label{eq:cov_mat_fourier}
\gamma_{\b{x}}(\lambda_\ell) =  \left( U^*\Sigma_{\b{x}} U \right)_{\ell,\ell}.
\end{equation} 
Table \ref{tab:summary_stationarity} presents the differences and the similarities between the classical and the graph case. For a regular cyclic graph (ring), the localization operator is equivalent to the traditional translation and we recover the classical cyclic-stationarity results by setting $\eta_{\b{x}} = \T_0\gamma_{\b{x}}$. Our framework is thus a generalization of stationarity to irregular domains.


\begin{example}[Gaussian i.i.d. noise (see also~\cite{girault2015stationary})]
Normalized Gaussian i.i.d. noise is GWSS for any graph. Indeed, the first moment is $\Esp\big\{\b{x}[k]\big\} = 0 $. Moreover, the covariance matrix can be written as $I = \Sigma_{\b{x}} = U I U^*$ with any orthonormal matrix $U$ and thus is diagonalizable with any graph Laplacian. We also observe that the PSD is constant, which implies that similar to the classical case, white noise contains all "graph frequencies".
\end{example}

When $\gamma_{\b{x}}$ is a bijective function, the covariance matrix contains an important part of the graph structure: the Laplacian eigenvectors\footnote{If the laplacian contains eigenvalues with multiplicity, then the covariance matrix contains all its eigenspaces.}. 
On the contrary, if $\gamma_{\b{x}}$ is not bijective, some of the graph structure is lost as it is not possible to recover all eigenvectors. This is for instance the case when the covariance matrix is low-rank.
As another example, let us consider completely uncorrelated centered samples with variance $1$. In this case, the covariance matrix becomes $\Sigma_{\b{x}} = I$ and loses all graph information, even if by definition the stochastic signal remains stationary on the graph. 

One of the crucial benefits of stationarity is that it is preserved by  filtering, while the PSD is simply reshaped by the filter. The same property holds on graphs.

\begin{theorem} \label{theo:graph_psd_trans}
When a graph filter $g$ is applied to a GWSS signal, the result remains GWSS, the mean becomes $m_{{g(\Larg)\b{x}}} = m_{\b{x}}g(0)$ and the PSD satisfies:
\begin{equation}
\gamma_{g(\Larg)\b{x}}(\ll) = |g(\ll)|^2 \cdot \gamma_{\b{x}} (\ll).
\end{equation}
\begin{proof}
The output of a filter $g$ can be written as $\b{x}' = g(L)\b{\tilde{\b{x}}} + g(L) m_{\b{x}}$. If the input signal $\b{x}$ is GWSS, we can check easily that the first moment of the filter's output is constant, $\Esp \{g(L)\b{x}[i]\} = g(L)\Esp \{m_{\b{x}}\} = g(0) m_{\b{x}} $. The computation of the second moment gives:
\begin{eqnarray*}
\Esp \left\{g(\Larg)\b{\tilde{\b{x}}} \big(g(\Larg)\b{\tilde{\b{x}}}\big)^*\right\} 
& = & g(\Larg) \Esp \left\{ \b{\tilde{\b{x}}} \b{\tilde{\b{x}}}^* \right\} g(\Larg)^* \\ 
& = & g(\Larg) \Sigma_{\b{x}} g(\Larg)^* \\
& = & U g^2(\Lambda) \gamma_{\b{x}}(\Lambda) U^*,
\end{eqnarray*}
which is equivalent to our claim.
\end{proof}
\end{theorem}

Theorem~\ref{theo:graph_psd_trans} provides a simple way to artificially produce stationary signals with a prescribed PSD by simply filtering white noise~:
\begin{center}
\includegraphics[width=0.15\textwidth]{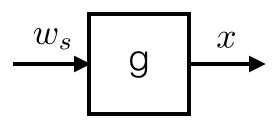}
\end{center}
The resulting signal will be stationary with PSD $g^2$. In the sequel, we assume for simplicity that the signal is centered at $0$, i.e: $m_\b{x} = 0$. Note that the input white noise could well be non-Gaussian.


\begin{table*}[ht!]
\begin{center}
\renewcommand{\arraystretch}{1.6}

\begin{tabular}{|l|c|c|}
\hline
 & Classical & Graph \\
\hline
\hline
Stationary with respect to & Translation & The localization operator\\ 
\hline
 First moment & $\Esp \big\{ \b{x}[t] \big\} = m_{\b{x}} = c \in \Rbb$ & $\Esp\big\{\b{x}[i]\big\} = m_{\b{x}} = c \in \Rbb$  \\
 \hline
 Second moment  & $\Sigma_{\b{x}}[t,s] = \Esp \big\{ \tilde{\b{x}}[t]\tilde{\b{x}}^*[s] \big\} = \eta_{{\bf x }}[t-s]$ & $\Sigma_{\b{x}}[i,n] =\Esp \big\{ \tilde{\b{x}}[i]\tilde{\b{x}}^*[n] \big\} = \gamma_{\b{x}}(\Larg)_{i,n}$ \\
 (We use $ \tilde{\b{x}}=\b{x}-m_{\b{x}}$) & $\Sigma_{\b{x}}$ Toeplitz & $\Sigma_{\b{x}}$ diagonalizable with $\Larg$\\
 \hline
Wiener Khintchine & $\gamma_{\b{x}}(\ll) = \frac{1}{\sqrt{N}}\sum_{i=1}^N \eta_{\b{x}}[t]e^{-j 2\pi \frac{t \ell }{N}}$ & $ \gamma_{\b{x}}(\lambda_\ell) = \left(\Gamma_{\b{x}}\right)_{\ell,\ell} = \left(U^* \Sigma_{\b{x}} U\right)_{\ell,\ell} $\\
 \hline
Result of filtering & $\gamma_{\check{g} \ast \b{x}}(\ll) = |g(\ll)|^2 \cdot \gamma_{\b{x}} (\ll)$  & $\gamma_{g(\Larg)\b{x}}[\ell] = |g(\ll)|^2 \cdot \gamma_{\b{x}} (\ll)$ \\
 \hline
\end{tabular}
\renewcommand{\arraystretch}{1}
\end{center}
\caption{Comparison between classical and graph stationarity. In the classical case, we work with a $N$ periodic discrete signal and we use $\check{g}$ to denote the inverse Fourier transform of $g$. }
\label{tab:summary_stationarity}
\vspace{-0.5cm}
\end{table*}

\subsection{Comparison with the work of B. Girault}
\label{subsec:Girault}
Stationarity for graph signals has been defined in the past~\cite{girault2015signal,girault2015stationary}. The proposed definition is based on an isometric graph translation operator defined for a graph signal $s$ as: 
$$T_B s := \exp\left(j 2\pi \sqrt{\frac{\Larg}{\rho_\G}} \right) s = b(\Larg)s,$$
where $b(x) = \exp \left( j 2 \pi \sqrt{ \frac{x}{\rho_\G} }\right)$ and $\rho_\G $ is an upper bound\footnote{$\rho_\G = \max_{i\in\V} \sqrt{ 2 d[i]( d[i]+\bar{d}[i] ) }$ where $\bar{d}[i]=\frac{\sum_n=1^N W[i,n]d[n]}{d[i]}$} on $\lmax$. While this operator conserves the energy of the signal ($\|T_B s\|_2 =\|s\|_2$), it does not have localization properties. In a sense, one trades localization for isometry. Using this operator, the stationarity definition of Girault~\cite{girault2015signal,girault2015stationary} is a natural extension of the classical case (Definition~\ref{def:time-stationarity}).
\begin{definition}{\normalfont \cite[Definition 16]{girault2015signal}} \label{def:girault}
A stochastic signal $\b{x}$ on the graph $\G$ is Wide-Sense Stationary (WSS) if and only if
\begin{enumerate}
\item $\Esp \left\{T_{B}\b{x}\right\}=\Esp \left\{\b{x}\right\}$
\item $\Esp \left\{T_{B}\b{x}\left(T_{B}\b{x}\right)^{*}\right\}=\Esp\left\{\b{x}\b{x}^{*}\right\}$
\end{enumerate}
\end{definition}
While this definition is based on a fairly different construction, the resulting notion of stationarity is similar. We distinguish two cases: 1) In the case where all eigenvalues are disjoint, they are equivalent. Indeed, \cite[Theorem 7]{girault2015signal} says that if a signal is stationary with Definition~\ref{def:girault}, then its first moment is constant and the covariance matrix in the spectral domain $U^* \Sigma_{\b{x}} U$ has to be diagonal. Using Theorem~\ref{theo:diag_L}, we therefore recover Definition~\ref{def:GWSS}. 
2) In the case where the graph has at least an eigenvalue with multiplicity, e.g., a ring graph, our Definition~\ref{def:GWSS} is more restrictive than Definition~\ref{def:girault}, since we need for every $\lambda_{\ell_1}=\lambda_{\ell_2}$, that $\gamma_\b{x}(\lambda_{\ell_1})=\gamma_\b{x}(\lambda_{\ell_2})$. As a result, there exist signals that are only stationary according to the Definition~\ref{def:girault} of Girault, but not according to our Definition~\ref{def:GWSS}. 
As a consequence, for real signals on a ring graph, our definition forces the autocorrelation function to be symmetric, matching exactly the classical stationarity Definition~\ref{def:time-stationarity}, whereas this is not true for Girault's definition. 

Let us consider as an example the ring graph with real sine/cosine Fourier basis. Note that a different basis choice leads to the same conclusion. The stochastic signal $\b{x}[i] = \b{w} \cos(2\pi i \frac{k}{N})$, where $\b{w}\sim\mathcal{N}(0,1)$. This signal is made of a single graph Fourier mode, i.e., $\b{x}=\b{w} u_{2k-1}$. The first moment is given by $\Esp \{\b{x}\}=u_{2k-1} \Esp \{\b{w}\} = 0$ and the covariance matrix reads:
$$
\Sigma_{\b{x}}[n,i]
= \Esp \{\b{x}[n]\b{x}[i]\} 
=  \cos\left(2\pi n\frac{k}{N}\right)\cos\left(2\pi i\frac{k}{N}\right).
$$
To verify the stationary property of this signal, let us observe this quantity in the spectral domain: 
$$u_{\ell_1}^* \Sigma_{\b{x}} u_{\ell_2} = \begin{cases}
\frac{N}{4} & \text{if }\ell_{1}=\ell_{2} = 2k-1 \\
0 & \text{otherwise}
\end{cases} .$$
This signal is not stationary according to the classical definition. Indeed it is not invariant with respect to translation. To observe it, just compute $ 1 = \Esp \{\b{x}[N]\b{x}[N]\}\neq\Esp \{\b{x}[\frac{N}{4k}]\b{x}[\frac{N}{4k}]\}=0$.
Our definition agrees to this: Applying Theorem~\ref{theo:diag_L}, even if the covariance matrix in the spectral domain is diagonal, we find that the signal is not stationary. Indeed, we cannot find a kernel satisfying $g(\lambda_{\ell}) = u_{\ell}^* \Sigma_{\b{x}} u_{\ell}$ for all $\ell$ as we have $\lambda_{2k}=\lambda_{2k-1}$ and $0=u_{2k}^* \Sigma_{\b{x}} u_{2k} \neq u_{2k-1}^* \Sigma_{\b{x}} u_{2k-1} = \frac{N}{4}$.
However, according to the definition by Girault, this signal is stationary \cite[Theorem 7]{girault2015signal}. 


Another key difference is that our definition allows us to generalize the notion of PSD to the graph setting in a simpler manner. 
To extend the notion of PSD using Girault's definition, one would have to deal with a block diagonal structure of the covariance matrix in the spectral domain that changes depending on the choice of eigenvectors at eigenvalue multiplicities\footnote{For a subspace associated with an eigenvalue with multiplicity greater than one, there exist multiple possible sets of eigenvectors.}.

\subsection{Gaussian random field interpretation}

The framework of stationary signals on graphs can be interpreted  using Gaussian Markov Random Field (GMRF). Let us assume that the signal $\b{x}$ is drawn from a distribution
\begin{equation}
\mathbb{P}( \b{x} ) = \frac{1}{Z_p}  e^{-  (\b{x}-m_{\b{x}})^* p(\Larg) (\b{x}-m_{\b{x}})},
\end{equation}
where $Z_p = \int_{\mathbb{R}^N} e^{-  (\b{x}-m_{\b{x}})^* p(\Larg) (\b{x}-m_{\b{x}})} \text{d}\b{x} $.
If we assume that $p(\Larg) $ is invertible, drawing from this distribution will generate a stationary  $\b{x}$ with covariance matrix given by: 
$$
\Sigma_{\b{x}} = \left(p(\Larg) \right)^{-1} = p^{-1}(\Larg).
$$
In other words, assuming a GRF probabilistic model with inverse covariance matrix $p(\Larg)$ leads to a stationary graph signal with a $\textrm{PSD} = p^{-1}$. However a stationary graph signal is not necessarily a GRF. Indeed, stationarity assumes statistical properties on the signal that are not necessarily based on Gaussian distribution.

In Section 3 of~\cite{gadde2015probabilistic}, Gadde and Ortega have presented a GMRF model for graph signals. But they restrict themselves to the case where $p(\Larg) = \Larg +\delta I$. Following a similar approach Zhang et al.~\cite{zhang2015graph} link the inverse covariance matrix of a GMRF with the Laplacian. Our approach is much broader than these two contributions since we do not make any assumption on the function $p(\Larg)$. Finally, we exploit properties of stationary signals, such as the characterization of the PSD, to explicitly solve signal processing problems in Section~\ref{sec:wiener}.


\section{Estimation of the signal PSD} \label{sec:psd_estimation}
As the PSD is central to our method, we need a reliable and scalable way to compute it. Equation~\eqref{eq:cov_mat_fourier} suggests a direct estimation method using the Fourier transform of the covariance matrix.  We could thus estimate the covariance $\Sigma_{\b{x}}$ empirically from $N_s$ realizations $\{x_n\}_{n=1\dots,N_s}$ of the stochastic graph signal $\b{x}$, as
$$
\bar{\Sigma}_{\b{x}}[i,j] = \frac{1}{N_s-1}\sum_{n=1}^{N_s} (x_n[i]-\bar{m}_{\b{x}}[i]) )(x_n[j]-\bar{m}_{\b{x}}[j])^*,
$$
where $\bar{m}_{\b{x}}[i] = \frac{1}{N_s} \sum_{n=1}^{N_s} x_n[i]$. Then our estimate of the PSD would read
$$
\bar{\gamma}_{\b{x}}(\ll) = U^*\bar{\Sigma}_{\b{x}}U[\ell,\ell].
$$
Unfortunately, when the number of nodes is considerable, this method requires the diagonalization of the Laplacian, an operation whose complexity in the general case scales as $O(N^3)$ for the number of operations and $O(N^2)$ for memory requirements. Additionally, when the number of available realizations $N_s$ is small, it is not possible to obtain a good estimate of the covariance matrix. To overcome these issues, inspired by Bartlett~\cite{bartlett1950periodogram} and Welch~\cite{welch1967use}, we propose to use a graph generalization of the Short Time Fourier transform~\cite{shuman2016vertex} to construct a scalable estimation method.

Bartlett's method can be summarized as follows. After removing the mean, the signal is first cut into equally sized segments without overlap. Then, the Fourier transform of each segment is computed. Finally, the PSD  is obtained by averaging over segments the squared amplitude of the Fourier coefficients. Welch's method is a generalization that works with overlapping segments.

\begin{figure}[htb!]
\begin{center}
\includegraphics[width=0.45\linewidth]{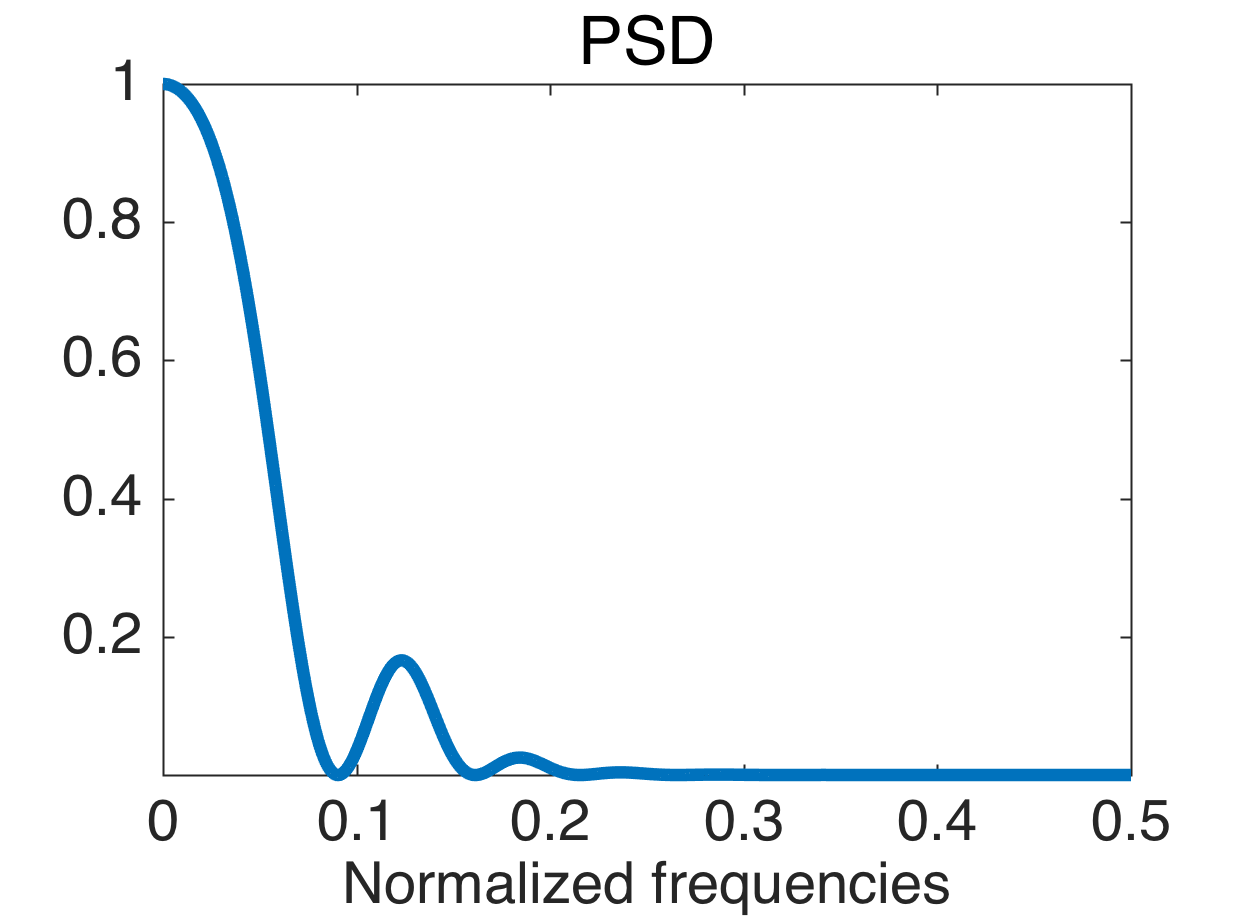} 
\includegraphics[width=0.45\linewidth]{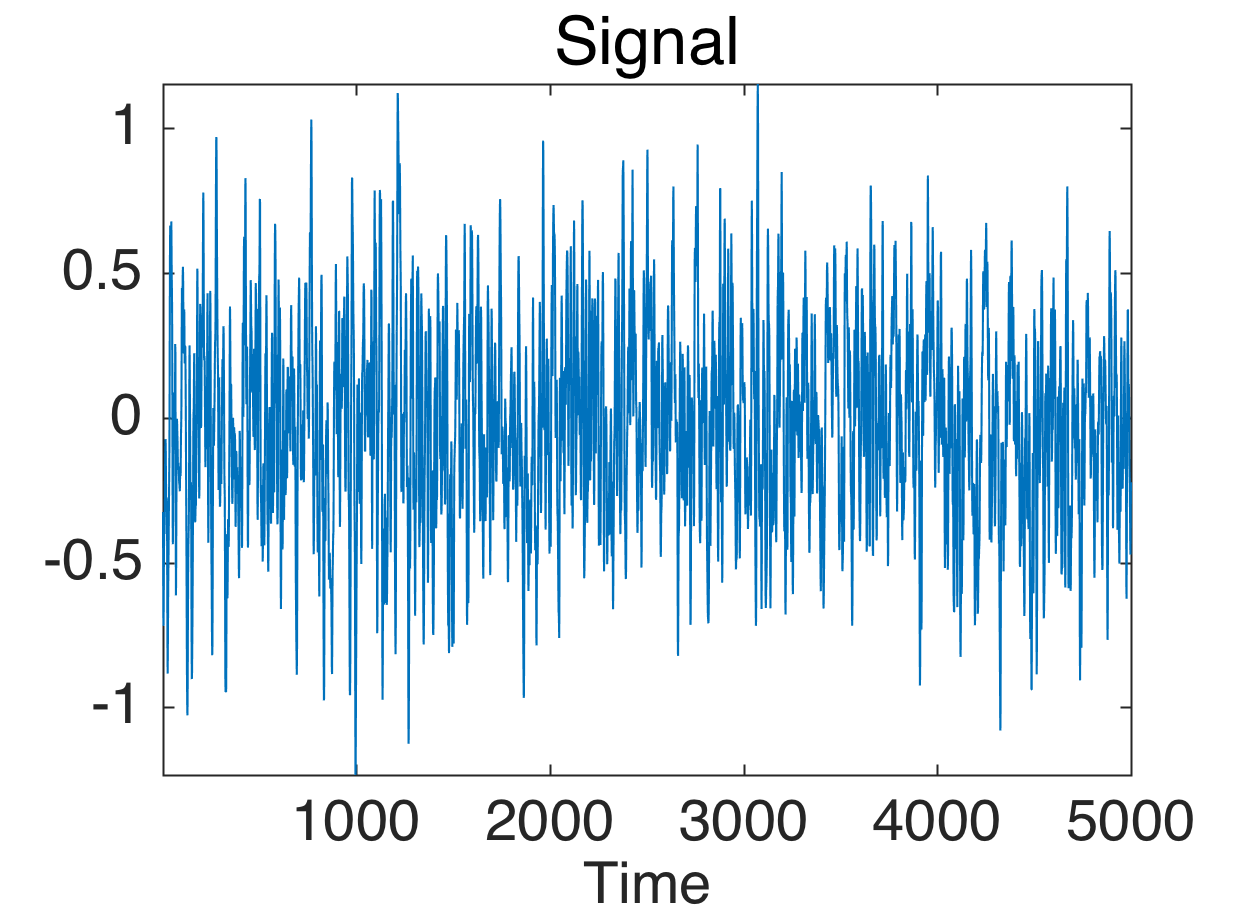} \\
\includegraphics[width=0.45\linewidth]{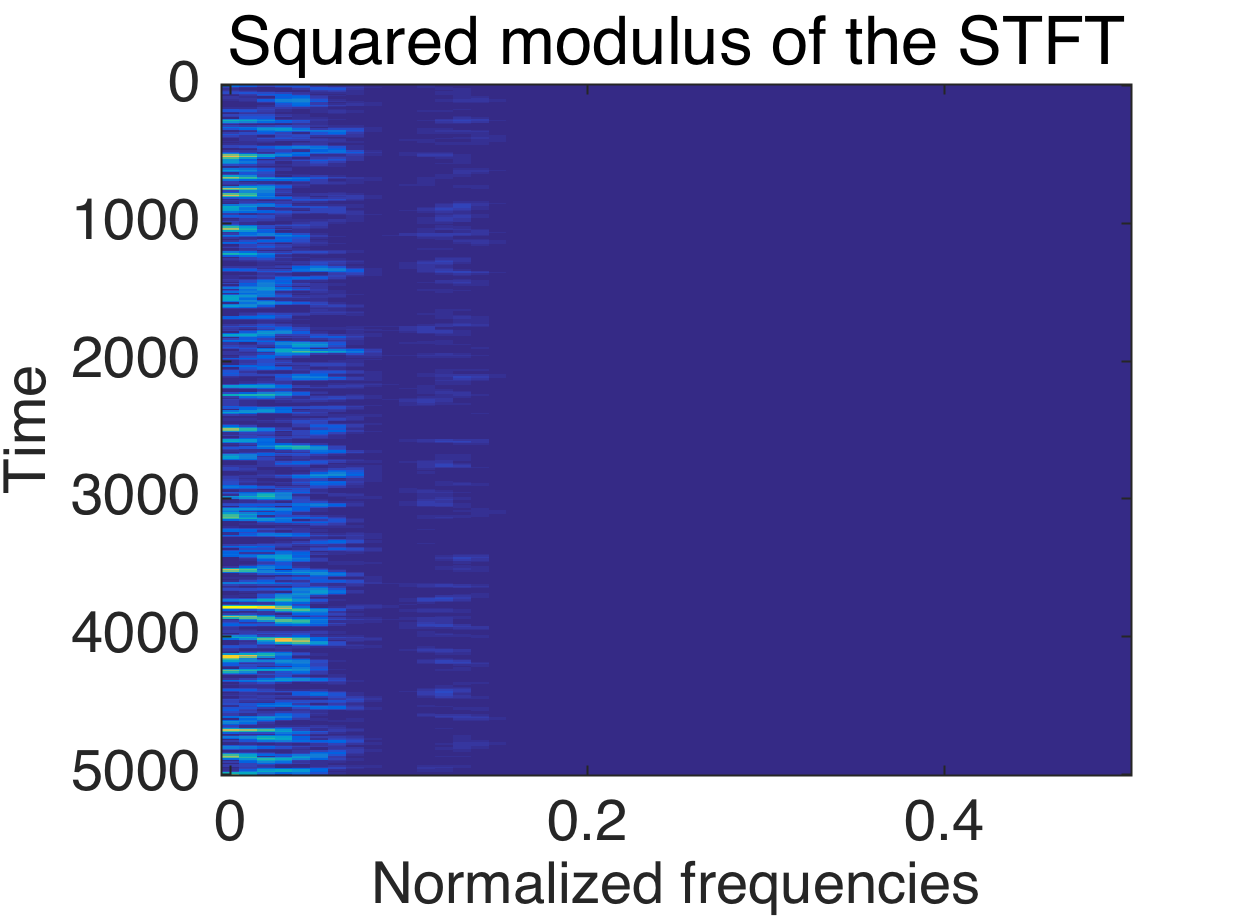} 
\includegraphics[width=0.45\linewidth]{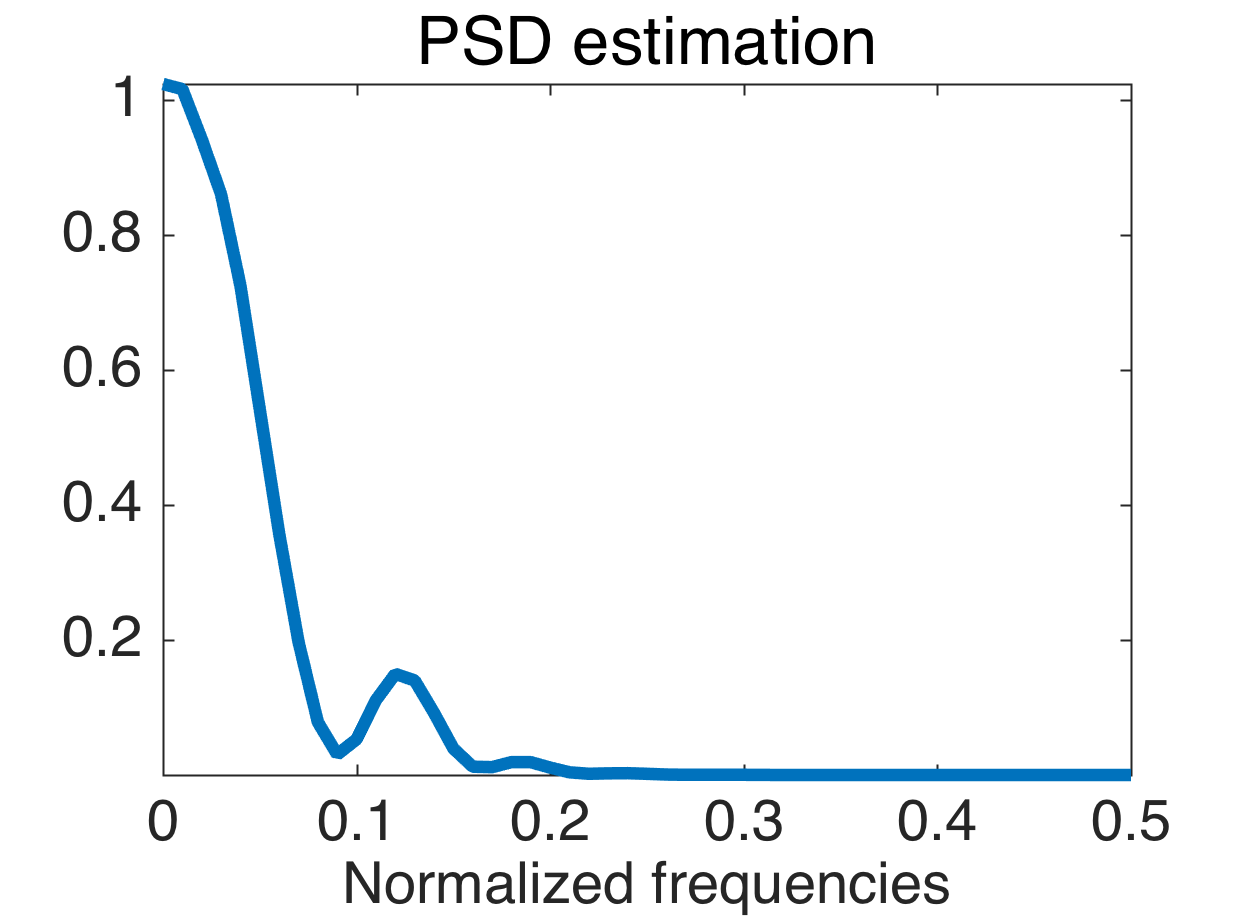} \\
\end{center}
\caption{Illustration of the PSD estimation process for a temporal signal. Top right: original PSD. Top left: a stationary signal. Bottom right: The squared modulus of the STFT of the signal. Bottom left: Sum of the STFT squared coefficients over time. We observe that averaging the squared STFT coefficients approximate well the PSD. This method is a version of the Welch method that is generalizable to graphs.}
\label{fig:psd_explanation}
\end{figure}

On the other hand, we can see the PSD estimation of both methods as the averaging over time of the squared coefficients of a Short Time Fourier Transform (STFT). Let $\tilde{x}$ be a zero mean stochastic graph signal, the classical PSD estimator can thus be written as 
$$
\bar{\gamma}_{\b{x}}[\ell] = \frac{  \sum_{n=1}^N \left( \text{STFT}\{ \tilde{x} \}[\ell,n] \right)^2 }{ N\|g\|_2^2 },
$$
where $g$ is the window used for the STFT. This is shown in Figure~\ref{fig:psd_explanation}.

\paragraph{Method}
Our method is based on this idea, using the windowed graph Fourier transform \cite{shuman2016vertex}. Instead of a translated rectangular window in time, we use a kernel $g$ shifted by multiples of a step $\tau$ in the spectral domain, i.e. 
\[g_m(\ll) = g(\ll-m\tau),\hspace{0.5cm} m=1\dots M,\hspace{0.25cm} \tau=\frac{\lmax}{M}.\]
We then localize each spectral translation at each individual node of the graph.
The coefficients of the graph windowed Fourier transform can be seen as a matrix with elements
\[C[i, m]  = \langle x,\T_ig_m \rangle = \left[g_m(\Larg)x\right]_i.\]
\textit{Our algorithm consists in averaging the squared coefficients of this transform over the vertex set}.
Because graphs have an irregular spectrum, we additionally need a normalization factor which is given by the norm of the window $g_m$: $\sum_\ell g(\ll-m\tau)^2 = \|g_m(\Larg)\|_F^2$, where $\|\cdot\|_F$ is used for the Frobenius norm. Note that this norm will vary for the different $m$. Our final estimator reads~:
\begin{equation}
	\label{eq:PSD_estimator}
	\bar{\gamma}_{\b{x}}(m\tau) = \frac{  \| g_m(\Larg) x \|_2^2 }{ \|g_m(\Larg)\|_F^2 } =\frac{  \sum_{i=1}^N C[i,m]^2 }{ \|g_m(\Larg)\|_F^2 },
	\end{equation}
where $x$ is a single realization of the stationary stochastic graph signal $\b{x}$. This estimator provides a discrete approximation of the PSD. Interpolation is used to obtain a continuous estimator. This approach avoids the computation of the eigenvectors and the eigenvalues of the Laplacian.

Our complete estimation procedure is as follows. First, we design a filterbank by choosing a mother function $g$ (for example a Gaussian $g(\lambda) = e^{-\lambda^2/\sigma^2}$). A frame is then created by shifting uniformly $M$ times $g$ in the spectral domain: $g_m(\lambda) = g(\lambda-m\tau) = e^{-(\lambda-m\tau)^2/\sigma^2}$.
Second, we compute the estimator $\bar{\gamma}_{\b{x}}(m\tau)$ from the stationary signal $\b{x}$. Note that if we have access to $K_1$ realizations $\{x_k\}_{k=1,\dots, K_1}$ of the stationary signal, we can, of course, average the estimator to further reduce the variance using $\Esp \left\{\| g_m(\Larg)  \tilde{\b{x}} \|_2^2 \right\}\approx 1/K_1 \sum_k \| g_m(\Larg) \tilde{x}_k  \|_2^2$. Third we use the following trick to quickly approximate $\| g_m (\Larg) \|_F^2$. Using $K_2$ randomly-generated Gaussian normalized zero centered white signals, we estimate 
$$ \Esp \left\{ \| g_m(\Larg) {\bf w } \|_2^2 \right\} = \|g_m(\Larg)\|_F^2.$$
Finally, the last step consists in computing the ratio between the two quantities and interpolating the discrete points $(m\tau, \big(g \ast \gamma_{\b{x}} \big)(m\tau))$. 

\paragraph{Variance of the estimator}
Studying the bias of \eqref{eq:PSD_estimator} reveals its interest~:
\begin{equation} \label{eq:psd_estimation}
\frac{ \Esp \left\{\| g_m(\Larg) \tilde{\b{x}} \|_2^2 \right\}}{ \|g_m(\Larg)\|_F^2 } = \frac{\sum_{\ell=0}^{N-1} \left(g(\ll-m\tau) \right)^2 \gamma_{\b{x}}(\ll)}{\sum_{\ell=0}^{N-1} \left( g(\ll-m\tau)\right)^2},
\end{equation}
where $\b{x}$ is the stationary stochastic graph signal.
For a filter $g$ well concentrated at the origin, \eqref{eq:psd_estimation} gives a smoothed estimate of $\gamma_{\b{x}}(m \tau)$. This smoothing corresponds to the windowing operation in the vertex domain: the less localized the kernel $g$ in the spectral domain, the more pronounced the smoothing effect in~\eqref{eq:psd_estimation} and the more concentrated the window in the vertex domain. It is very interesting to note we recover the traditional trade-off between bias and variance in non-parametric spectral estimation. Indeed, if $g$ is very sharply localized on the spectrum, ultimately a Dirac delta, the estimator \eqref{eq:PSD_estimator} is unbiased. Let us now study the variance. Intuitively, if the signal is correlated only over small regions of the vertex set, we could isolate them with localized windows of a small size and averaging those uncorrelated estimates together would reduce the variance. These small size windows on the vertex set correspond to large band-pass kernel $g_m$ and therefore large bias. However, if those correlated regions are large, and this happens when the PSD is localized in low-frequencies, we cannot hope to benefit from vertex-domain averaging since the graph is finite. Indeed the corresponding windows $g_m$ on the vertex set are so large that a single window spans the whole graph and there is no averaging effect: the variance increases precisely when we try to suppress the bias.

\begin{figure}[htb!]
\begin{center}
\includegraphics[width=0.45\linewidth]{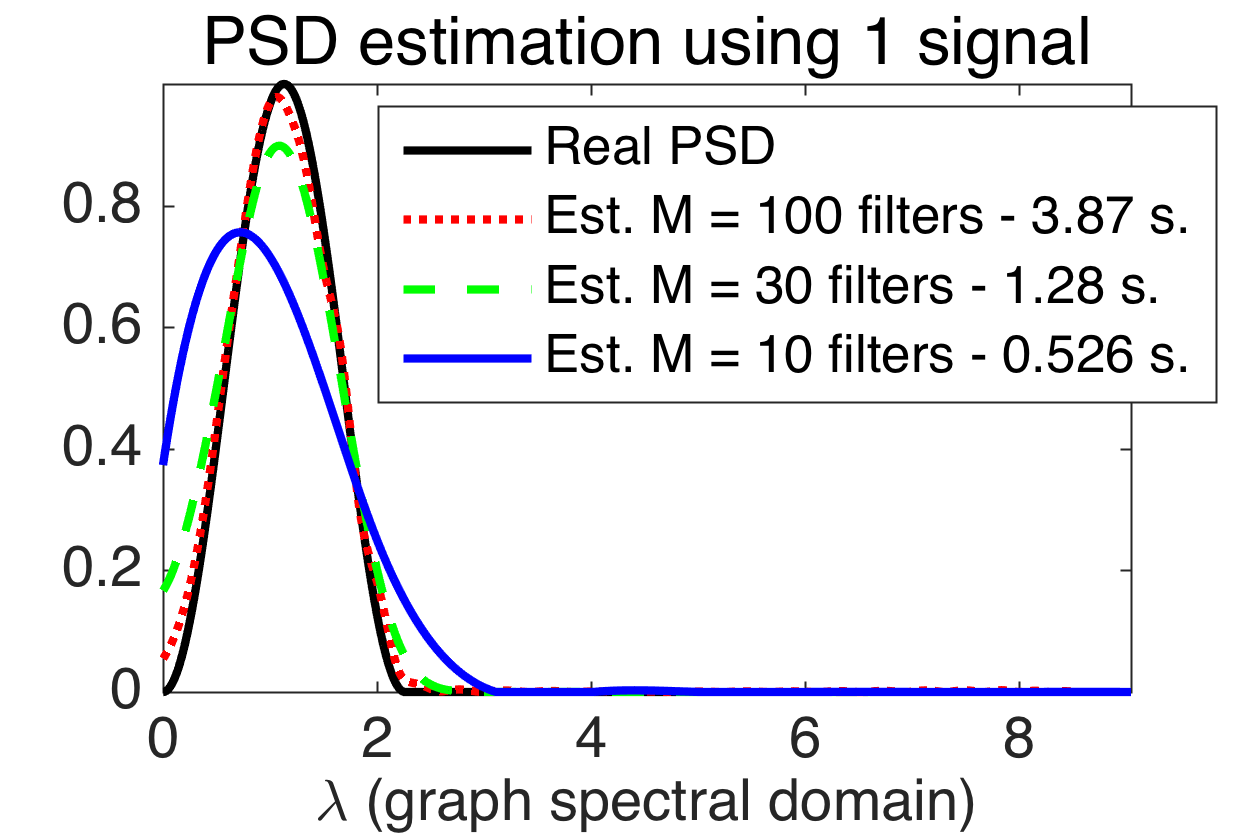}
\includegraphics[width=0.45\linewidth]{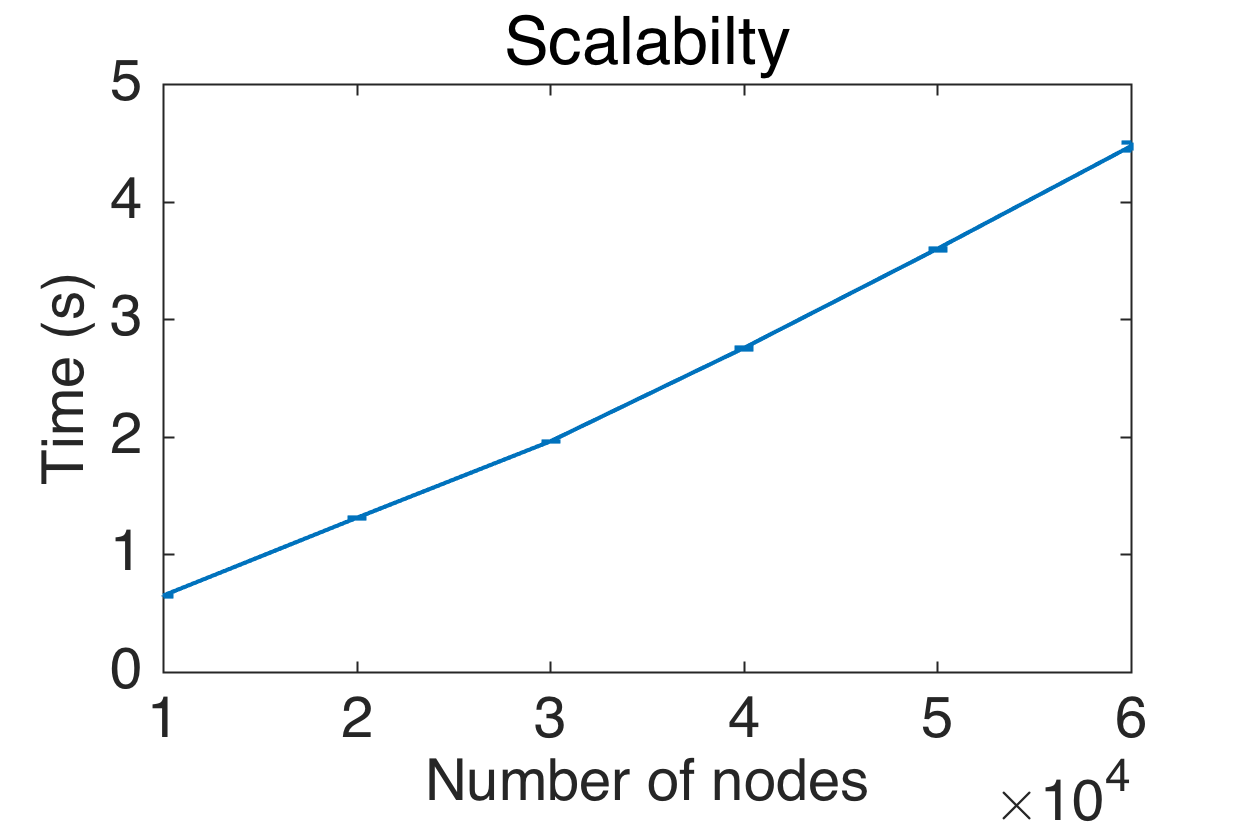} 
\end{center}
\caption{Left: PSD estimation on a graph of $20'000$ nodes with $K = 1$ measurements. Our algorithm is able to successively estimate the PSD of a signal. Right: Computation time versus size of the graph (average over $10$ runs.). We use $m=30$ filters. The algorithm scales linearly with the number of edges.}
\label{fig:psd_estimation}
\end{figure}

\paragraph{Experimental assessment of the method} 
Figure~\ref{fig:psd_estimation} shows the results of our PSD-estimation algorithm on a $10$-nearest neighbors graph of $20'000$ nodes (random geometric graph, weighted with an exponential kernel) and only $K = 1$ realization of the stationary graph signal. We compare the estimation using frames of $M=$ $10$, $30$, $100$ Gaussian filters. The parameters $\sigma$ and $\tau$ are adapted to the number of filters such that the shifted windows have an overlap of approximately $2$ ($\tau = \sigma^2 = \frac{(M+1)\lambda_{\max }}{M^2}$). For this experiment $K_2$ is set to $4$ and the Chebysheff polynomial order is $30$ 
The estimated curves are smoothed versions of the PSD. 

\paragraph{Complexity analysis} 
The approximation scales with the number of edges of the graph $\mathcal{O}(|\E|)$, (which is proportional to $N$ in many graphs). Precisely, our PSD estimation method necessitates $ (K + K_2) M $ filtering operations (with $M$ the number of shifts of $g$). A filtering operation costs approximately $O_c |E|$, with $O_c$ the order of the Chebysheff polynomial \cite{susnjara2015accelerated}. The final computational cost of the method is thus $\mathcal{O}\left( O_c (K + K_2) M |\E| \right)$.

\paragraph{Error analysis} The difference between the approximation and the exact PSD is caused by three different factors.
\begin{enumerate}
\item The inherent bias of the estimator, which is now directly controlled by the parameter $\sigma$.
\item We estimate the expected value using $K_1$ realization of the signal (often $K_1=1$). For large graphs $N\gg K_1$ and a few filters $M \oldll N$, this error is usually low because the variance of $\| g_m(\Larg) \tilde{\b{x}} \|_2^2$ is inversely proportional to bias. The estimation error improves as $\frac{1}{K_1}$.
\item We use a fast-filtering method based on a polynomial approximation of the filter. For a rough approximation, $\sigma \gg \frac{\lmax}{N}$, this error is usually negligible. However, in the other cases, this error may become large. 
\end{enumerate}


\section{Graph Wiener filters and optimization framework} \label{sec:wiener}
\begin{figure}[htb!]
\begin{center}
\includegraphics[width=0.30\textwidth]{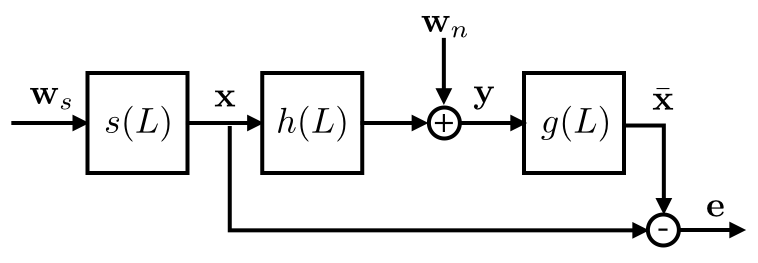}
\end{center}
\caption{Wiener estimation scheme. $\b{w}_s$ is a centered random variable with covariance $I$. $\b{w}_s$ generates the stationary stochastic signal $\b{x}$ thanks to the filter $s(\Larg)$. The random variable $\b{y}$ is then generated by filtering $\b{x}$ through $h(\Larg)$ and adding uncorrelated noise $\b{w}_n$ with PSD $n(\lambda)$. The estimator of $\b{x}$ given $\b{y}$: $\bar{\b{x}}|\b{y}$  is obtained with the Wiener filter $g(\Larg)$. The estimation error is denoted $\b{e}$. For simplicity, we present the case where $m_{\b{x}}=0$.}
\label{fig:winner_filter}
\end{figure}

Using stationary signals, we can naturally extend the framework of Wiener filters~\cite{Wiener1949extrapolation} largely used in signal processing for Mean Square Error (MSE) optimal linear prediction. Wiener filters for graphs have already been succinctly proposed in~\cite[pp 100]{girault2015signal}. Since the construction of Wiener filters is very similar for non-graph and graph signals, we present only the latter here. The main difference is that the traditional frequencies are replaced by the graph Laplacian eigenvalues\footnote{The graph eigenvalues are equivalent to classical squared frequencies.} $\ll$. Figure~\ref{fig:winner_filter} presents the Wiener estimation scheme.

\paragraph{Graph Wiener filtering}
The Wiener filter can be used to produce a mean-square error optimal estimate of a stationary signal under a linear but noisy observation model.
Let us consider the GWSS stochastic signal $\b{x}$ with PSD of $s^2(\ll)$. For simplicity, we assume $m_{\b{x}}=0$ in this subsection. 
The measurements $\bf{y}$ are given by:
\begin{equation}
\b{y} = h(\Larg)  \b{x} + \b{w}_n,
\end{equation}
where $h(\Larg)$ is a graph filter and $\b{w}_n$ additive uncorrelated noise of PSD $n(\ll)$.

To recover $\b{x}$, Wiener filters can be extended to the graph case:
\begin{equation} \label{eq:Wiener_filter}
g(\ll) = \frac{h(\ll)s^2(\ll)}{h^2(\ll)s^2(\ll) + n(\ll)}.
\end{equation}
The expression above can be derived by exactly mimicking the classical case and minimizes the expected quadratic error, which can be written as:
$$
e[\ell] = \Esp \left\{ \left( \hat{\b{x}}[\ell] - \hat{\bar{\b{x}}}[\ell]\right)^2\right\}= \Esp \left\{ \hat{\b{x}}[\ell] - g(\ll)\hat{\b{y}}[\ell]\right\}^2,
$$
where $\bar{\b{x}} = g(\Larg) \b{y}$ is the estimator of $\b{x}$ given $\b{y}$.
Theorem~\ref{theo:optimality} proves the optimality of this filter for the graph case. 

\paragraph{Wiener optimization}
In this contribution, we would like to address a  more general problem. 
Let us suppose that our measurements are generated as:
\begin{equation} \label{eq:general_data_model}
\b{y} = H\b{x}+\b{w}_n,
\end{equation}
where the GWSS stochastic graph signal $\b{x}$ has a PSD denoted $s^2(\ll)$ and the noise $\b{w}_n$ a PSD of $n(\ll)$. We assume $\b{x}$ and $\b{w}_n$ to be uncorrelated.
$H$ is a general linear operator not assumed to be diagonalizable with $\Larg$. As a result, we cannot build a Wiener filter that constructs a direct estimation of the signal $x$.
If $\b{x}$ varies smoothly on the graph, i.e is low frequency based, a classic optimization scheme would be the following:
\begin{equation} \label{prob:graph_tik}
\bar{\b{x}}|\b{y} = \argmin_{\b{x}} \|H\b{x}-\b{y}\|_2^2 + \beta \b{x}^*L\b{x} .
\end{equation}
This optimization scheme presents two main disadvantages. Firstly, the parameter $\beta$ must be tuned in order to remove the best amount of noise. Secondly, it does not take into account the data structure characterized by the PSD $s^2(\ll)$.

Our solution to overcome these issues is to solve the following optimization problem that we suggestively call \emph{Wiener optimization}
\begin{equation} \label{prob:Wiener-opt}
 \bar{\b{x}}|\b{y} = \argmin_{\b{x}}  \|H\b{x} - \b{y}\|_2^2 + \|w(\Larg) (\b{x}-m_{\b{x}}) \|_2^2,
 \end{equation}
where $w(\ll)$ is the Fourier penalization weights. These weights are defined as 
$$
w(\ll)=\left|\frac{\sqrt{n(\ll)}}{s(\ll)}\right|=\frac{1}{\sqrt{SNR(\ll)}}.
$$
Notice that compared to \eqref{prob:graph_tik}, the parameter $\beta$ is exchanged with the PSD of the noise. As a result, if the noise parameters are unknown, Wiener optimization does not solve completely the issue of finding the regularization parameter.
In the noise-less case, one can alternatively solve the following problem
\begin{equation} \label{prob:Wiener-opt-noiseless}
 \bar{\b{x}} = \argmin_x  \|s^{-1}(\Larg) (\b{x}-m_{\b{x}}) \|_2^2, \hspace{1cm} \text{s. t. } H\b{x} = \b{y} .
 \end{equation}
  For both Problems~\ref{prob:Wiener-opt} and~\ref{prob:Wiener-opt-noiseless} we assume that $0\times\infty=0$. It forces $\hat{\b{x}}[\ell] =0$ when $s(\lambda_\ell)=0$.
Problem~\eqref{prob:Wiener-opt} generalizes Problem~\eqref{prob:graph_tik} which assumes implicitly a PSD of $\frac{1}{\ll}$ and a constant noise level of $\gamma$ across all frequencies.
Note that this framework generalizes two main assumptions made on the data in practice:
\begin{enumerate}
	\item The signal is smooth on the graph, i.e: the edge derivative has a small $\ell_2$-norm. As seen before this is done by setting the PSD as $\frac{1}{\ll}$. This case is studied in \cite{girault2014semi}.
	\item The signal is band-limited, i.e it is a linear combination of the $k$ lowest graph Laplacian eigenvectors. This class of signal simply have a null PSD for $\lambda_\ell>\lambda_k$.
\end{enumerate}

\paragraph{Theoretical motivations for the optimization framework}
The first motivation is intuitive. The weight $w(\ll)$ heavily penalizes frequencies associated with low SNR and vice versa.

The second and main motivation is theoretical. If we have a Gaussian Random multivariate signal with i.i.d Gaussian noise, then Problem~\eqref{prob:Wiener-opt} is a MAP estimator.
\begin{theorem} \label{theo:map}
If $\b{x}\sim\mathcal{N}\left(0,s^{2}(\Larg)\right)$ and $\b{w}_n\sim\mathcal{N}\left(0,\sigma^2 I \right)$, i.e: $\b{x}$ is GWSS and Gaussian, then problem~\eqref{prob:Wiener-opt} is a MAP estimator for $\b{x}|\b{y}$ 
\end{theorem}
The proof is given in Appendix \ref{sec:proof map}. 
\begin{theorem} \label{theo:lmmse}
If $\b{x}$ is GWSS with PSD $s^2(\Larg)$ and $\b{w}_n$ is i.i.d white noise, i.e: $n(\ell) = \sigma^2$, then problem~\eqref{prob:Wiener-opt} leads to the linear minimum mean square estimator:
\begin{eqnarray}
\b{x}|\b{y} 
&= & \Sigma_{\b{x}\b{y}}  \Sigma_{\b{y}}^{-1}\b{y} + \left( I- \Sigma_{\b{x}\b{y}}  \Sigma_{\b{y}}^{-1}H\right) m_{\b{x}}
\end{eqnarray}
with $\Sigma_{\b{x}\b{y}}  = s^2(\Larg) H^{*} $ and $\Sigma_{\b{y}} = H s^2(\Larg) H^{*}$
\end{theorem}
The proof is given in Appendix \ref{sec:proof lmmse}.

Additionally, when $H$ is jointly diagonalizable with $\Larg$, Problem~\eqref{prob:Wiener-opt} can be solved by a single filtering operation.
\begin{theorem} \label{theo:optimality}
If the operator $H$ is diagonalizable with $\Larg$, (i.e: $H = h(\Larg) = U a(\Lambda)U^*$), then problem \eqref{prob:Wiener-opt} is optimal with respect to the weighting $w$ in the sense that its solution minimizes the mean square error:
\begin{equation*}
\Esp \left\{\| \b{e} \|_2^2 \right\} = \Esp \left\{\| \bar{\b{x}} - \b{x} \|_2^2 \right\} = \Esp \left\{ \sum_{i=1}^N \left(\bar{\b{x}}[i] - \b{x}[i] \right)^2 \right\}.
\end{equation*}
Additionally, the solution can be computed by the application of the corresponding Wiener filter.
\end{theorem}
The proof is given in Appendix \ref{sec:proof of optimality}. 

The last motivation is algorithmic and requires the knowledge of proximal splitting methods~\cite{combettes2011proximal,komodakis2015playing}. Problem~\eqref{prob:Wiener-opt} can be solved by a splitting scheme that minimizes iteratively each of the terms. The minimization of the regularizer, i.e the proximal operator of $\|w(\Larg) \tilde{x} \|_2^2 $, becomes a Wiener de-noising operation:
\begin{eqnarray*}
\mbox{prox}_{\frac{1}{2}\|w(\Larg)\tilde{x}\|_{2}^{2}}(y) 
& = & m_{\b{x}}+\argmin_{\tilde{x}}\|w(\Larg)\tilde{x}\|_{2}^{2}+\|\tilde{x}-\tilde{y}\|_{2}^{2} \\ 
& = & m_{\b{x}} + g(\Larg)\tilde{y} = m_{\b{x}} + g(\Larg) (y- m_{\b{x}})
\end{eqnarray*}
with 
$$
g(\ll)=\frac{1}{1+w^2(\ll)}=\frac{s^2(\ll)}{s^2(\ll)+n(\ll)}.
$$

\paragraph{Advantage of the Wiener optimization framework over a Gaussian MAP estimator}
Theorem~\ref{theo:map} shows that the optimization framework is equivalent to a Gaussian MAP estimator. In practice, when the data is only close to stationary, the true MAP estimator will perform better than Wiener optimization. So one could ask why we bother defining stationarity on graphs. Firstly, assuming stationarity allows us for a more robust estimate of the covariance matrix. This is shown is in Figure~\ref{fig:psd_estimation}, where only one signal is used to estimate the PSD (and thus the covariance matrix). Another example is the USPS experiment presented in the next section. We estimate the PSD by using only $20$ digits. The final result is much better than a Gaussian MAP based on the empirical covariance. Secondly, we have a scalable solution for Problem~\eqref{prob:Wiener-opt} (See Algorithm~\ref{CHalgorithm} below). On the contrary the classical Gaussian MAP estimator requires the explicit computation of a large part of the covariance matrix and it's inverse, which are both not scalable operations.

\paragraph{Solving Problem~\eqref{prob:Wiener-opt}}
Note that Problem~\eqref{prob:Wiener-opt} can be solved with a simple gradient descent. However, for a large number of nodes $N$, the matrix $w(\Larg)$ requires $\O(N^3)$ operations to be computed and $\O(N^2)$ bits to be stored. This difficulty can be overcome by applying its corresponding filter operator at each iteration. As already mentioned, the cost of the approximation scale with the number of edges $\O(O_c|E|)$~\cite{susnjara2015accelerated}.

When $s(\ll)\approx 0 $ for some $\ll$ the operator $w(\Larg)$ becomes badly conditioned. To overcome this issue, Problem~\eqref{prob:Wiener-opt} can be solved efficiently using a forward-backward splitting scheme~\cite{combettes2005signal,combettes2011proximal,komodakis2015playing}. The proximal operator of the function $\|w(\Larg)\tilde{x}\|_2^2 $ has been given above and we use the term $ \|Hx-y\|_2^2$ as the differentiable function. Algorithm~\ref{CHalgorithm} uses an accelerated forward backward scheme~\cite{beck2009fast} to solve Problem~\eqref{prob:Wiener-opt} where $\beta$ is the step size (we select $\beta = \frac{1}{2\lmax(H)^2}$), $\epsilon$ the stopping tolerance, $J$ the maximum number of iterations and $\delta$ is a very small number to avoid a possible division by $0$.
\begin{algorithm}[ht!]
\caption{Fast Wiener optimization to solve~\eqref{prob:Wiener-opt}}
\label{CHalgorithm}
\begin{algorithmic}
\State INPUT: $z_1 = x$, $u_0 = x$, $t_1 = 1$, $\epsilon > 0$, $\beta \leq \frac{1}{2\lmax(H)^2}$
\State SET: $g(\lambda) = \frac{s^2(\lambda)}{s^2(\lambda)+\beta n(\lambda)}$ \Comment{Wiener filter}
\For{ $j = 1,\dots J$ }
\State  $v = z_j - \beta H^*(Hz_{j}-y)$ \Comment{Gradient step}
\State  $u_{j+1} = g(L) v$ \Comment{Proximal step}
\State  $t_{j+1} = \frac{1+\sqrt{1+4t_j^2}}{2}$ \Comment{FISTA scheme}
\State  $z_{j+1} = z_j +\frac{t_j-1}{t_{j+1}} (u_j-u_{j-1})$ \Comment{Update step}
\If{$\frac{\|z_{j+1} - z_{j}\|_F^2}{\| z_{j}\|_F^2+\delta}<\epsilon$} \Comment{Stopping criterion}
\State BREAK
\EndIf
\EndFor
\State SOLUTION: $z_J$
\end{algorithmic}
\end{algorithm}
\vspace{-0.5cm}

\section{Evidence of graph stationarity: illustration with USPS} \label{sec:USPS}

Stationarity may not be an obvious hypothesis for a general dataset, since our intuition does not allow us to easily capture the kind of shift invariance that is really implied. In this section we give additional insights on stationarity from a more experimental point of view. To do so, we will show that the well-known USPS dataset is close to stationary on a nearest neighbor graph. We show similar results with a dataset of faces.

Images can be considered as signals on the 2-dimensional euclidean plane and, naturally, when the signal is sampled, a grid graph is used as a discretization of this manifold. The corresponding eigenbasis is the 2 dimensional DCT\footnote{This is a natural extension of \cite{strang1999discrete}}. Many papers have exploited the fact that natural texture images are stationary 2-dimensional signals~\cite{heine1955models,jain1978partial,chellappa1985texture}, i.e stationary signals on the grid graph. In \cite{roux2008learning}, the authors go one step further and ask the following question: suppose that pixels of images have been permuted, can we recover their relative two-dimensional location? Amazingly, they answer positively adding that only a few thousand images are enough to approximately recover the relative location of the pixels. The grid graph seems naturally encoded within images.

The observation of~\cite{roux2008learning} motivates the following experiment involving stationarity on graphs. Let us select the USPS data set which contains $9298$ digit images of $16 \times 16$ pixels. We create 5 classes of data: (a) the circularly shifted digits\footnote{We performed all possible shifts in both directions. Because of this, the covariance matrix becomes Toeplitz}, (b) the original digits and (c), (d) and (e) the classes of digit $3$, $7$ and $9$. As a pre-processing step, we remove the mean of each pixel, thus forcing the first moment to be $0$, and focus on the second moment. For those 5 cases, we compute the covariance matrix $\Sigma$ and its "Fourier transform",
\begin{equation} \label{eq:Fourier_cov_matrix}
\Gamma = U^* \Sigma U,
\end{equation}
for 2 different graphs: (a) the grid and (b) the $20$ nearest neighbors graph. In this latter case, each node is a pixel and is associated to a feature vector containing the corresponding pixel value of all images. We use the squared euclidean distance between feature vectors and an exponential kernel to define edge weights\footnote{$W[i,n]=e^{\frac{-\|x_i-x_n\|_2^2}{\sigma^2}}$ if $x_i$ is in the $20$ nearest neighbors of $x_n$.}. We then compute the stationarity level of each class of data with both graphs using the following measure:
\begin{equation}
s_r(\Gamma) = \left(\frac{\sum_\ell \Gamma_{\ell,\ell}^2}{\sum_{\ell_1}\sum_{\ell_2} \Gamma_{\ell_1,\ell_2}^2} \right)^{\frac{1}{2}} =\frac{\| \rm{diag}(\Gamma) \|_2}{\| \Gamma \|_F} .
\end{equation}
The closer $s_r(\Gamma)$ is to $1$, the more diagonal the matrix $\Gamma$ is and the more stationary the signal. Table \ref{tab:stationarity measures} shows the obtained stationarity measures. The less universal the data, the less stationary it is on the grid. Clearly, specificity inside the data requires a finer structure than a grid. This is confirmed by the behavior of the nearest neighbors graph. When only one digit class is selected, the nearest neighbors graph still yields very stationary signals. 
\begin{table}
\begin{center}
\begin{tabular}{|l | c | c |}
 \hline
 Data \textbackslash Graph  & 2-dimensional grid & $20$ nearest neighbors graph \\
 \hline
 \hline
 Shifted all digits  & $0.86$ & $1$ 
 \\
 \hline
 All digits  & $0.66$ & $0.79$ \\
 \hline
 Digit 3  & $0.64$ & $0.83$ \\
 \hline
 Digit 7  & $0.52$ & $0.79$\\
 \hline
 Digit 9  & $0.52$ & $0.81$ \\
 \hline
\end{tabular}
\caption{$s_r(\Gamma) =\frac{\| \rm{diag}(\Gamma) \|_2}{\| \Gamma \|_F}$: stationarity measures for different graphs and different datasets. The nearest neighbors graph adapts to the data. The individual digits are stationary with the nearest neighbor graph.}
\label{tab:stationarity measures}
\end{center}
\vspace{-0.75cm}
\end{table}

\begin{figure}[htb!]
\begin{center}
\includegraphics[width=0.45\linewidth]{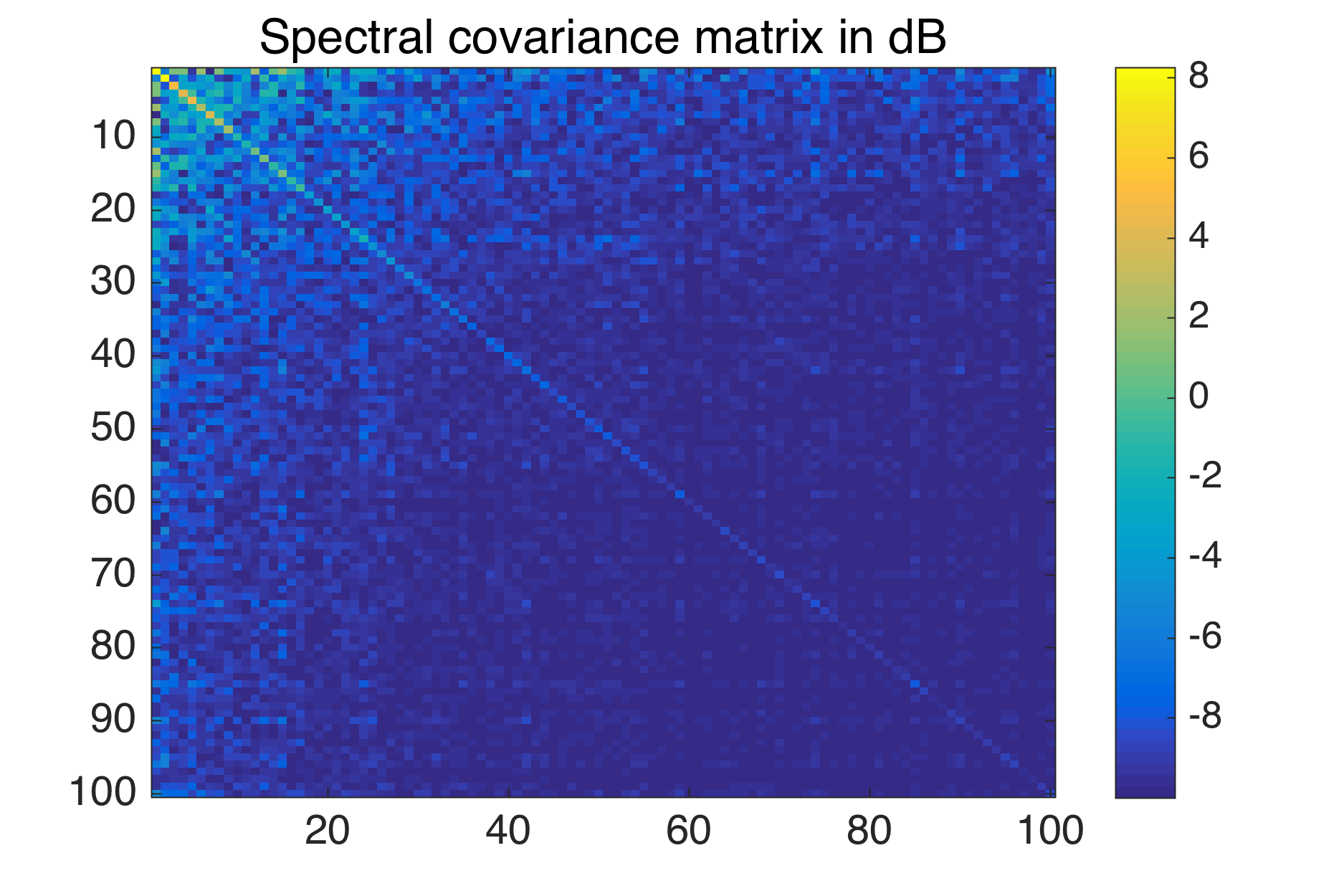} 
\includegraphics[width=0.45\linewidth]{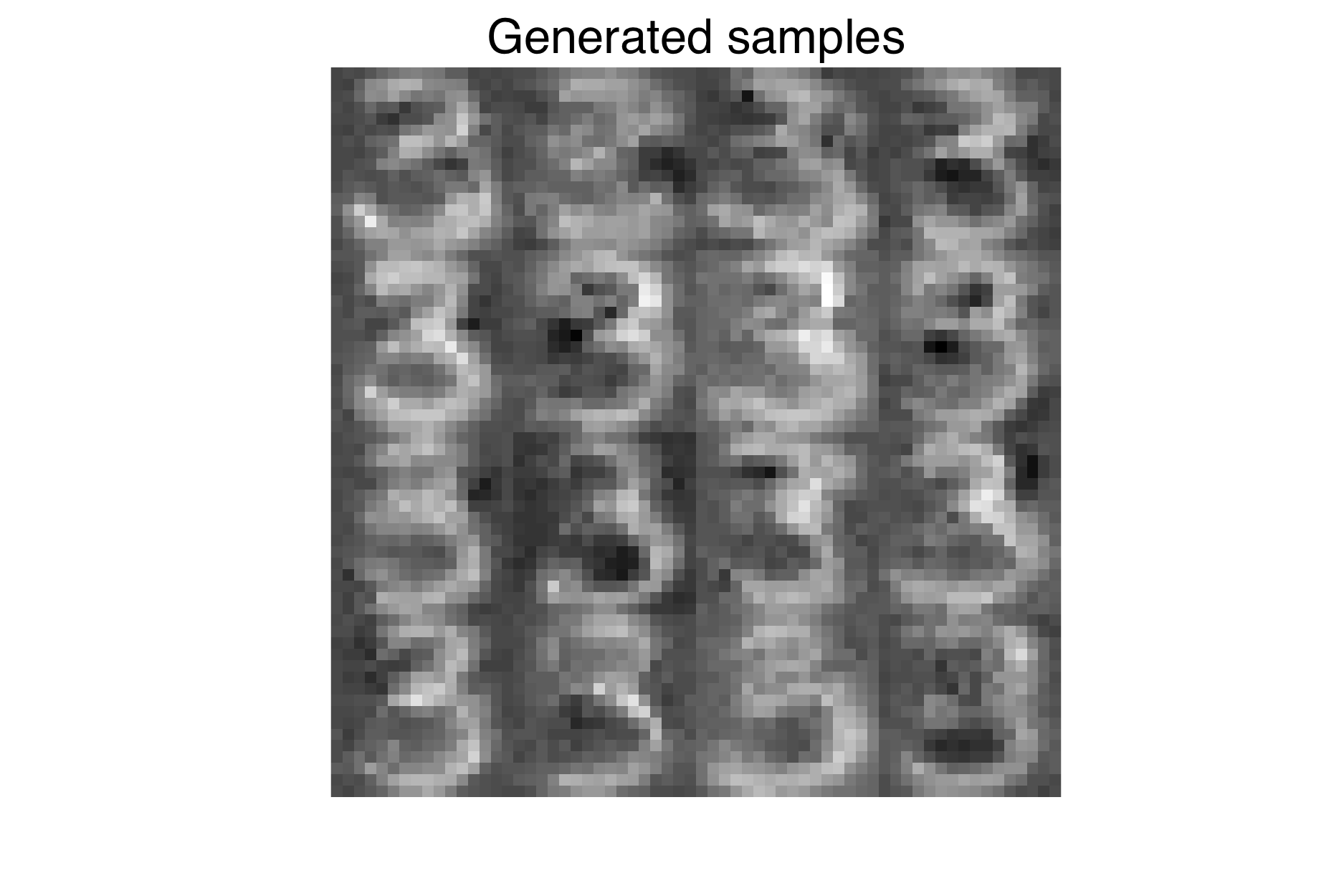} \\
\includegraphics[width=0.45\linewidth]{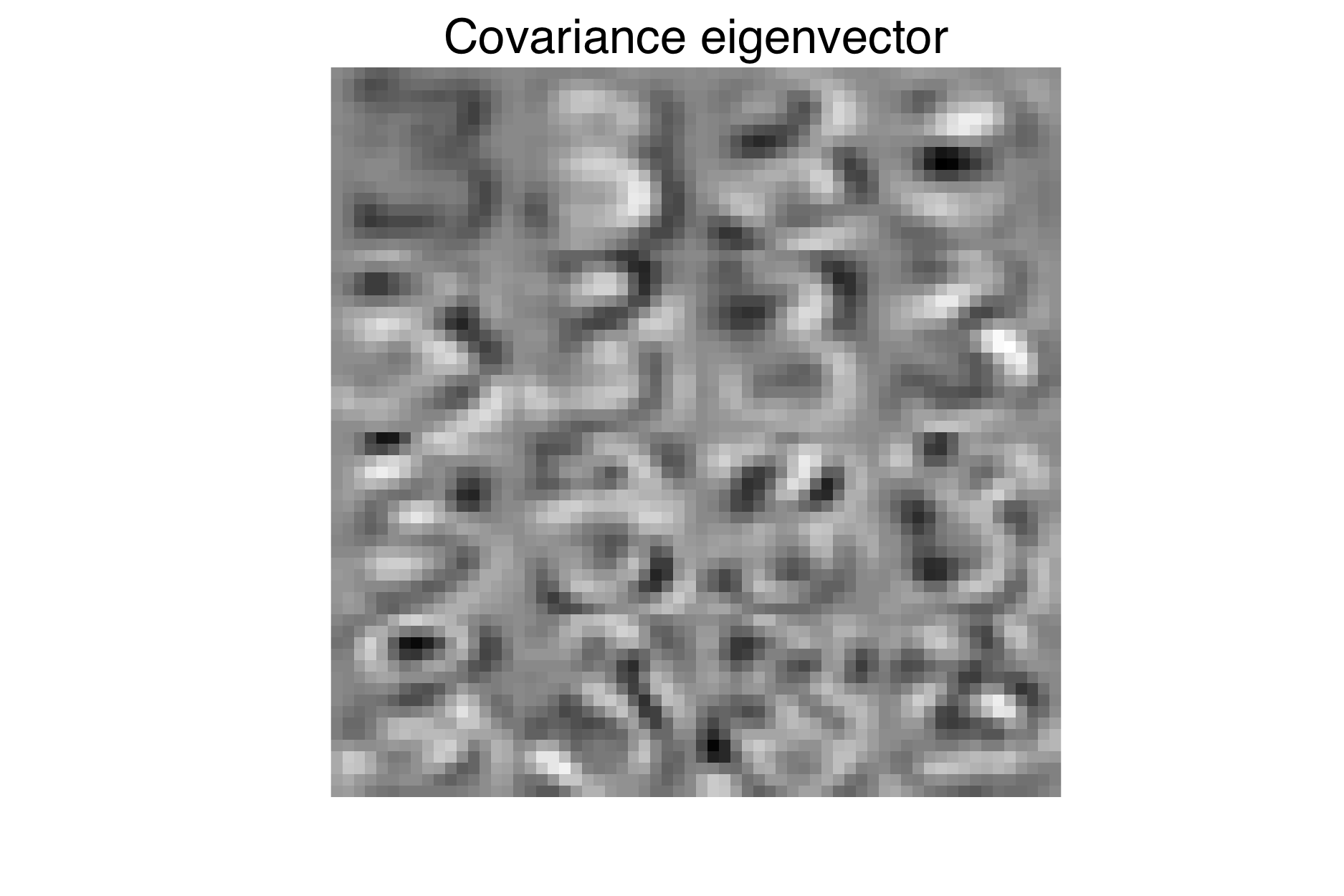} 
\includegraphics[width=0.45\linewidth]{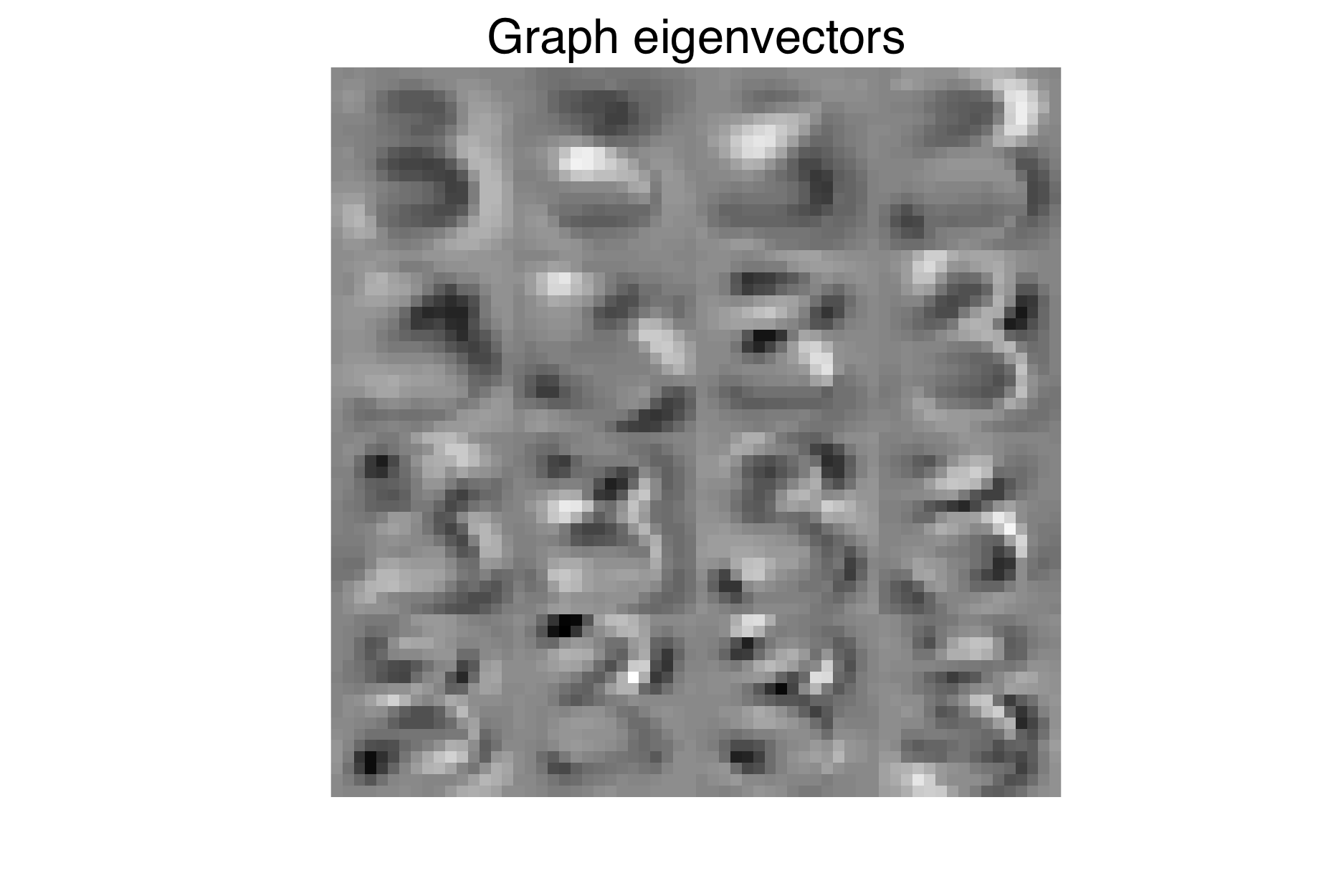} \\
\end{center}
\caption{Studying the number $3$ of USPS using a $20$-neighbors graph. Top left: Spectral covariance matrix of the data (Note the diagonal shape of the matrix). We only display the upper left part for better visibility. Top right: generated samples by filtering Gaussian random noise on the graph. Bottom left: Covariance eigenvectors associated with the $16$ highest eigenvalues. Bottom right: Laplacian eigenvectors associated with the $16$ smallest non-zero eigenvalues. Because of stationarity, Laplacian eigenvectors are similar to the covariance eigenvectors.}
\label{fig:exp3}
\vspace{-0.5cm}
\end{figure}

Let us focus on the digit $3$. For this experiment, we build a $20$ nearest neighbors graph with only $50$ samples. Figure~\ref{fig:exp3} shows the eigenvectors of the Laplacian and of the covariance matrix. Because of stationarity, they are very similar. Moreover, they have a $3$-like shape. Since the data is almost stationary, we can use the associated graph and the PSD to generate samples by filtering i.i.d Gaussian noise with the following PSD based kernel: $g(\lambda_\ell) = \sqrt{\Gamma_{\ell,\ell}}$. The resulting digits have a $3$-like shape confirming that the class is stationary on the nearest neighbors graph.

To further illustrate this phenomenon on a different dataset, we use the CMUPIE set of cropped faces. With a nearest neighbor graph we obtained a stationarity level of $s_r = 0.92$. This has already been observed in~\cite{he2005face} where the concept of Laplacianfaces is introduced. Finally, in \cite{shahid2016fast} the authors successfully use the graph between features to improve the quality of a low-rank recovery problem. The reason seems to be that the principal components of the data are the lowest eigenvectors of the graph, which is again a stationarity assumption.

To intuitively motivate the effectiveness of nearest neighbors at producing stationary signals, let us define the centering operator $J = I - \boldsymbol{1}\boldsymbol{1}^{\top}/N$. Given $K$ signal $x_k$,
the matrix of average squared distances between the centered features ($\sum_{i=1}^N x_k[i]=0$) is directly proportional to the covariance matrix~:
\begin{equation} \label{eq:link_gram}
\bar{\Sigma}_{\b{x}} = - \frac{1}{2} J D J,
\end{equation}
where $D[i,n] = \frac{1}{K}\sum_{k=1}^K \left( x_k[i] -x_k[n] \right)^2$ and $ \bar{\Sigma}_{\b{x}}[i,n] = \frac{1}{K} \sum_{k=1}^K x_k[i] x_k[n] $. The proof is given in Appendix~\ref{sec:proof_gram}.
The nearest-neighbors graph can be seen as an approximation of the original distance matrix, which pleads for using it as a good proxy destined to leverage the spectral content of the covariance. Put differently, when using realizations of the signal as features and computing the k-NN graph we are connecting strongly correlated variables via strong edge weights.


\section{Experiments} \label{sec:experiments}
All experiments were performed with the GSPBox~\cite{perraudin2014gspbox} and the UNLocBoX~\cite{perraudin2014unlocbox} two open-source software library. The code to reproduce all figures of the paper can be downloaded at: \url{https://lts2.epfl.ch/rrp/stationarity/}. As the stationary signals are random, the reader may obtain slightly different results. However, conclusions shall remain identical. The models used in our comparisons are detailed in the Appendix~\ref{sec:convex_model} for completeness, where we also detail how the tuning of the parameters is done. 
All experiments are evaluated with respect to the Signal to Noise Ratio (SNR) measure:
$$
\text{SNR}(x,\dot{x}) = - 10 \log \left( \frac{\text{var}(x-\dot{x}) }{\text{var}(x)} \right)
$$

\subsection{Synthetic dataset}
In order to obtain a first insight into applications using stationarity, we begin with some classical problems solved on a synthetic dataset. Compared to real data, this framework allows us to be sure that the signal is stationary on the graph. 
\paragraph{Graph Wiener deconvolution}
We start with a de-convolution example on a random geometric graph. This can model an array of sensors distributed in space or simply a mesh. The signal is chosen with a low frequency band-limited PSD. To produce the measurements, the signal is convolved with the heat kernel $h(\lambda) = e^{-\tau \lambda}$. Additionally, we add some uncorrelated i.i.d Gaussian noise. The heat kernel is chosen because it simulates a heat diffusion process. Using de-convolution we aim at recovering the original signal before  diffusion. For this experiment, we put ourselves in an ideal case and suppose that both the PSD of the input signal and the noise level are known.

\begin{figure}[htb!]
\begin{center}
\includegraphics[width=0.8\linewidth]{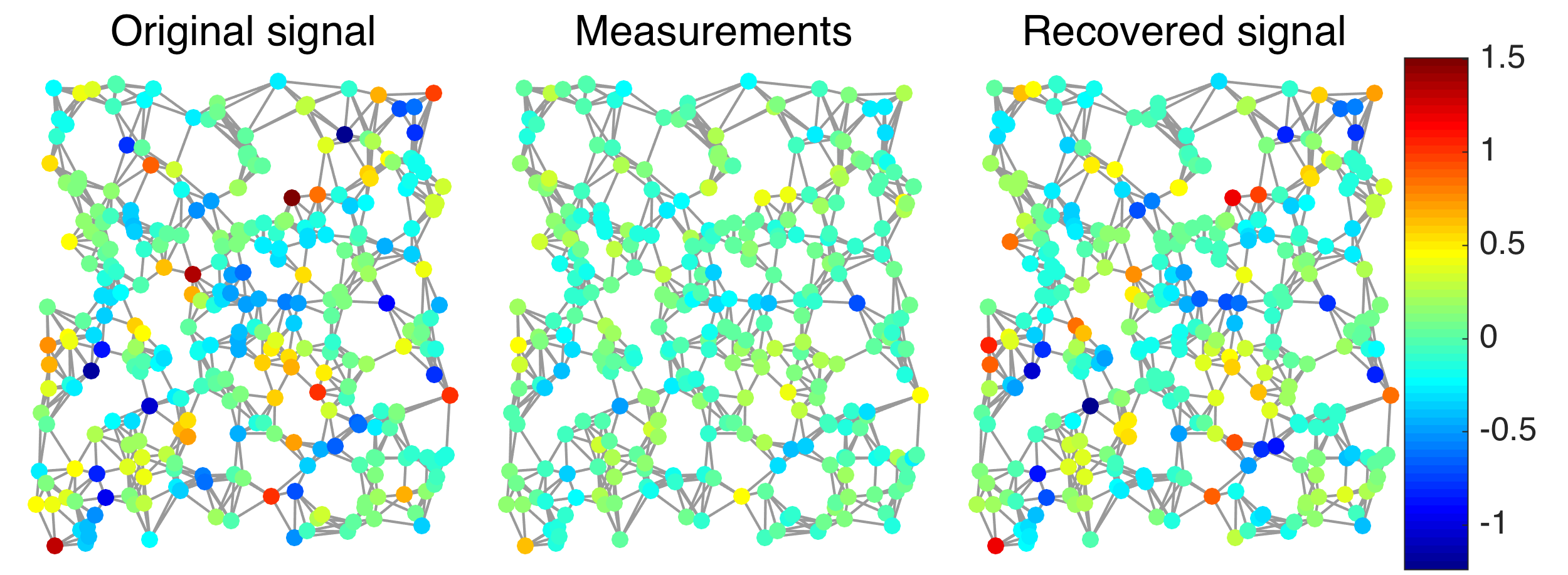} 
\includegraphics[width=0.4\linewidth]{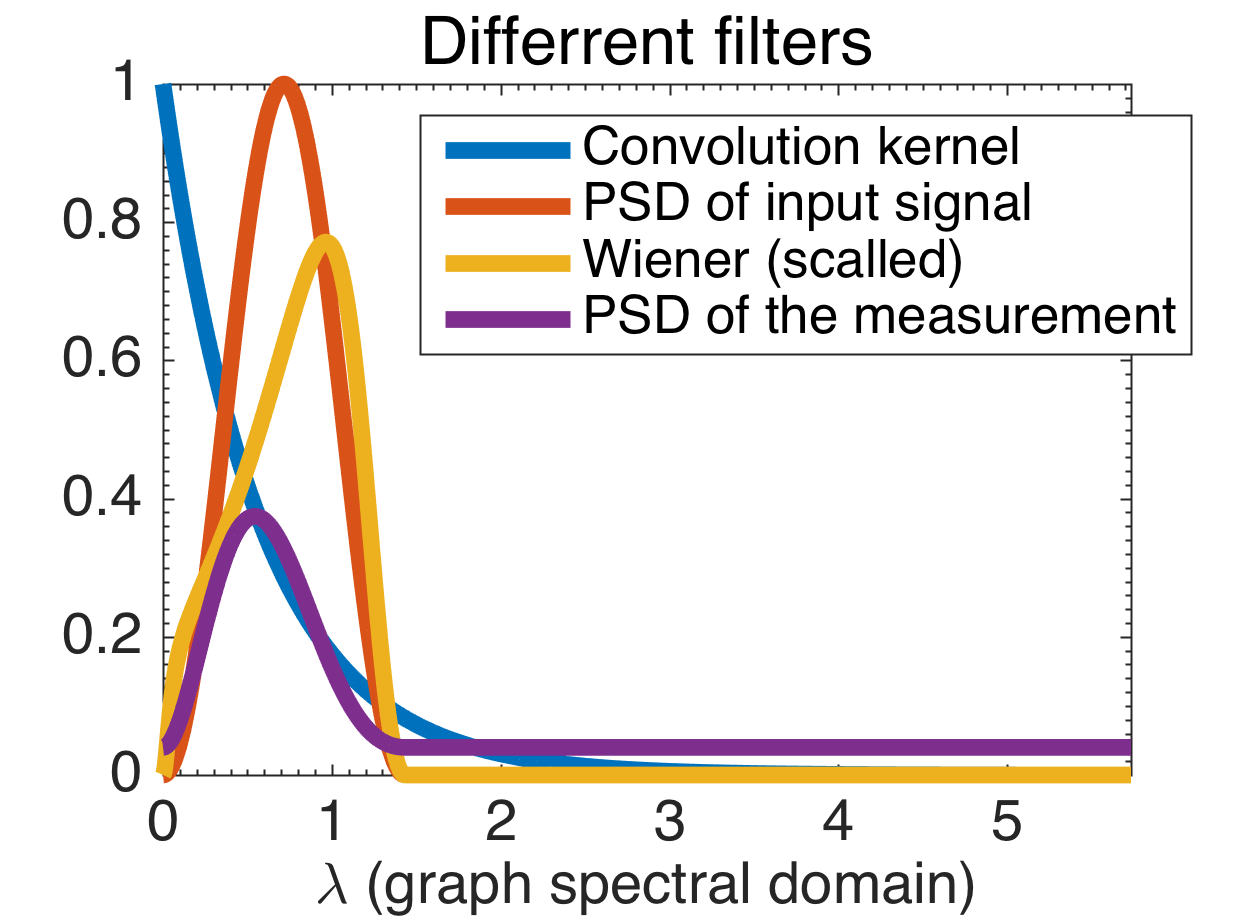} 
\includegraphics[width=0.4\linewidth]{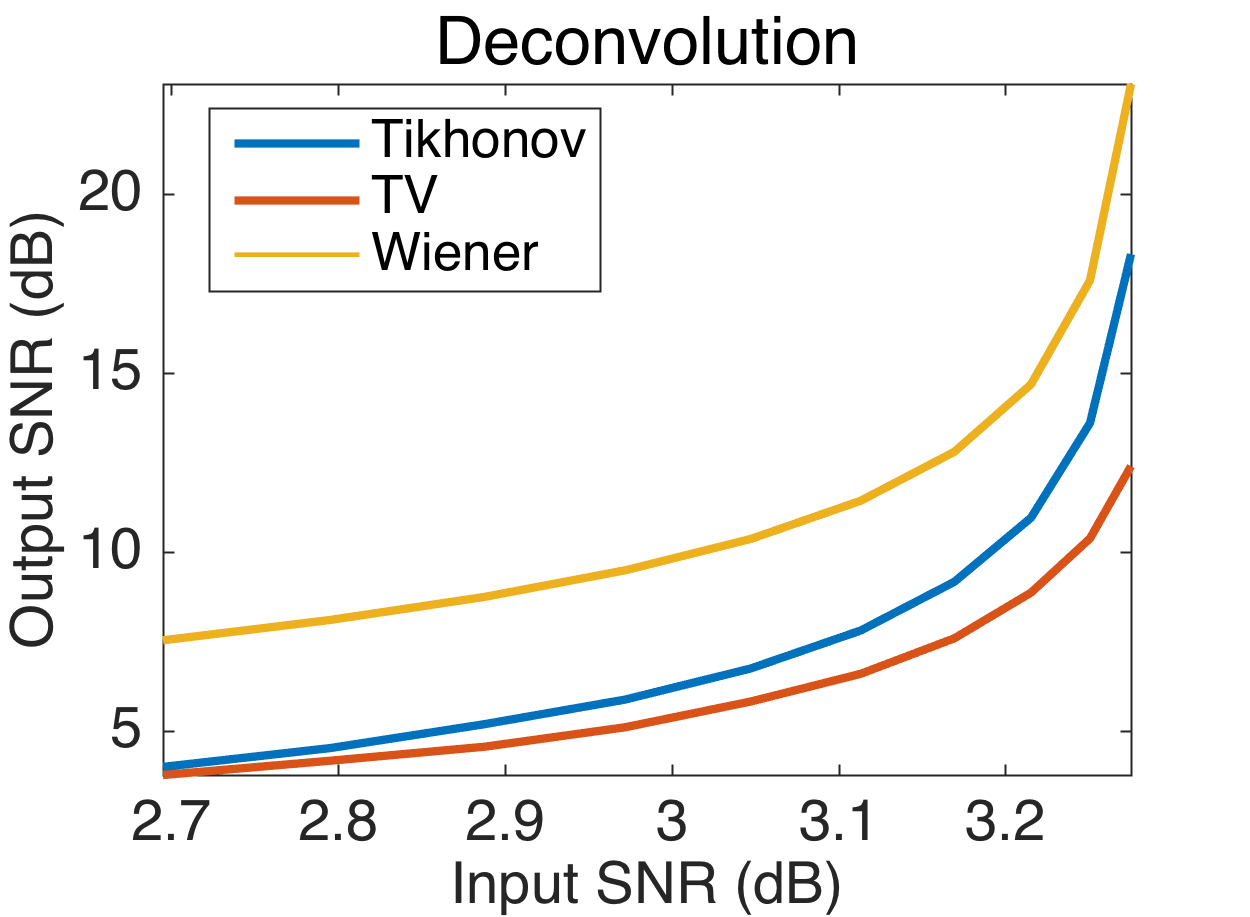} 
\end{center}
\caption{Graph de-convolution on a random geometric graph. The convolution kernel is $e^{-\frac{10 x}{\lmax}}$. Top: Signal and filters for a noise level of $0.16$. Bottom: evolution of the error with respect to the noise.}
\label{fig:synthetic-deconvolution}
\vspace{-0.25cm}
\end{figure}

Figure~\ref{fig:synthetic-deconvolution} presents the results. We observe that Wiener filtering is able to de-convolve the measurements. The second plot shows the reconstruction errors for three different methods: Tikhonov presented in problem \eqref{prob:deconvolution-tik}, TV in \eqref{prob:dec-in-den-tv} and Wiener filtering in \eqref{eq:Wiener_filter}. Wiener filtering performs clearly much better than the other methods because it has a much better prior assumption. 

\begin{figure}[htb!]
\begin{center}
\includegraphics[width=0.4\linewidth]{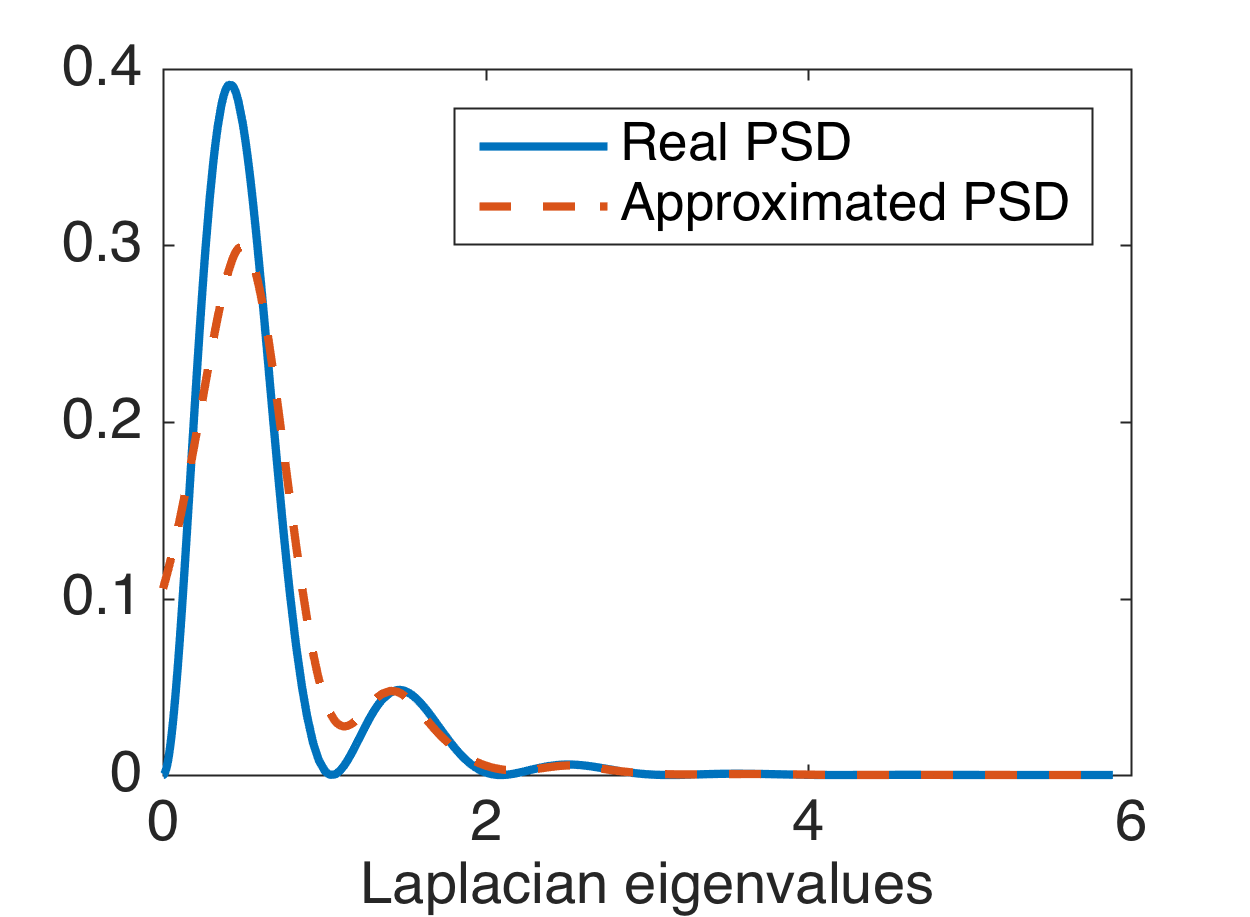} 
\includegraphics[width=0.4\linewidth]{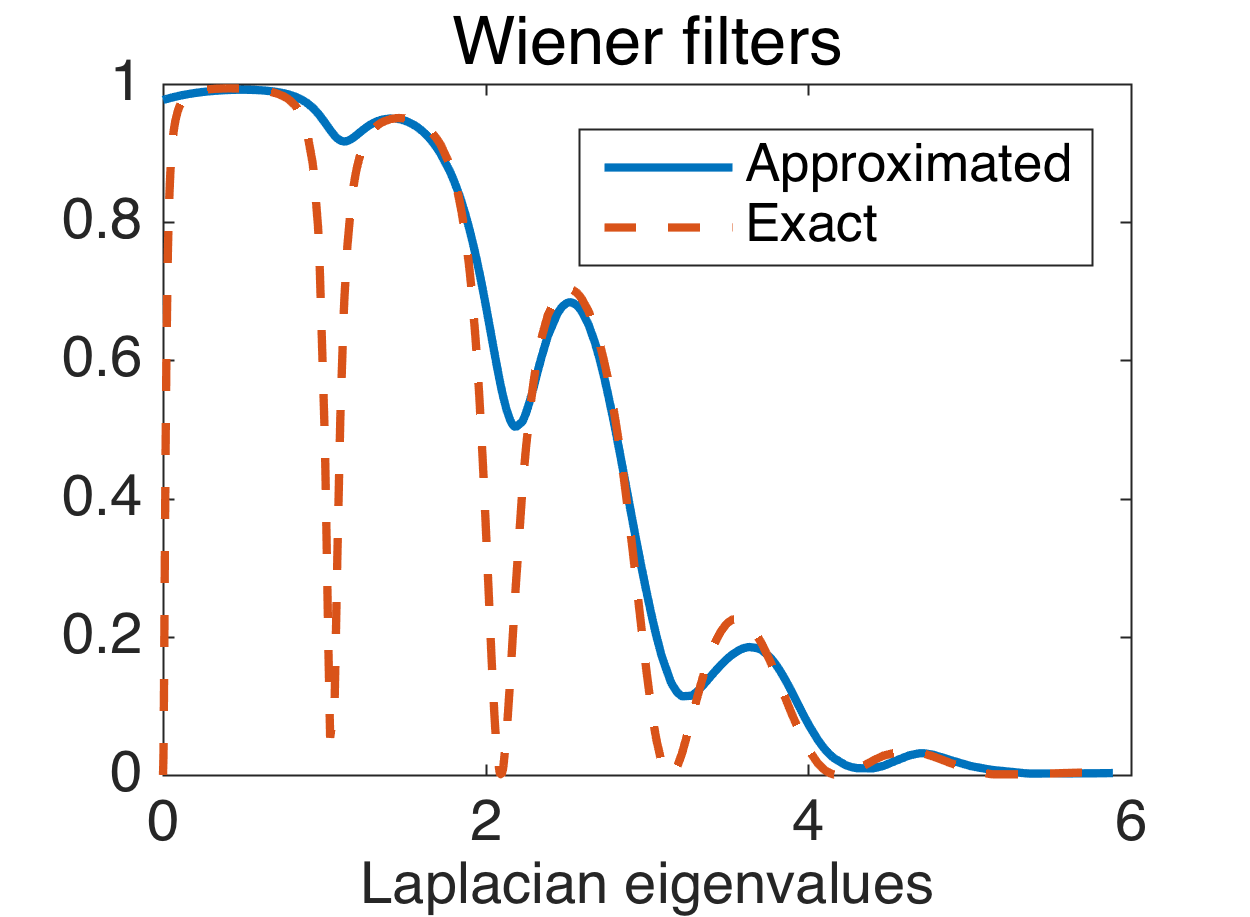} 
\includegraphics[width=0.8\linewidth]{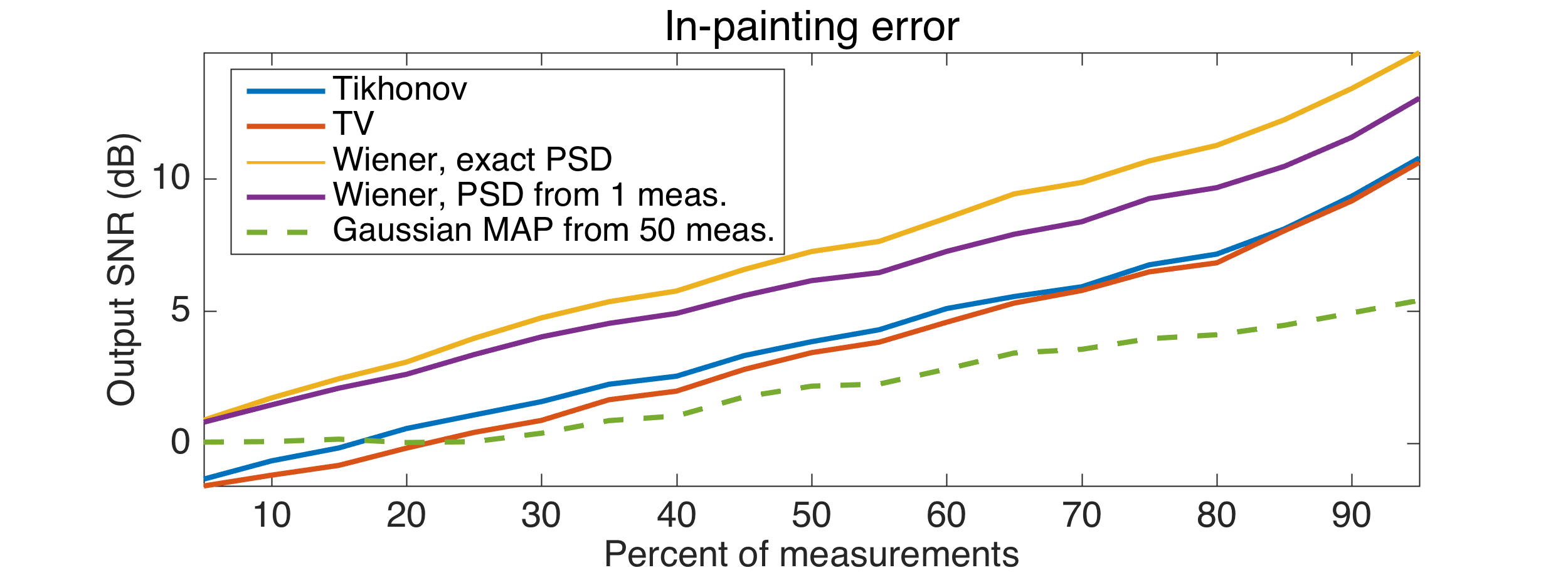} 
\end{center}
\caption{Wiener in-painting on a geometric graph of 400 nodes. Top: true VS approximated PSD and resulting Wiener filters. Bottom: in-painting relative error with respect to the number of measurements.}
\label{fig:synthetic-in-painting}
\vspace{-0.5cm}
\end{figure}

\paragraph{Graph Wiener in-painting}
In our second example, we use Wiener optimization to solve an in-painting problem. This time, we suppose that the PSD of the input signal is unknown and we estimate it using $50$ signals. Figure~\ref{fig:synthetic-in-painting} presents quantitative results for the in-painting. Again, we compare three different optimization methods: Tikhonov \eqref{prob:inpainting-tik}, TV \eqref{prob:dec-in-den-tv} and Wiener \eqref{prob:Wiener-opt}. Additionally we compute the classical MAP estimator based on the empirical covariance matrix (see \cite{rasmussen2004gaussian} 2.23). Wiener optimization performs clearly much better than the other methods because it has a much better prior assumption. Even with $50$ measurements, the MAP estimator performs poorly compared to graph methods. The reason is that the graph contains a lot of the covariance information. Note that the PSD estimated with only one measurement is sufficient to outperform Tikhonov and TV.

\subsection{Meteorological dataset}
We apply our methods to a weather measurements dataset, more precisely to the temperature and the humidity. Since intuitively these two quantities are correlated smoothly across  space, it suggests that they are more or less stationary on a nearest neighbors geographical graph. 

The French national meteorological service has published in open access a dataset\footnote{Access to the raw data is possible directly through our code or through the link \url{https://donneespubliques.meteofrance.fr/donnees_libres/Hackathon/RADOMEH.tar.gz}} with hourly weather observations collected during the Month of January 2014 in the region of Brest (France). From these data, we wish to ascertain that our method still performs better than the two other models (TV and Tikhonov) on real measurements. The graph is built from the coordinates of the weather stations by connecting all the neighbors in a given radius with a weight function $W[i,n] = e^{-d_{in}^2 \tau}$ where $\tau$ is adjusted to obtain an average degree around $3$ ($\tau$, however, is not a sensitive parameter). For our experiments, we consider every time-step as an independent realization of a GWSS signal. As sole pre-processing, we remove the temperature mean of each station independently. This is equivalent to removing the first moment. Thanks to the $744$ time observation, we can estimate the covariance matrix and check whether the signal is stationary on the graph. 

\begin{figure}[htb!]
\begin{center}
\includegraphics[width=0.8\linewidth]{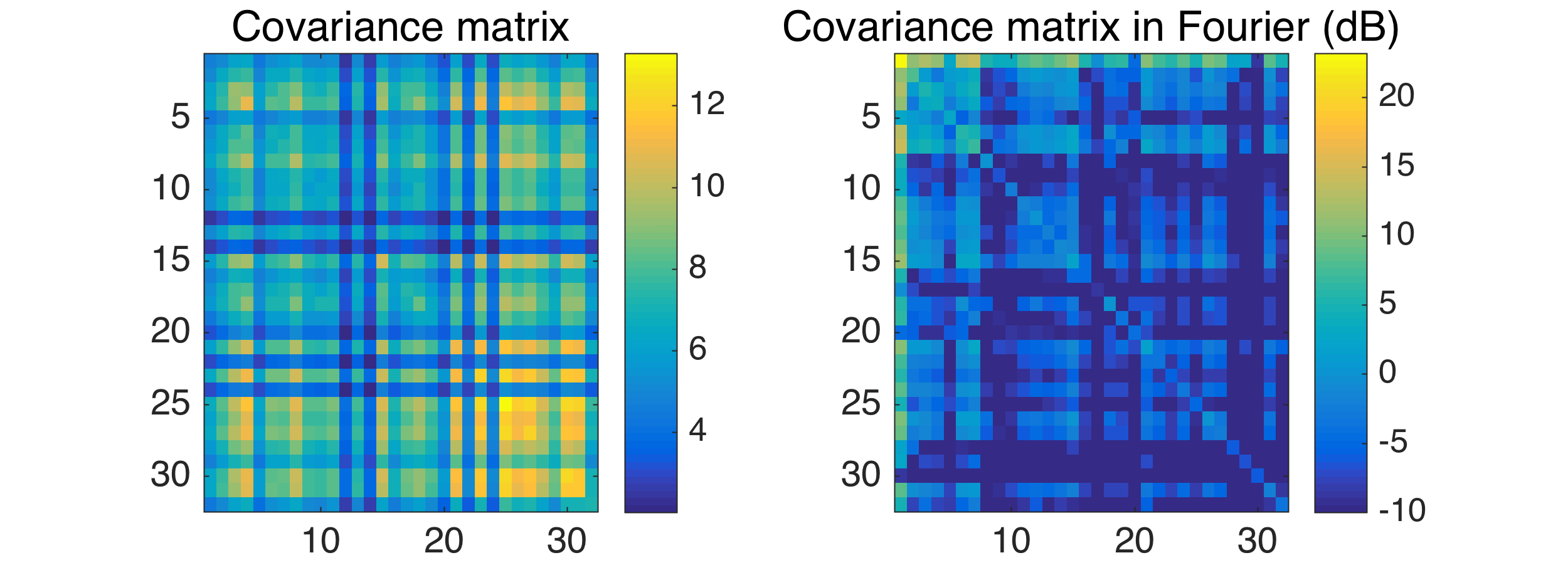} \\
\includegraphics[width=0.7\linewidth]{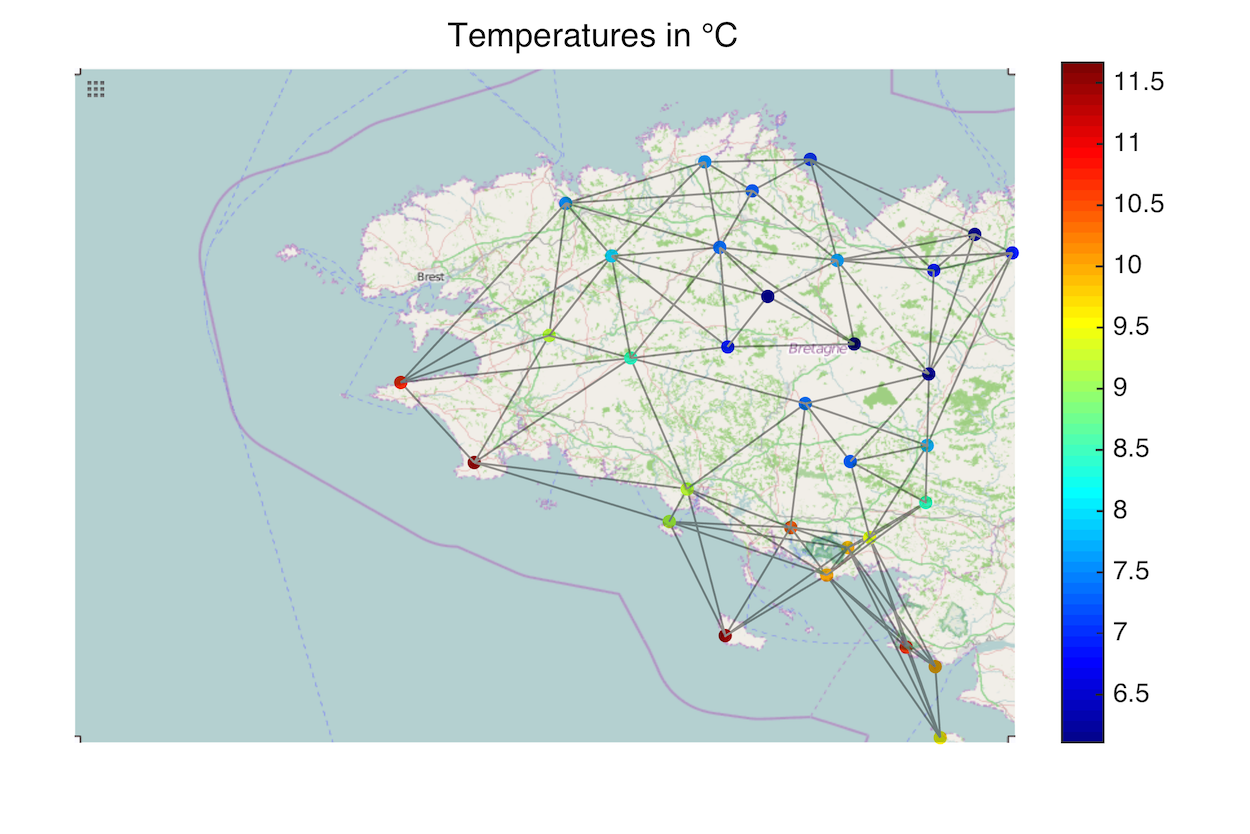} \\
\includegraphics[width=0.4\linewidth]{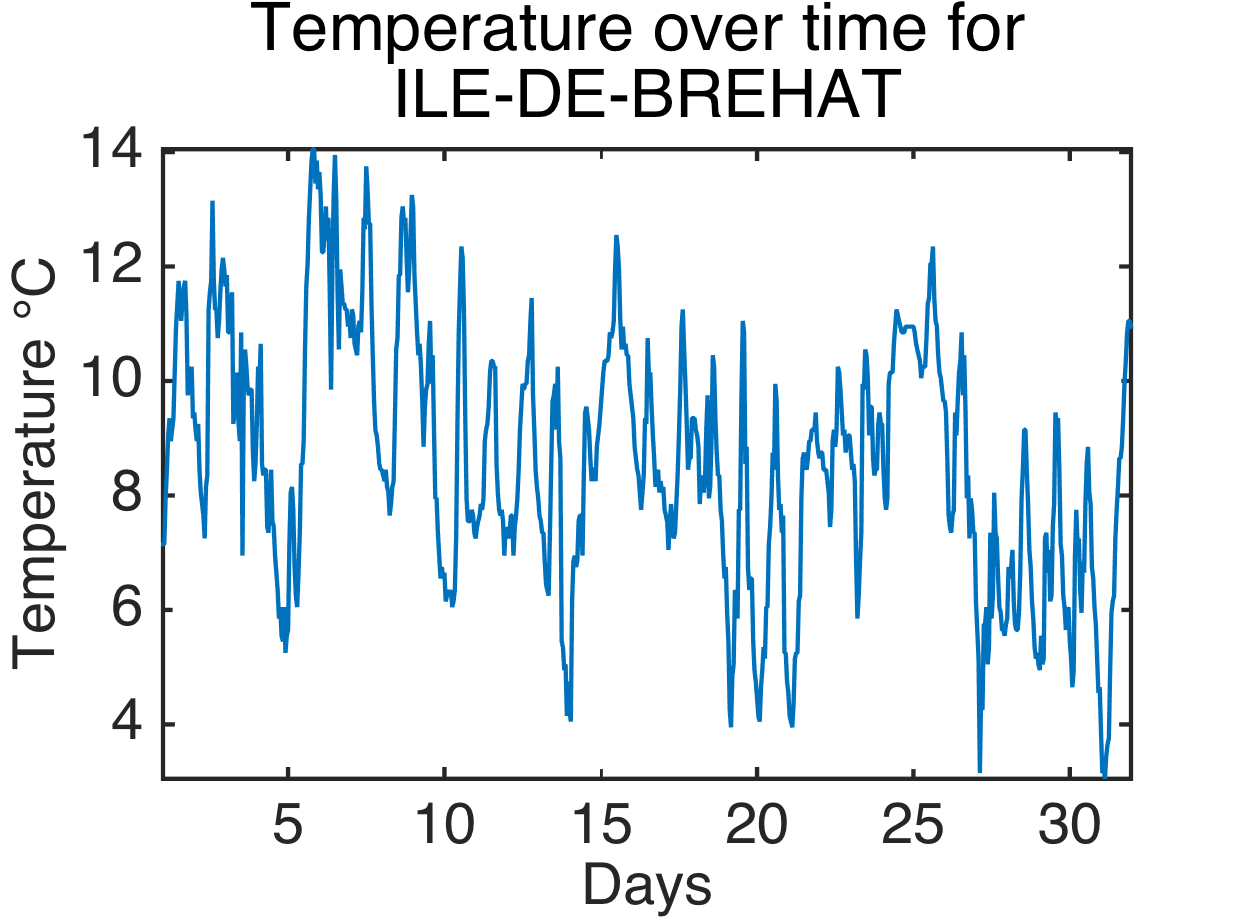} 
\includegraphics[width=0.4\linewidth]{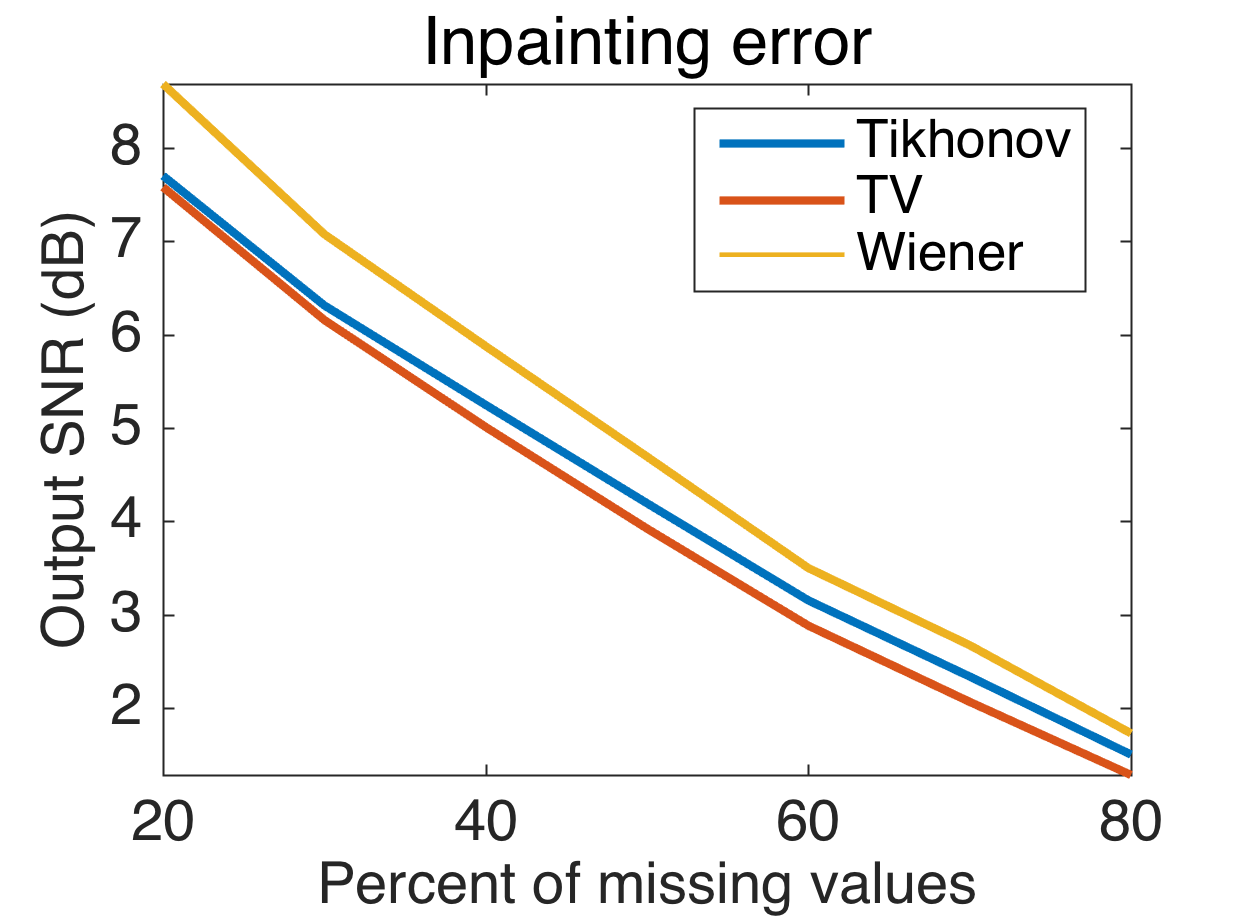} 
\end{center}
\caption{Experiments on the temperature of Molene. Top: Covariance matrices. Bottom left: A realization of the stochastic graph signal (first measure). Bottom center: the temperature of the Island of Brehat. Bottom right: Recovery errors for different noise levels.}
\label{fig:molene-temperature}
\vspace{-0.5cm}
\end{figure}

\paragraph{Prediction - Temperature} The result of the experiment with temperatures is displayed in Figure~\ref{fig:molene-temperature}. The covariance matrix shows a strong correlation between the different weather stations. Diagonalizing it with the Fourier basis of the graph shows that the meteorological instances are not really stationary within the distance graph as the resulting matrix is not really diagonal. However, even in this case, Wiener optimization still outperforms graph TV and Tikhonov models, showing the robustness of the proposed method.
In our experiment, we solve a prediction problem with a mask operator covering 50 per cent of measurements and an initial average SNR of $13.4$ dB . We then average the result over $744$ experiments (corresponding to the $744$ observations) to obtain the curves displayed in Figure~\ref{fig:molene-temperature}. 
We observe that Wiener optimization always performs better than the two other methods. 

\begin{figure}[htb!]
\begin{center}
\includegraphics[width=0.9\linewidth]{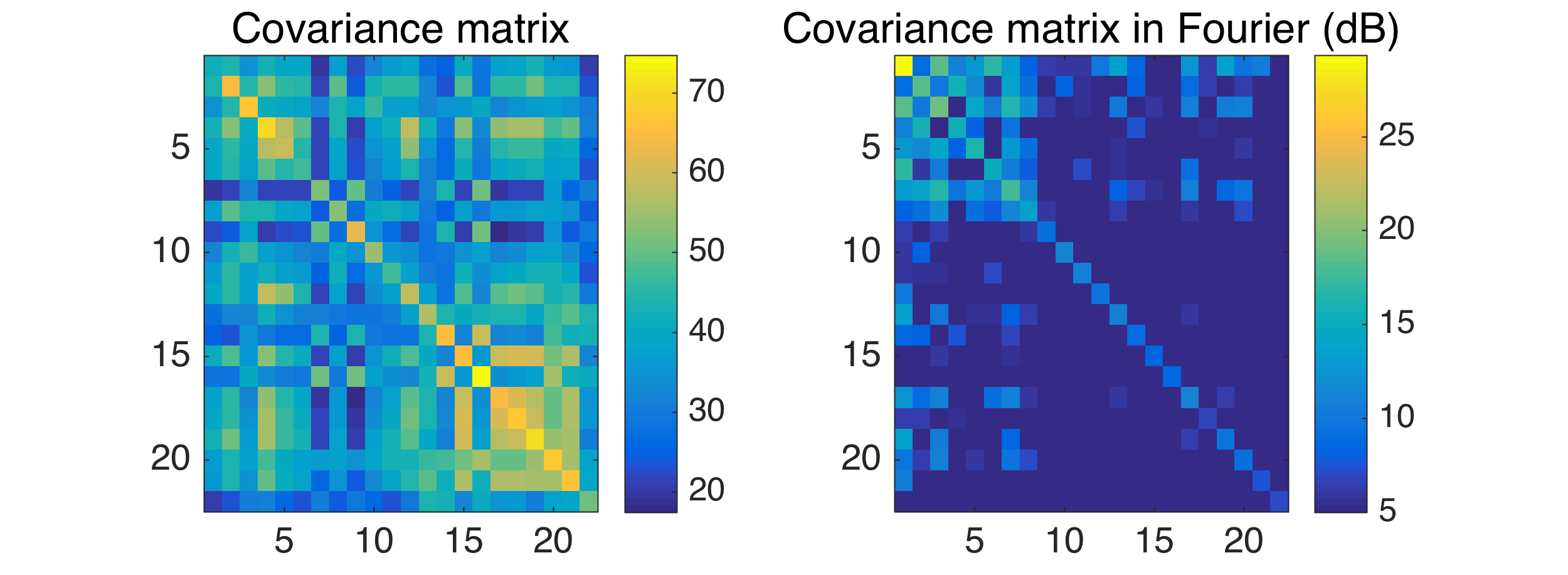} \\
\includegraphics[width=0.45\linewidth]{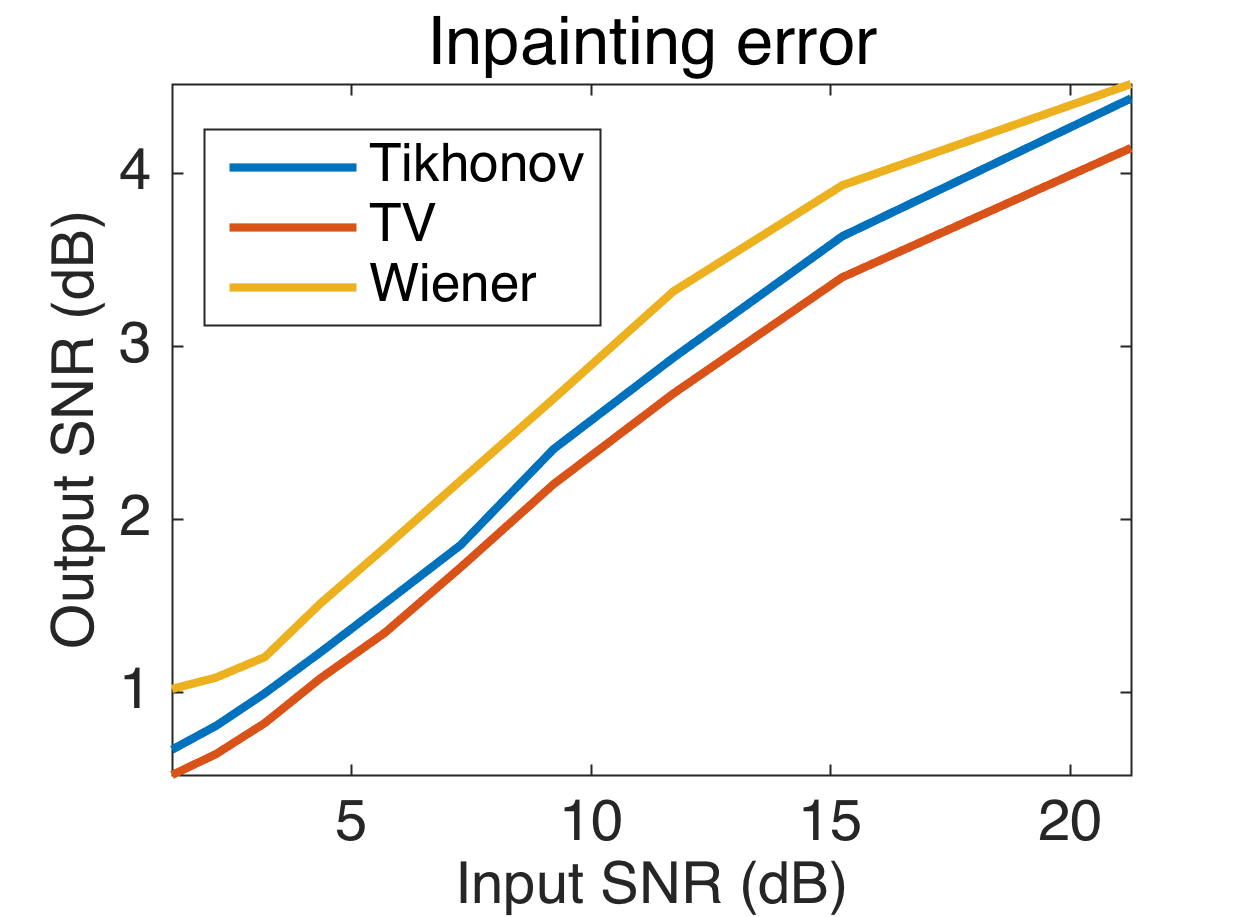} 
\end{center}
\caption{Experiments on the humidity of Molene. Top: Covariance matrices. Bottom: Recovery errors for different noise levels. }
\label{fig:molene-humidity}
\vspace{-0.5cm}
\end{figure}

\paragraph{Prediction - Humidity} Using the same graph, we have performed another set of experiments on humidity observations. The results are displayed in Figure~\ref{fig:molene-humidity}. In our experiment, we solve a prediction problem with a mask operator covering $50\%$ of measurements and various amount of noise. The rest of the testing framework is identical as for the temperature and the conclusions are similar.
 
\subsection{USPS dataset}
\begin{figure}[htb!]
\begin{center}
\includegraphics[width=0.45\linewidth]{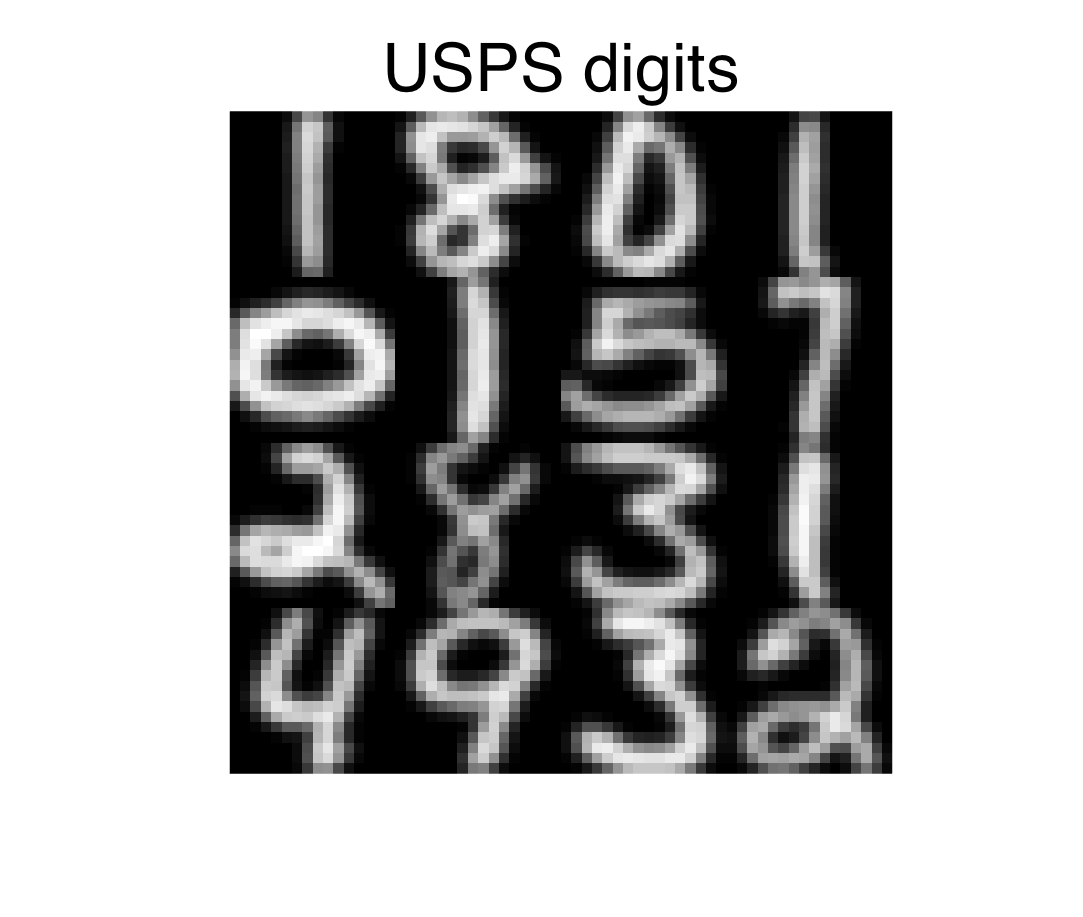} 
\includegraphics[width=0.45\linewidth]{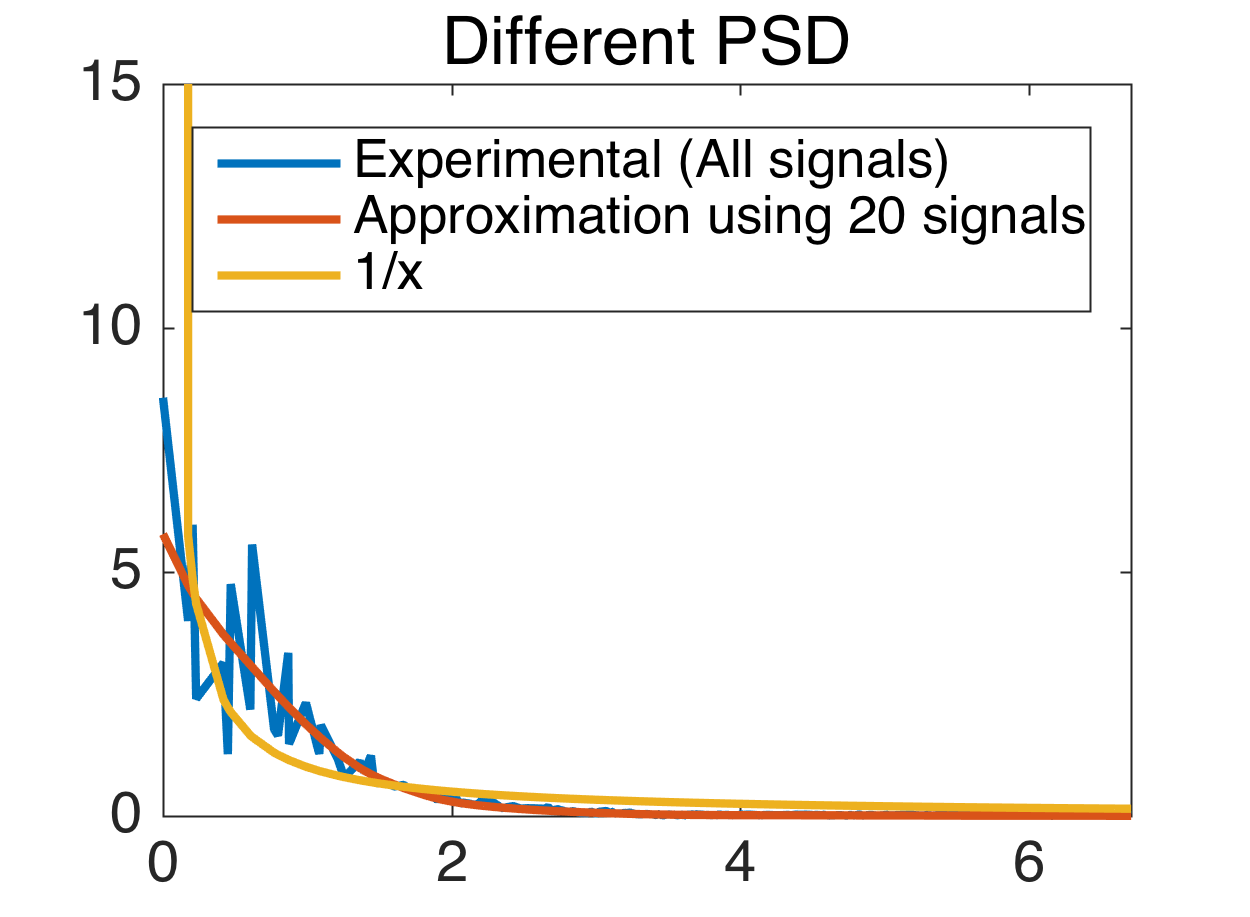} \\
\includegraphics[width=0.9\linewidth]{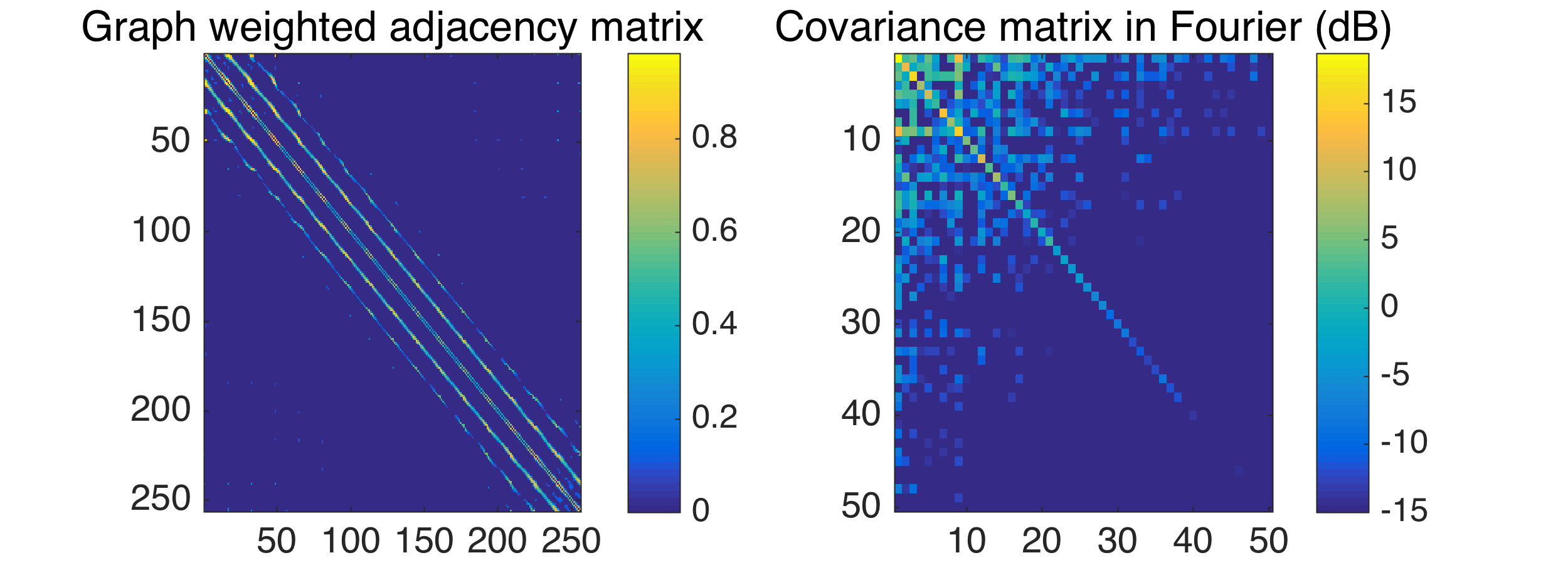} \\
\includegraphics[width=0.9\linewidth]{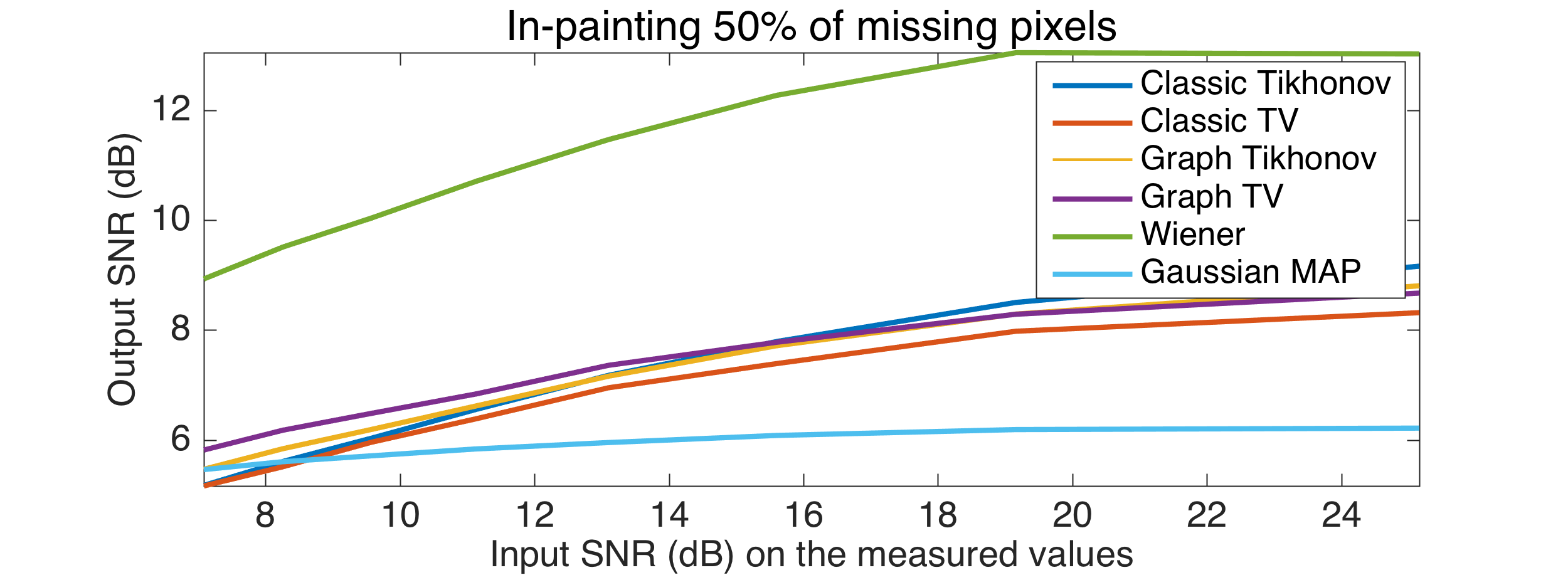} 
\end{center}
\caption{Top left: Some digits of the USPS dataset. Top right: Different PSDs. Compared to $\frac{1}{\lambda}$, the approximation is a smoothed version of the experimental PSD. Middle left: Weights matrix of the $10$ nearest neighbors (patch) graph (The diagonal shape indicates the grid base topology of the graph). Middle right: spectral covariance matrix for the first $50$ graph frequencies. Since we only use $20$ digits for the graph construction, the stationarity level is low. Nevertheless, Wiener optimization outperforms other methods. Bottom: Recovery errors for different noise levels. Methods using the graph perform better. Even if the data is not stationary on the graph, the stationarity assumption helps a lot in the recovery.}
\label{fig:usps_inpainting}
\vspace{-0.5cm}
\end{figure}

We perform the same kind of in-painting/de-noising experiments with the USPS dataset. For our experiments, we consider every digit as an independent realization of a GWSS signal. As sole pre-processing, we remove the mean of each pixel separately. This ensures that the first moment is $0$. We create the graph\footnote{The graph is created using patches of pixels of size $5 \times 5$. The pixels' patches help because we have only a few digits available. When the size of the data increases, a nearest neighbor graph performs even better.} and estimate the PSD using only the first $20$ digits and we use $500$ of the remaining ones to test our algorithm. We use a mask covering $50 \%$ of the pixel and various amount of noise. We then average the result over $500$ experiments (corresponding to the $500$ digits) to obtain the curves displayed in Figure~\ref{fig:usps_inpainting}\footnote{All parameters have been tuned optimally in a probabilistic way. This is possible since the noise is added artificially. The models presented in Appendix~\ref{sec:convex_model} have only one parameter to be tuned: $\epsilon$ which is set to $\epsilon = \sigma \sqrt{ \# y }$, where $\sigma$ is the variance of the noise and $\# y$ the number of elements of the vector $y$. In order to be fair with the MAP estimator, we construct the graph with the only $20$ digits used in the PSD estimation. }.
For this experiment, we also compare with traditional TV de-noising~\cite{chambolle2004algorithm} and Tikhonov de-noising. The optimization problems used are similar to \eqref{prob:inpainting-tik}. Additionally we compute the classical MAP estimator based on the empirical covariance matrix for the solution see (\cite{rasmussen2004gaussian} 2.23).
The results presented in Figure~\ref{fig:usps_inpainting} show that graph optimization is outperforming classical techniques, meaning that the grid is not the optimal graph for the USPS dataset. Wiener once again outperforms the other graph-based models. Moreover, this experiment shows that our PSD estimation is robust when the number of signals is small. In other words, using the graph allows us for a much better covariance estimation than a simple empirical average. When the number of measurements increases, the MAP estimator improves in performance and eventually outperforms Wiener because the data is close to stationary on the graph.

\subsection{ORL dataset}
\begin{figure}[htb!]
\begin{center}
\includegraphics[width=0.9\linewidth]{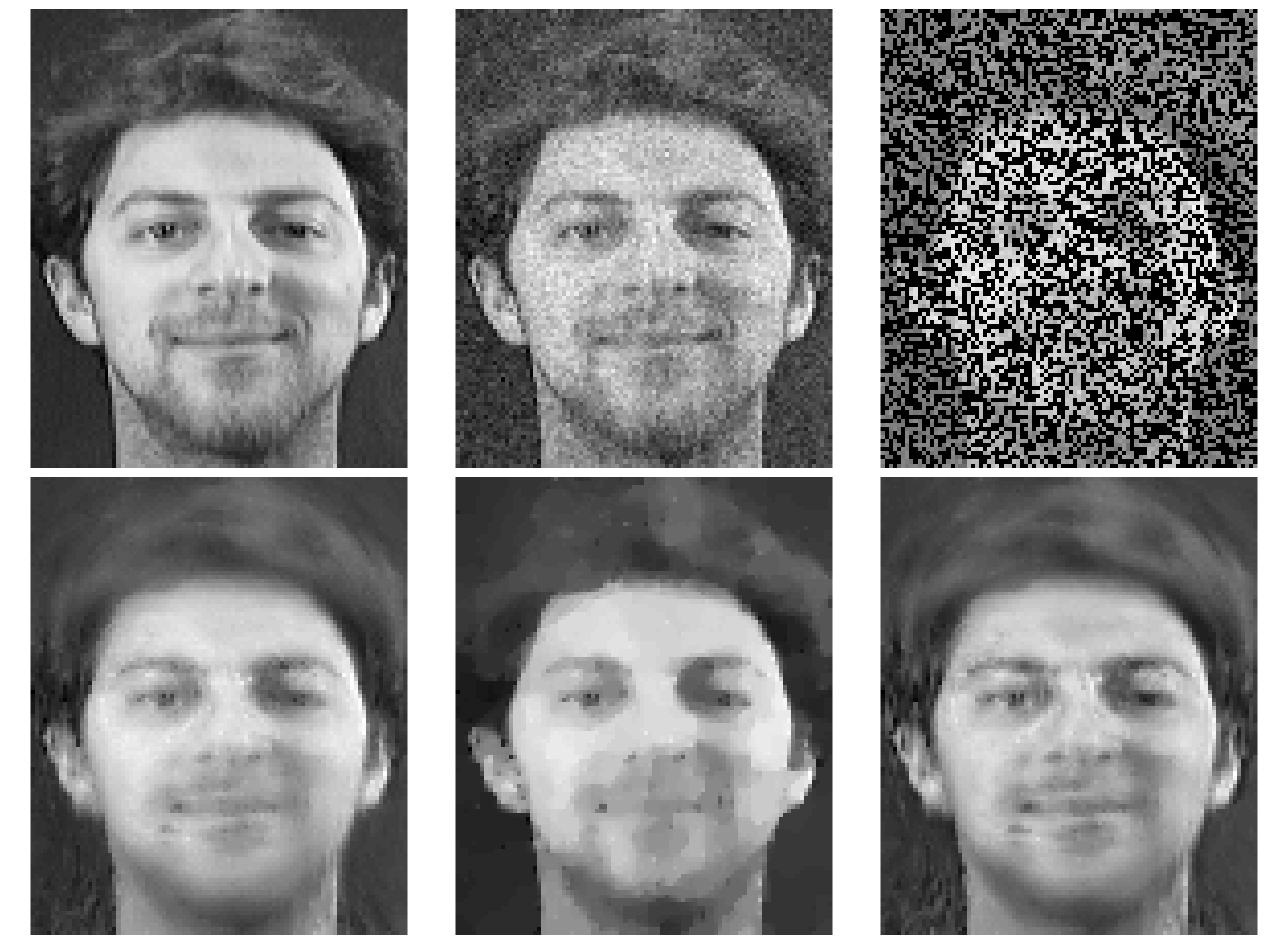} \\
\end{center}
\caption{ORL dataset, single in-painting experiment. Top left: Original image. Top center: Noisy image (SNR $12.43$ dB). Top right: Measurements $50 \% $ of the noisy image. Bottom left: Reconstruction using Tikhonov prior (SNR $12.12$ dB). Bottom center: Reconstruction using classic TV prior (SNR $13.53$ dB). Bottom right: Reconstruction using Wiener optimization (SNR $14.42$ dB). }
\label{fig:orl_single}
\vspace{-0.5cm}
\end{figure}

\begin{figure}[htb!]
\begin{center}
\includegraphics[width=0.45\linewidth]{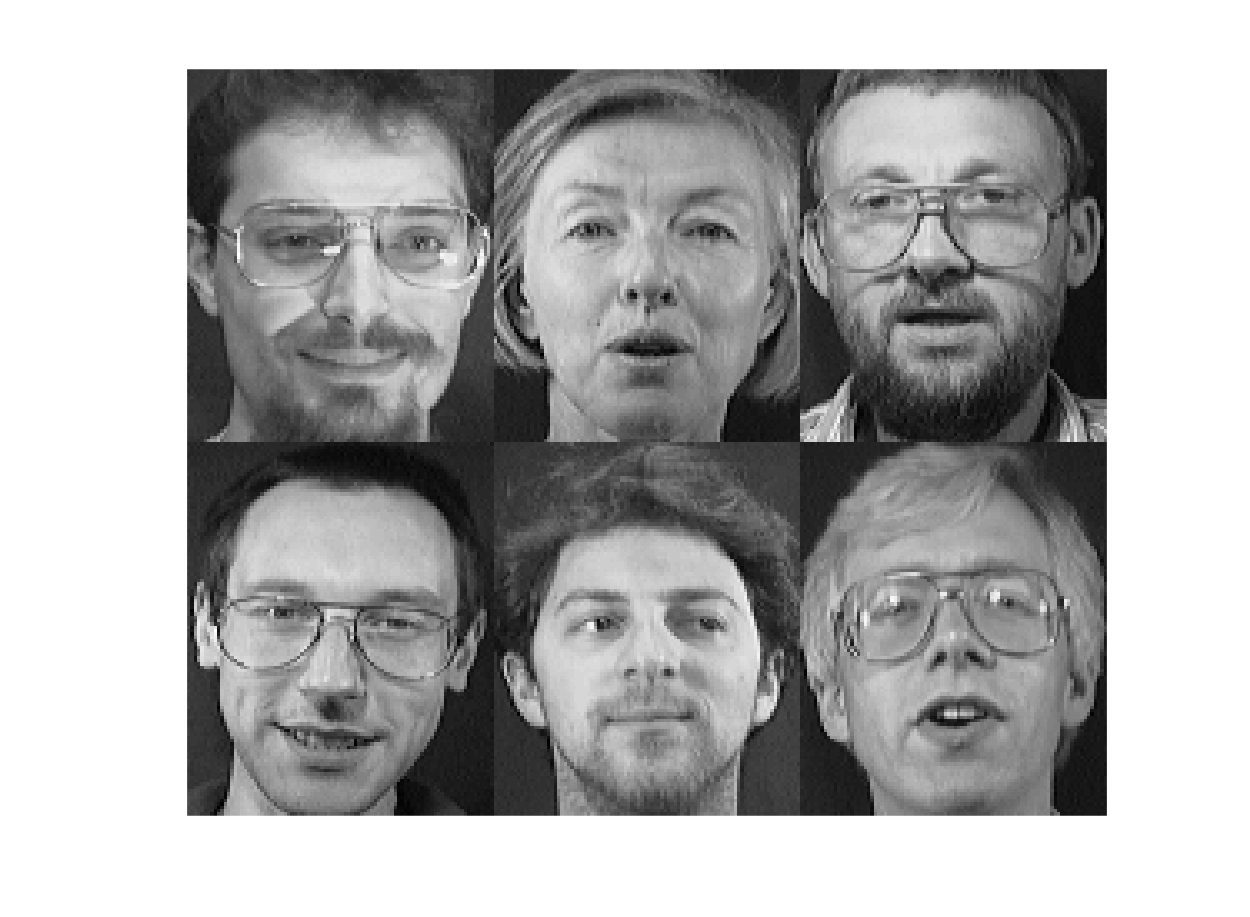}
\includegraphics[width=0.45\linewidth]{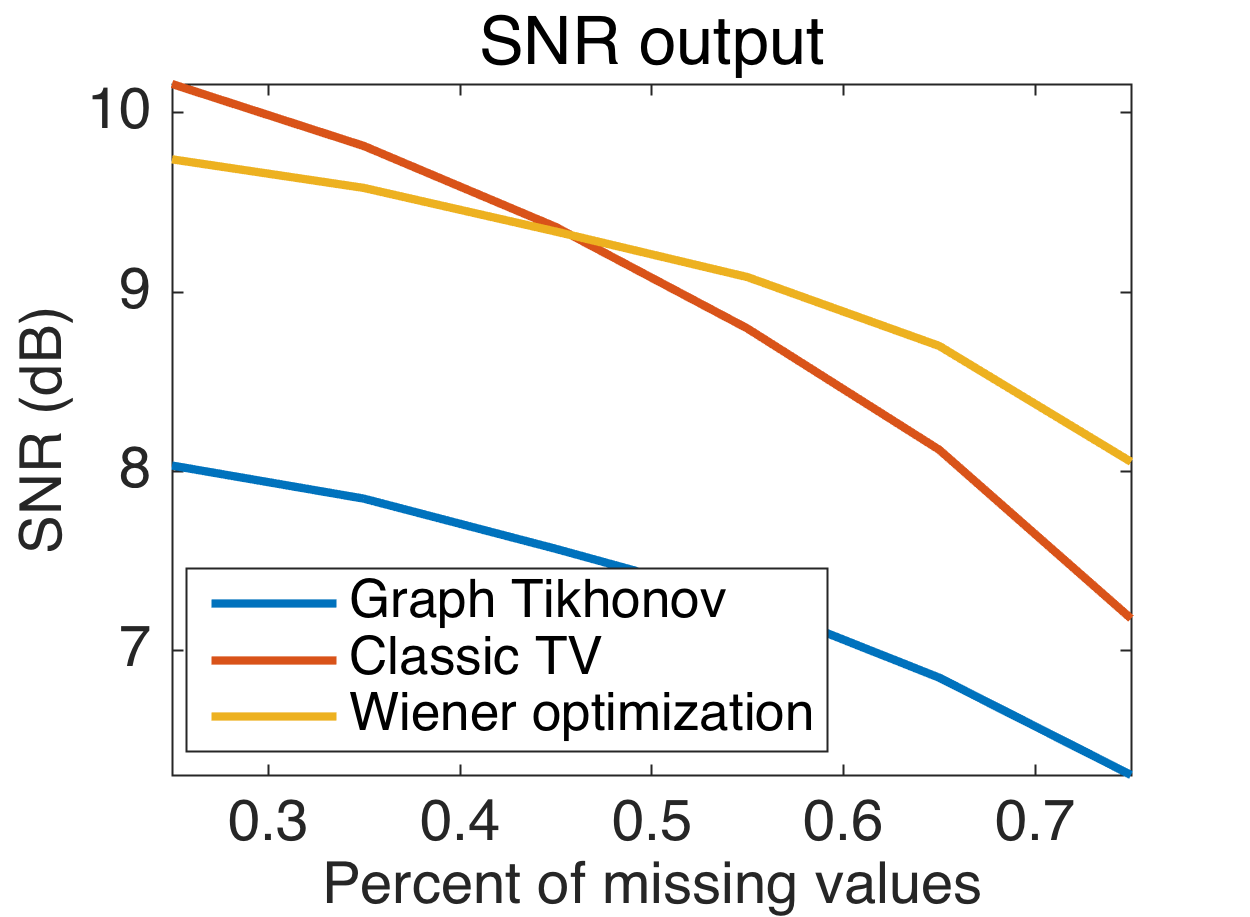} 
\end{center}
\caption{Inpainting experiments on ORL dataset. Left: some images of the dataset. Right: reconstruction error. }
\label{fig:orl_inpainting}
\vspace{-0.5cm}
\end{figure}

For this last experiment, we use the ORL face dataset. We have a good indication that this dataset is close to stationary since CMUPIE (a smaller faces dataset) is also close to stationary. Each image has $112 \times 92=10304$ pixels making it complicated to estimate the covariance matrix and to use a Gaussian MAP estimator. Wiener optimization, on the other hand, does not necessitate an explicit computation of the covariance matrix. Instead, we estimate the PSD using the algorithm presented in Section~\ref{sec:psd_estimation}. A detailed experiment is performed in Figure~\ref{fig:orl_single}. After adding Gaussian noise to the image, we remove randomly a percentage of the pixels. We consider the obtained image as the measurement and we reconstruct the original image using TV, Tikhonov and Wiener priors. In Figure~\ref{fig:orl_inpainting}, we display the reconstruction results for various noise levels. We create the graph with $300$ faces\footnote{We build a nearest neighbor graph based on the pixels values.} and estimate the PSD with $100$ faces. We test the different algorithms on the $100$ remaining faces. 

\section{Conclusion} \label{sec:conclusion}
In this contribution, we have extended the common concept of stationarity to graph signals. Using this statistical model, we proposed a new regularization framework that leverages the stationarity hypothesis by using the Power Spectral Density (PSD) of the signal. Since the PSD can be efficiently estimated, even for large graphs, the proposed Wiener regularization framework offers a compelling way to solve traditional problems such as denoising, regression or semi-supervised learning. We believe that stationarity is a natural hypothesis for many signals on graphs and showed experimentally that it is deeply connected with the popular nearest neighbor graph construction. As future work, it would be very interesting to clarify this connection and explore if stationarity could be used to infer the graph structure from training signals, in the spirit of~\cite{kalofolias2016learn}.

\section*{Acknowledgments}
We thank the anonymous reviewers for their constructive comments that helped us improve the structure of the paper, especially Section III B. We also thank Andreas Loukas, Vassilis Kalofolias and Nauman Shahid for their useful suggestions.

This work has been supported by the Swiss National Science Foundation research project \textit{Towards Signal Processing on Graphs}, grant number: 2000\_21/154350/1.

\appendix

\section{Convex models} \label{sec:convex_model}
Convex optimization has recently become a standard tool for problems such as de-noising, de-convolution or in-painting. Graph priors have been used in this field for more than a decade~\cite{smola2003kernels,zhou2004regularization,peyre2008non}. The general assumption is that the signal varies smoothly along the edges, which is equivalent to saying that the signal is low-frequency-based. Using this assumption, one way to express mathematically an in-painting problem is the following:
\begin{equation} \label{prob:inpainting-tik}
\bar{x} = \argmin_x x^T\Larg x \hspace{0.5cm} \text{ s.t. } \hspace{0.5cm} \|M x - y\|_2\leq \epsilon
\end{equation}
where $M$ is a masking operator and $\epsilon$ a constant computed thanks to the noise level. We could also rewrite the objective function as $x^T \Larg x + \gamma \|M x - y\|_2^2$, but this implies a greedy search of the regularization parameter $\gamma$ even when the level of noise is known. 
For our simulations, we use Gaussian i.i.d. noise of standard deviation $n$. It allows us to optimally set the regularization parameter $\epsilon = n \sqrt{\# y}$, where $\# y$ is the number of elements of the measurement vector. 

Graph de-convolution can also be addressed with the same prior assumption leading to 
\begin{equation} \label{prob:deconvolution-tik}
\bar{x} = \argmin_x x^T\Larg x \hspace{0.5cm} \text{ s.t. } \hspace{0.5cm} \|h(\Larg) x - y\|_2\leq \epsilon
\end{equation}
where $h$ is the convolution kernel. To be as generic as possible, we combine problems \eqref{prob:inpainting-tik} and \eqref{prob:deconvolution-tik} together leading to a model capable of performing de-convolution, in-painting and de-noising at the same time:
\begin{equation} \label{prob:dec-in-den-tik}
\bar{x} = \argmin_x x^T\Larg x \hspace{0.5cm} \text{ s.t. } \hspace{0.5cm} \|M h(\Larg) x - y\|_2\leq \epsilon.
\end{equation}

When the signal is piecewise smooth on the graph, another regularization term can be used instead of $x^T\Larg x = \|\nabla_\G x\|_2^2$, which is the $\ell_2$-norm of the gradient on the graph\footnote{The gradient on the graph is defined as $\nabla_\G x (e) = \frac{1}{2}\sqrt{W(i,j)} \left(x(i)-x(j)\right)$, where $e$ is the index corresponding the edge linking the nodes $i$ and $j$.}. 
Using the $\ell_1$-norm of the gradient favors a small number of major changes in the signal and thus is better for piecewise smooth signals. The resulting model is: 
\begin{equation} \label{prob:dec-in-den-tv}
\bar{x} = \argmin_x \|\nabla_\G x\|_1 \hspace{0.5cm} \text{ s.t. } \hspace{0.5cm} \|M h(\Larg) x - y\|_2\leq \epsilon
\end{equation}

In order to solve these problems, we use a subset of convex optimization tools called proximal splitting methods. Since we are not going to summarize them here, we encourage a novice reader to consult~\cite{combettes2011proximal,komodakis2015playing} and the references therein for an introduction to the field.

\section{Proof of Theorem \ref{theo:map}} \label{sec:proof map}
\begin{proof}
The proof is a classic development used in Bayesian machine learning. 
By assumption $\b{x}$ is a sample of a Gaussian random multivariate signal $\b{x}\sim\mathcal{N}\left(m_{\b{x}},s^{2}(\Larg)\right)$.
The measurements are given by 
\[
\b{y}=H\b{x}+\b{w}_n,
\]
where $\b{w}_n\sim\mathcal{N}\left(0,\sigma^2\right)$ and thus have the following first and second moments:
$\b{y}|\b{x}\sim\mathcal{N}\left(H \b{x},\sigma^{2}I)\right)$. 
For simplicity, we assume $s^{2}(\Larg)$ to be invertible. However this assumption is not necessary.
We can write the
probabilities of $\b{x}$ and $\b{y}|\b{x}$ as: 
\[
\mathbb{P}(\b{x})=\frac{1}{Z_{H\b{x}}}e^{-\|s^{-1}(\Larg)(\b{x}-m_{\b{x}})\|_{2}^{2}}
=\frac{1}{Z_{H\b{x}}}e^{-\|s^{-1}(\Larg)\tilde{\b{x}}\|_{2}^{2}},
\]
\[
\mathbb{P}(\b{y}|\b{x})=\frac{1}{Z_{s\b{x}}}e^{-\sigma^{2}\left\Vert (H\b{x}-\b{y})\right\Vert_{2}^{2}}.
\]
Using Bayes law, we find 
\[
\mathbb{P}(\b{x}|\b{y})=\frac{\mathbb{P}(\b{y}|\b{x})\mathbb{P}(\b{x})}{\mathbb{P}(\b{y})}.
\]
The MAP estimator is
\begin{eqnarray*}
\bar{\b{x}}|\b{y} & = & \argmax_{\b{x}}\mathbb{P}(\b{x}|\b{y})\\
 & = & \argmax_{\b{x}}\log\left(\mathbb{P}(\b{x}|\b{y})\right)\\
 & = & \argmin_{\b{x}}-\log\left(\mathbb{P}(\b{y}|\b{x})\right)-\log\left(\mathbb{P}(\b{x})\right)+\log\left(\mathbb{P}(\b{y})\right)\\
 & = & \argmin_{\b{x}}\|s^{-1}(\Larg)\tilde{\b{x}}\|_{2}^{2}+\sigma^{-2}\|(H\b{x}-\b{y})\|_{2}^{2}\\
 & = & \argmin_{\b{x}}\|w(\Larg)\tilde{\b{x}}\|_{2}^{2}+\|(H\b{x}-\b{y})\|_{2}^{2},
\end{eqnarray*}
where $w(\Larg)=\sigma s^{-1}(\Larg).$ 
\end{proof}

\section{Proof of Theorem \ref{theo:optimality}} \label{sec:proof of optimality}
The following is a generalization of the classical proof. For simplicity, we assume that $m_{\tilde{\b{x}}} = 0$, i.e $\tilde{\b{x}}=\b{x}$.
\begin{proof}
Since by hypothesis $H=h(\Larg) = U h(\Lambda) U^*$, we can rewrite the optimization problem \eqref{prob:Wiener-opt} in the graph Fourier domain using the Parseval identity $\|\b{x}\|_2 = \|U\b{x}\|_2 = \|\hat{\b{x}}\|_2$: 
\[
\hat{\bar{\b{x}}}|\hat{\b{y}} = \arg\min_{\hat{\b{x}}}\|w(\Lambda)\hat{\b{x}}\|_{2}^{2}+\|h(\Lambda)\hat{\b{x}}-\hat{\b{y}}\|_{2}^{2}.
\]
Since the matrix $h(\Lambda)$ is diagonal, the solution of this problem satisfies for all graph eigenvalue $\ll$
\begin{equation} \label{eq:app_optimality_constraint}
w^{2}(\ll)\hat{\bar{\b{x}}}[\ell]+h^{2}(\ll)\hat{\bar{\b{x}}}[\ell]-h(\ll)\hat{\b{y}}[\ell]=0 .
\end{equation}
For simplicity, we drop the notation $(\ll)$ and $[\ell]$. The previous equation is transformed in   
\[
\bar{\b{x}}=\frac{h}{w^{2}+h^{2}}\hat{\b{y}}.
\]
As a next step, we use the fact that
$\hat{\b{y}}=h\hat{\b{x}}+\hat{\b{w}}_n$ to find:
\[
\hat{\bar{\hat{\b{x}}}}=\frac{h^{2}\hat{\b{x}}+h\hat{\b{w}}_n}{w^{2}+h^{2}}.
\]
The error performed by the algorithm becomes
\[
\hat{\b{e}} = \bar{\hat{\b{x}}}-\hat{\b{x}} 
=\frac{-w^{2}\hat{\b{x}}}{w^{2}+h^{2}}+\frac{h\hat{\b{w}}_n}{w^{2}+h^{2}} .
\]
The expectation of the error can thus be computed:
\begin{eqnarray*}
\Esp\left\{\hat{\b{e}}^{2}\right\}
 & = & \frac{w^{4}\Esp\left\{\hat{\b{x}}^{2}\right\}}{\left(w^{2}+h^{2}\right)^{2}}+\frac{h^{2}\Esp\left\{\hat{\b{w}}_n^{2}\right\}}{\left(w^{2}+h^{2}\right)^{2}}-\frac{hw^{2}\Esp\left\{\hat{\b{x}}\hat{\b{w}}_n\right\}}{\left(w^{2}+h^{2}\right)^{2}}\\
 & = & \frac{w^{4}s^2+h^{2}n}{\left(w^{2}+h^{2}\right)^{2}},
\end{eqnarray*}
with $s^2$ the PSD of $\b{x}$ and $n$ the PSD of the noise $\b{w}_n$. Note that $\Esp\left\{\hat{x} \hat{w}_n\right\} = 0$ because $\b{x}$ and $\b{w}$ are uncorrelated. Let us now substitute $w^{2}$ by $z$ and minimize the expected error (for each $\ll$) with respect to $z$:
\begin{eqnarray*}
\frac{\partial}{\partial z}\Esp\left\{\hat{\b{e}}^{2}\right\} & = & \frac{\partial}{\partial z}\frac{z^2 s^2+h^{2}n}{\left(z+h^{2}\right)^{2}}\\
 & = & \frac{2zs^2\left(z+h^{2}\right)-2\left(z^{2}s^2+h^{2}n\right)}{\left(z+h^{2}\right)^{3}} = 0.
\end{eqnarray*}
From the numerator, we get:
\[
2zs^2h^{2}-2h^{2}n=0
\]
The three possible solutions for $z$ are $z_{1} =  \frac{n}{s^2}$, $z_{2}=\infty$ and $z_{3}=-\infty$. $z_3$ is not possible because $z$ is required to be positive. $z_2$ leads to $\dot{x} = 0$ which is optimal only if $s^2=0$. The optimal solution is therefore $z(\ll) = \frac{n(\ll)}{s^2(\ll)}$, resulting in 
$$
w(\ll) =\sqrt{\frac{n(\ll)}{s^2(\ll)}}.
$$
This finishes the first part of the proof. To show that the solution to \eqref{prob:Wiener-opt} is a Wiener filtering operation, we replace $w^2(\ll)$ by $\frac{n(\ll)}{s^2(\ll)}$ in \eqref{eq:app_optimality_constraint} and find
$$
\hat{\b{x}}[\ell] = \frac{s^2(\ll)h(\ll)}{h^2(\ll)s^2(\ll)+n(\ll)}\hat{\b{y}}[\ell],
$$
which is the Wiener filter associated with the convolution $h(\Larg)=H$.
\end{proof}

\section{Proof of Theorem \ref{theo:lmmse}} \label{sec:proof lmmse}

\begin{proof}
Let $\b{x}$ be GWSS with covariance matrix $\Sigma_{\b{x}} = s^2(\Larg)$ and mean $m_\b{x}$. The measurements satisfy
$$
\b{y} = H\b{x}+\b{w}_n,
$$
where $\b{w}_n$ is i.i.d noise with PSD $\sigma^2$.
The variable $\b{y}$ has a covariance matrix $\Sigma_{\b{y}} = Hs^2(\Larg)H^*+\sigma^2I$ and a mean $m_{\b{y}} = Hm_\b{x}$. The covariance between $\b{x}$ and $\b{y}$ is $\Sigma_{\b{x}\b{y}} = \Sigma_{\b{y}\b{x}}^* = s^2(\Larg)H^*$. 
For simplicity, we assume $s^2(\Larg)$ and $Hs^2(\Larg)H^*+\sigma^2I$ to be invertible. However this assumption is not necessary. 
The Wiener optimization framework reads:
\[
\bar{\b{x}}=\argmin_{\b{x}}\|H\b{x}-\b{y}\|_{2}^{2}+\sigma^{2}\|s^{-1}(\Larg)(\b{x}-m_{\b{x}})\|_{2}^{2}.
\]
We perform the following change of variable $\tilde{\b{x}}=\b{x}-m_{\b{x}}$, $\tilde{\b{y}}=\b{y}-m_{\b{x}}$ and we obtain:
\[
\bar{\tilde{\b{x}}}=\argmin_{\tilde{\b{x}}}\|H\tilde{\b{x}}-\tilde{\b{y}}\|_{2}^{2}+\sigma^{2}\|s^{-1}(\Larg)\tilde{\b{x}})\|_{2}^{2}.
\]
The solution of the problem satisfies
\[
H^{*}H\bar{\tilde{\b{x}}}-H^{*}\tilde{\b{y}}+\sigma^{2}s^{-2}(\Larg)\bar{\tilde{\b{x}}}=0.
\]
From this equation we get $\bar{\tilde{\b{x}}}$ and transform it as:
\begin{eqnarray}
\bar{\tilde{\b{x}}} & = & \left(H^{*}H+\sigma^{2}s^{-2}(\Larg)\right)^{-1}H^{*} \tilde{\b{y}} \nonumber \\
 & = & s(\Larg)\left(\sigma^{2}I+s(\Larg)H^{*}Hs(\Larg)\right)^{-1}s(\Larg)H^{*}\tilde{\b{y}}\nonumber \\
 & = & {\scriptstyle \left(\frac{1}{\sigma^{2}}s^{2}(\Larg)H^{*}-\frac{1}{\sigma^{2}}s^{2}(\Larg)H^{*}\left(\sigma^{2}I+H^{*}s^{2}(\Larg)H\right)^{-1}Hs^{2}(\Larg)H^{*}\right)\tilde{\b{y}}\label{eq:demo-woodbury} }\\
 & = & {\displaystyle \frac{1}{\sigma^{2}}s^{2}(\Larg)H^{*}\left(I-\left(\sigma^{2}I+H^{*}s^{2}(\Larg)H\right)^{-1}Hs^{2}(\Larg)H^{*}\right)\tilde{\b{y}} }\nonumber\\
 & = & s^{2}(\Larg)H^{*} \left(\sigma^{2}I+H^{*}s^{2}(\Larg)H\right)^{-1} \tilde{\b{y}}\nonumber\\
 & = & \Sigma_{\b{x}\b{y}}  \Sigma_{\b{y}}^{-1} \tilde{\b{y}}\nonumber
\end{eqnarray}
where \eqref{eq:demo-woodbury} follows from the Woodbury, Sherman and Morrison formula. 
The linear estimator of $\b{x}$ corresponding to Wiener optimization is thus:
\begin{eqnarray*}
\bar{\b{x}} & =  & \bar{\tilde{\b{x}}} + m_{\b{x}} \\
& =  & \Sigma_{\b{x}\b{y}}  \Sigma_{\b{y}}^{-1} (\b{y} - H m_{\b{x}} ) + m_{\b{x}} \\
& =  & \Sigma_{\b{x}\b{y}}  \Sigma_{\b{y}}^{-1}\b{y} + \left( I- \Sigma_{\b{x}\b{y}}  \Sigma_{\b{y}}^{-1}H\right) m_{\b{x}} \\
& =  & Q \b{y} + \left( I- Q H\right) m_{\b{x}} 
\end{eqnarray*}
We observe that it is equivalent to the solution of the linear minimum mean square error estimator:
\begin{equation*}
\argmin_{Q,b} \Esp \left\{ \| Q \b{y}+ b - \bar{x} \|\right\}^2
\end{equation*}
with $\b{y} = H \b{x} + \b{w}_n$. See \cite[Equation 12.6]{kay1993fundamentals}.
\end{proof}
Using similar arguments, we can prove that 
\[
\bar{\b{x}}=\arg\min_{x}\|s^{-1}(\Larg)(\b{x}-m_{\b{x}}\|_{2}^{2}\hspace{1em}\mbox{s.t. }y=Xx
\]
leads to
$$
\bar{\b{x}} = s^2(\Larg) \left(Hs^2(\Larg)H^*\right)^{-1} \b{y} + \left( I- s^2(\Larg) \left(Hs^2(\Larg)H^*\right)^{-1} H\right) m_{\b{x}} 
$$
and is thus a linear minimum mean square estimator too.

\section{Development of equation~\ref{eq:link_gram} } \label{sec:proof_gram}
\begin{proof}
Let us denote the matrix of squared distances $D[i,j] = \tfrac{1}{K} \sum_k |x_k[i]-x_k[j]|^2 $ for the samples $\{x_1,x_2,\ldots x_{K}\}$ of the random multivariate variable $\b{x}$ on a $N$ vertex graph. Let us assume further that $m[k] =  \sum_{n=1}^{N} x_k[n] =0$. We show then that $ \bar{\Sigma}_{\b{x}} = - \tfrac{1}{2} J D_{\b{x}} J$ where $\bar{\Sigma}_{\b{x}}$ is the covariance (Gram) matrix defined as $\bar{\Sigma}_{\b{x}}[i,j] = \tfrac{1}{K} \sum_{k=1}^K x_k[i] x_k[j]$ and $J$ is centering matrix $J[k,l]=\delta_k[l] - \tfrac{1}{N}$. 

We have
\begin{footnotesize}
$$
(JD J) [i,j] = D[i,j]+N^{-2}\sum_{k,l=1}^{N} D[k,l]-N^{-1} \sum_{k=1}^{N} (D[i,k]+D[k,j])
$$
\end{footnotesize}
Let us substitute $D[i,j]=\bar{\Sigma}_{\b{x}}[i,i]+\bar{\Sigma}_{\b{x}}[j,j] -2 \bar{\Sigma}_{\b{x}}[i,j]$, then we find
\begin{footnotesize}
\begin{eqnarray*}
& & (JD J) [i,j] \\
& = & \bar{\Sigma}_{\b{x}}[i,i] + \bar{\Sigma}_{\b{x}}[j,j] - 2 \bar{\Sigma}_{\b{x}}[i,j] + N^{-2} \left(2 N \sum_{n=1}^N \bar{\Sigma}_{\b{x}}[n,n] - 2 m^* m \right) \\
 &- & N^{-1}\left(N \bar{\Sigma}_{\b{x}}[i,i] + N \bar{\Sigma}_{\b{x}}[j,j] + 2 \sum_{n=1}^N \bar{\Sigma}_{\b{x}}[n,n] - 2 m^* (x[j] + x[i]) \right) \\
& = & - 2 \bar{\Sigma}_{\b{x}}[i,j] - 2 N^{-2} m^* m + 2 N^{-1} m^* (x[j] +x[i]).
\end{eqnarray*}
\end{footnotesize}
Under the assumption $m[k] = \sum_{n=1}^{N}x_k[n]=0$, we recover the desired result $ \bar{\Sigma}_{\b{x}} = - \tfrac{1}{2} J D_{\b{x}} J$.
\end{proof}

\bibliographystyle{IEEEtran}
\bibliography{bibliography}

\end{document}